\documentclass[11pt]{article}
\usepackage[latin9]{inputenc}
\usepackage{geometry}
\usepackage[english]{babel}
\usepackage{array}
\usepackage{float}
\usepackage{multirow}
\usepackage{amsmath}
\usepackage{amsthm}
\usepackage{amssymb}
\usepackage{graphicx}
\usepackage[authoryear]{natbib}
\usepackage[unicode=true,pdfusetitle,
 bookmarks=true,bookmarksnumbered=false,bookmarksopen=false,
 breaklinks=false,pdfborder={0 0 0},pdfborderstyle={},backref=false,colorlinks=false]
 {hyperref}

\makeatletter

\providecommand{\tabularnewline}{\\}
\floatstyle{ruled}
\newfloat{algorithm}{tbp}{loa}
\providecommand{\algorithmname}{Algorithm}
\floatname{algorithm}{\protect\algorithmname}

\theoremstyle{definition}
\newtheorem{defn}{\protect\definitionname}
\theoremstyle{definition}
 \newtheorem{example}{\protect\examplename}
\theoremstyle{plain}
\newtheorem{prop}{\protect\propositionname}
\theoremstyle{plain}
\newtheorem{lem}{\protect\lemmaname}

\bibpunct{(}{)}{,}{a}{,}{,}
\usepackage{url}
\theoremstyle{plain}

\usepackage{lscape}
\usepackage[bottom]{footmisc}

\newtheorem{assumption}{Assumption}

\newtheorem{ass}{Assumption}
\newenvironment{assbis}[1]
  {\renewcommand{\theass}{\ref{#1}$'$}%
   \addtocounter{ass}{-1}%
   \begin{ass}}
  {\end{ass}}
\makeatother

\providecommand{\definitionname}{Definition}
\providecommand{\examplename}{Example}
\providecommand{\lemmaname}{Lemma}
\providecommand{\propositionname}{Proposition}

\begin{document}
\title{When do firms sell high durability products?\\
The case of light bulb industry\thanks{This paper is based on Chapter 2 of my PhD dissertation at Graduate School of Economics, the University of Tokyo. I thank my adviser Hiroshi Ohashi for sharing proprietary market data used in this study, and for his continuous support and comments. I also appreciate my dissertation committee members, and the comments from and discussions with Gautam Gowrisankaran, Mitsuru Igami, Toshiaki Iizuka, Hiroyuki Kasahara, Kohei Kawaguchi, John Rust, Bertel Schjerning, Katsumi Shimotsu, Yuta Toyama, Yangkeun Yun, and the participants of JEMIOW fall 2023, APIOC 2023, IIOC 2024, JEA 2024 Spring, AEW, and Hosei.}}
\author{Takeshi Fukasawa\thanks{Waseda Institute for Advanced Study, Waseda University. 1-21-1, Nishiwaseda, Shinjuku, Tokyo, Japan. E-mail: fukasawa3431@gmail.com.}}
\maketitle
\begin{abstract}
This study empirically investigates firms' incentives on the choice of product durability, and its social optimality, by developing a dynamic structural model of durable goods with forward-looking consumers and oligopolistic multi-product firms. Based on the observations of the light bulb market, it specifies a model where firms produce multiple products with different durability levels and set product prices based on dynamic incentives. It proposes and applies novel estimation algorithms that alleviate the computational burden and data requirement for estimating demand and marginal cost parameters of dynamic demand models. Using light bulb market data in Japan, structural parameters are estimated.

This study obtains the following results. First, large firms have incentives to collude to eliminate high durability incandescent lamps, though it is profitable to sell them for each firm. In contrast, when they can collude on prices, they don't have incentives to eliminate high durability bulbs. Second, eliminating high durability incandescent lamps leads to larger producer and total surplus, though it leads to lower consumer surplus. %In the oligopolistic environment, firms do not internalize the effect of own firms' product durability on competitors' future demand.//

\end{abstract}
\pagebreak{}

\section{Introduction\label{sec:Introduction}}

Product durability affects not only consumers' purchase decisions but also durable goods producers' long-run profits through the change in future replacement demand. Since high durability of products implies smaller future replacement demand, durable goods firms might have incentives for not selling high durability products.

This study develops and estimates a dynamic structural model of durable goods with forward-looking consumers and forward-looking oligopolistic multi-product firms using light bulb market data in Japan. It applies novel methods for the estimation of demand and marginal cost parameters with smaller computational burden and data requirement, which have been the obstacles to the analysis of dynamic demand models. It then investigates firms' incentives mentioned above. More specifically, it mainly answers the following questions: 
\begin{enumerate}
\item Market structure / Collusion: When do firms have incentives for not selling high durability products?
\item Welfare: Are the durability levels of products set socially optimal? 
\end{enumerate}
Regarding the first research question, this study evaluates firms' incentives with / without collusion on durability / prices. Recent studies have pointed out the role of collusion in non-price dimensions (\citet{bourreau2021market}, \citet{ale2021colluding}),\footnote{\citet{ale2021colluding} pointed out the following:

\textsl{Collusion on prices is known to be illegal and frequently prosecuted, while collusion on technology choices is less well-defined and rarely prosecuted.}} and it is not unusual for durability, considering the existence of Phoebus cartel in 1920s-30s in the light bulb industry explained in detail in Section 3.2. The results show that large firms have incentives to collude to eliminate high durability incandescent lamps, though it is profitable to sell them for each firm. In contrast, when they can collude on prices, they don't have incentives to eliminate high durability bulbs. The former can be explained by each firm's decision not internalizing the effect of own product durability on competitors' profit. The latter can be explained by firms' incentives to raise product durability so as to increase ``service demand'' through the increase in consumer inventory. The incentives get large when they can set high prices, as in the case of collusion on prices. This study also develops a theoretical model extending the ones in the previous studies to explain the results.

Regarding the second question, this study develops a quantitative method to evaluate social optimality of product durability allowing for consumer surplus, producer surplus, and environmental externalities, based on the structural model. The results show that eliminating high durability incandescent lamps in the sample periods leads to larger producer and total surplus, though it leads to lower consumer surplus. In the oligopolistic environment, firms do not internalize the effect of own firms' product durability on competitors' profit, and it might lead to social overprovision of durability.

Understanding of firms' incentives on product durability is important for competition and consumer policies. So far, there have been several doubtful cases where firms with market power intentionally lowered product durability, which is known as ``planned obsolescence''. Examples include Phoebus cartel in the light bulb industry and Apple's iPhone. In the former example, the cartel was investigated by the competition authorities, and in the latter example Apple was sued by consumer groups and courts in some countries forced the company to pay fines.\footnote{Consumers found that running speed of old iPhones got slower after the Apple's release of a new product in 2017. Though Apple denied that it was intended to foster more replacement, Apple was fined 25 million euros in France. In the U.S., it had to pay up to 500 million dollars to compensate the consumers affected by the slowdown. For details, see \citet{BBC_Apple_news_2019}.} To design socially desirable policies related to durability, deeper understanding of firms' incentives and their economic consequences is essential.

The understanding is also important for environmental policy. When goods are less durable, products break more frequently and more wastes are emitted, causing negative externality.\footnote{Though recycling might be possible for some products, recycling process itself requires much energy.} In addition, lower durability implies more production, and it might lead to more CO2 emission especially for products emitting large amount of CO2 in the production stage.\footnote{Examples include electric vehicles.} Besides, lower durability and subsequent larger production generally leads to more use of scarce natural resources on the earth, especially when the products are not recyclable. Considering these negative effects of low durability products,\footnote{Note that higher durability of products is not always good to the environment for all the products, as discussed in \citet{cooper2010longer}. For instance, higher durability products which are harder to recycle than lower durability products might emit more wastes.} some countries introduced new policies aiming at circular economy. For instance, EU announced Circular Economy Action Plan in 2020, and it prohibits firms from intentionally lowering product durability levels. United Nations (\citet{UN_PLE}) also published a document describing policy instruments for extending product lifetimes. To appropriately design policies, understanding of firms' incentives on product durability and evaluation of social optimality of product durability are indispensable.

Light bulb industry is suitable for the analysis. First, light bulb products have clear measure of durability, in the form of product lifetime. They are shown in the label of product packages, and consumers can easily see how long they will last on average. Second, light bulb is treated as a typical example of durable goods in the literature (\citet{swan1970durability}, \citet{tirole1988theory}), probably because of its simple structure on durability. Third, Phoebus cartel existed, and firms actually lowered product durability levels in history. Finally, heterogeneity in durability exists in the light bulb market. For instance, average lifetimes of some bulbs are 1000 hours, but those of some others are 2000 hours. If no heterogeneity in durability levels exists, it is not easy to identify consumers' preferences on durability, and conduct counterfactual simulation on durability. For these reasons, light bulb market is ideal for the empirical analysis on durability. 

In short, the contribution of this study to the literature is threehold. First, this is the first empirical study explicitly studying oligopolistic firms' incentives on product durability. This study stresses the role of collusion on durability to increase firms' profits. Second, this study develops a quantitative method to assess the social optimality of product durability, allowing for consumer surplus, producer surplus, and externalities, which is becoming more important due to the growing interest in circular economy and product lifetime extension from environmental perspective. Third, this study proposes and applies novel algorithms for estimation of demand and marginal cost parameters with smaller computational burden and smaller data requirement. One large obstacle to the analysis of dynamic demand models is the computational cost, and the methods applied here ease the analysis.

\medskip{}

The rest of this paper is organized as follows. In Section \ref{sec:Literature}, we discuss in detail how the current study relates and contributes to the previous studies. In Section \ref{sec:Industry}, we describe the data and history of light bulb industry. In Section \ref{sec:Model}, we develop a dynamic structural model of durable goods with forward-looking consumers and oligopolistic multi-product firms. In Section \ref{sec:Estimation}, we discuss the estimation methods, and estimation results are shown in Section \ref{sec:Estimation-results}. In Section \ref{sec:Counterfactual}, we show the results of counterfactual simulations. Section \ref{sec:Conclusions} finally concludes. Appendix \ref{sec:Estimation_Counterfactuals_detail} describes the details of the estimations and counterfactuals. All the proofs are shown in Appendix \ref{sec:Proof}. Appendix \ref{sec:Data-details} describes the details of the data, Appendix \ref{sec:Additional-results} shows additional results, and Appendix \ref{sec:Considerations-specifications} discusses further considerations on the specifications. 

\section{Literature and Contributions\label{sec:Literature}}

This study relates and contributes to several strands of literature.

\subsection{Durable goods, durability, and firms}

First, this study contributes to the literature of durable goods firms' behavior, by empirically studying firms' incentives on product durability under an oligopolistic environment. So far, many theoretical studies have investigated how firms determine product durability, and discussed its welfare implications (e.g., \citet{swan1970durability}, \citet{bulow1986economic}, \citet{rust1986optimal}, and \citet{hendel1999interfering}).\footnote{Recently, \citet{Li2024} theoretically showed that the existence of present-biased consumer preferences leads to below-efficient durability levels under perfect competition and monopoly. Based on the theoretical results, they speculated that one reason behind the Phoebus cartel in the light bulb industry might be present-biased consumer preferences, implicitly assuming that the cartel led to below-efficient durability levels. \\
Though insightful, they did not explicitly consider the oligopolistic environment. The current study empirically and theoretically shows the possibility of incentives of collusion and overprovision of durability under oligopolistic environment, and discusses factors affecting the durability levels. } Though insightful, it is known that whether the durability level a firm chooses is socially optimal and affected by the market structure largely depends on the demand structure, as discussed in \citet{schmalensee1979market}. Besides, firms incentives under an oligopolistic market environment is less clear, as pointed out in \citet{waldman2003durable}.\footnote{In the literature, \citet{bulow1986economic} also analyzed oligopolistic firms' endogenous durability choice under the demand function applied in \citet{swan1970durability} and others, but the results are ambiguous. \citet{sasaki2008collusion} discussed the role of durability in the sustainability of price collusion.} This study contributes to the literature by developing an empirically relevant model and showing oligopolistic firms' incentives on durability, stressing the role of collusion on durability. 

In recent years, several empirical studies have investigated the supply-side behavior of durable goods firms, following the development of estimation and simulation of dynamic structural models. \citet{nair2007intertemporal} investigated the pricing decisions of monopolistic durable goods firms using market data of video games. \citet{goettler2011does} examined durable goods producers' innovation decisions in the microprocessor industry. \citet{chen2013secondary} investigated whether the existence of the used goods market harms durable goods firms in the automobile market.\footnote{Besides, \citet{Izuka2007empirical} showed reduced-form empirical evidence that publishers revise textbooks more frequently when competition from used textbooks increases, which is consistent with the conventional wisdom that durable goods producers introduce new products to kill off used products.} However, in these studies, oligopolistic firms' incentives on built-in durability are not necessarily explicitly investigated.\footnote{As additional simulations, \citet{goettler2011does} and \citet{chen2013secondary} investigated the counterfactual outcomes where the durability levels of all the products were exogenously changed. Nevertheless, whether such exogenous durability changes were empirically possible is not clear.} This study complements these studies by investigating oligopolistic firms' endogenous decisions on products' built-in durability and by proposing novel estimation algorithms that make the empirical analysis on durable goods easier. 

Note that the current study is not the first that investigates the light bulb industry from the viewpoint of firms' incentives on durability. \citet{prais1974electric} and \citet{swan1982less} investigated the industry, by using only light bulb product characteristics data. They estimated the technological relationship between lifetime, luminosity, and wattage of products, and investigated whether the durability level firms chose was socially optimal without explicitly specifying demand structure. \citet{avinger1981product} discussed electric lamps as a case study of the question whether market structure affected firms' durability choices. In contrast to these studies, the current study applies a fully structural approach following the recent development of dynamic structural econometrics, and derives deeper implications.

\medskip{}

In the engineering literature, many LCC (Life-cycle cost) and LCA (Life-cycle assessment) analyses have been conducted to assess the potential of product lifetime extension, because of the growing interest in the circular economy (See \citet{bakker2021understanding} for review). Though insightful, consumer demand and firms' profits have not necessarily been explicitly considered in most of the studies. Nevertheless, considering firms' profits is essential, if higher product durability harms firms. This study complements the literature by proposing a quantitative method for evaluating the social optimality of product durability, explicitly considering consumer surplus, producer surplus, and externalities. 

\subsection{Firms' endogenous product / quality choice and semi-collusion}

This study also builds on the literature of firms' endogenous product/quality choice (\citet{Fan2013}, \citet{sweeting2013dynamic}, \citet{Eizenberg2014}, \citet{Crawford2019}, \citet{Wollmann2018}, \citet{Reynaert2021}). There are some recent empirical studies investigating the role of collusion in non-price dimensions (\citet{sullivan2020split}, \citet{bourreau2021market}, \citet{ale2021colluding}), and the current study is in line with these studies. Unlike these studies, the current study introduces dynamics on the demand and the supply sides. In addition, durability, which can be thought of as a kind of quality, has an unique feature that it not only affects current but also future demand. This study contributes to the literature by discussing the role of durability, whose feature differs from other characteristics.

\subsection{Estimation methods of Structural dynamic demand models}

The model of durable goods can be thought of as one type of dynamic demand models, where the current and future demand are related. Other examples include storable goods, goods with switching costs, and network goods. Though many markets are characterized by dynamic demand structures, there are small number of empirical studies analyzing firms' dynamic pricing decisions under dynamic demand. Regarding firms' product choices or investment decisions under dynamic demand, there are only a few, to my knowledge.\footnote{\citet{goettler2011does} analyzed continuous investment decisions of durable single product firms by fully specifying firms' dynamic pricing decisions. \citet{carranza2010product} analyzed durable goods firms' product introduction decisions assuming that firms are monopolistic competitors, who do not consider the impacts of their decisions on market variables and other firms. \citet{lee2013vertical} investigated the role of vertical integration and exclusivity in the context of hardware and software products which are durable, by specifying software firms' endogenous product introduction decisions. Note that software firms' pricing decisions are not explicitly specified in the analysis. } Large obstacles to the analysis would be the computational costs and technical difficulties. Hence, to enhance our understanding on the real markets with dynamic demand structures, developing easier estimation methods is essential. The current study contributes to the literature by proposing novel algorithms for estimating demand and marginal cost parameters alleviating the problems.\footnote{See See Fukasawa (\citeyear{Fukasawa2022,fukasawa2024biases}) for the discussions on potential problems in applying static demand and static supply-side models under dynamic demand.}

\subsubsection{Marginal cost estimation}

This study contributes to the studies on the estimation of marginal cost parameters of dynamic demand models, by proposing a novel ``full-solution'' algorithm alleviating the computational burden and data requirement.

To my knowledge, previous studies have used the methods that do not require fully solving the equilibrium to infer marginal costs.\footnote{Besides the mentioned approaches, \citet{chen2013secondary} calibrated marginal cost parameters so that the data fits the structural model well.} Nevertheless, they are not always practical. For instance, \citet{nair2007intertemporal}, \citet{goettler2011does}, and \citet{conlon2012dynamic} used marginal cost data to analyze the behavior of durable goods producers. Nevertheless, marginal cost data are typically not available. \citet{gowrisankaran2013computing} and \citet{bollinger2019learning} proposed a ``two-step'' estimation method, where firms' pricing functions are approximated using data in the first step, and then marginal cost parameters are estimated in the second step. The method requires many observations to well approximate the functions in the first step. \citet{berry2000estimation} proposed the Euler equation method, using firms' dynamic optimality conditions to estimate supply-side parameters. Nevertheless, the method requires long time series data, and is only applicable to models where state transition probabilities are differentiable. \citet{cosguner2018dynamic}, empirically studying goods with switching costs, proposed an estimation method whose idea is to parameterize firms' optimal pricing decisions as functions of state variables, and search for the parameter values that minimize the objective function derived from firms' optimal pricing conditions and observed price data. Though the method does not necessarily require many observations as in the two-step estimation methods, it requires solving an optimization problem with a large number of nonlinear parameters. Especially when the number of products is large, the number of nonlinear parameters gets large, and it may take much time to solve the optimization problem.

The ``full-solution'' method I propose overcomes the drawbacks of the previous methods, regarding the data requirement and computational burden. The idea is to jointly solve the variables characterizing the equilibrium and marginal costs justifying the observed prices. As discussed in Section \ref{sec:Estimation}, it is relatively easy to implement once the algorithm to solve the equilibrium is available. Typically, solving the equilibrium is essential for conducting counterfactual simulation, and it is unavoidable to prepare a code. Also, the computational time of marginal cost estimation is mostly the same as the computational time of solving the equilibrium, at least in the current setting. In that sense, the proposed method would be useful for researchers planning to solve the equilibrium to examine counterfactual outcomes.\footnote{The full solution approach the current study proposes assumes that we can solve an equilibrium of the model. If it is not computationally practical to solve an equilibrium, the methods that do not require solving an equilibrium in the estimation step, such as the ones proposed in the previous studies, might be desirable.}

\subsubsection{Demand estimation}

The current study also contributes to the literature on dynamic demand estimation, by applying a computationally efficient inner-loop algorithm for estimating dynamic demand parameters, building on the findings in \citet{Fukasawa2024BLPalgorithm}. The model the current study considers can be classified as one of the dynamic BLP models, which is a dynamic extension of the static BLP model (\citet{berry1995automobile}). Dynamic BLP models have been applied to empirically study durable goods (e.g., \citet{schiraldi2011automobile}, \citet{gowrisankaran2012dynamics}) and goods with switching costs (e.g., \citet{shcherbakov2016measuring}). However, one of the obstacles to dynamic BLP estimations is the computational burden. A separate and methodological paper \citet{Fukasawa2024BLPalgorithm} recently proposed some simple ideas to accelerate the convergence of inner-loop iterations in BLP estimations, and the current study utilizes the ideas to develop a fast inner-loop algorithm, tailoring to the empirical model considered in the current study.

\section{Light bulb industry\label{sec:Industry}}

In Section \ref{subsec:Data}, we discuss the important characteristics of light bulb industry, based on the market data we use in the analysis. To discuss the industry, we should not ignore the existence of Phoebus cartel, where firms colluded on product durability. We discuss it in Section \ref{subsec:Phoebus-Cartel}.

\subsection{Data\label{subsec:Data}}

The main dataset used in the study is the light bulb market point-of-sale (POS) data in Japan from January 2009 to June 2009. The dataset includes the quantities and values of all the light bulb products sold in electronics retail stores and home-center stores each month.\footnote{Coverage rates are 98\% for electronics retail stores and 50\% for home-center stores.} Using value and quantity data, average prices of products in each month are computed. Since the dataset does not contain all the products' main characteristics, including average lifetimes, electricity usage, and colors, I manually collected the information from the Internet.

In this study, I will focus on light bulb products with E26 sockets for ordinary use. Regarding lighting equipment, households have to renovate their houses when installing new ones in some cases. In contrast, regarding light bulbs, what households have to do is to buy a new bulb and screw it into a socket. Note that light bulbs with E26 sockets have been mainly used in the residential sector. For details of the dataset, see Appendix \ref{sec:Data-details}.

In the sample period, there were two types of light bulb products: incandescent lamps (Inc.) and compact fluorescent lamps (CFLs). Though CFLs are more durable and more energy efficient,\footnote{Wattages of 60W equivalent incandescent lamps are 54$\sim$60W, while that of CFLs are 10$\sim$20W.} they emit light light based on different mechanisms, and the tastes of light are not the same. In addition, CFLs contain mercury, which is harmful if it leaks from the products. 

\begin{figure}[H]
\begin{centering}
\includegraphics[scale=0.5]{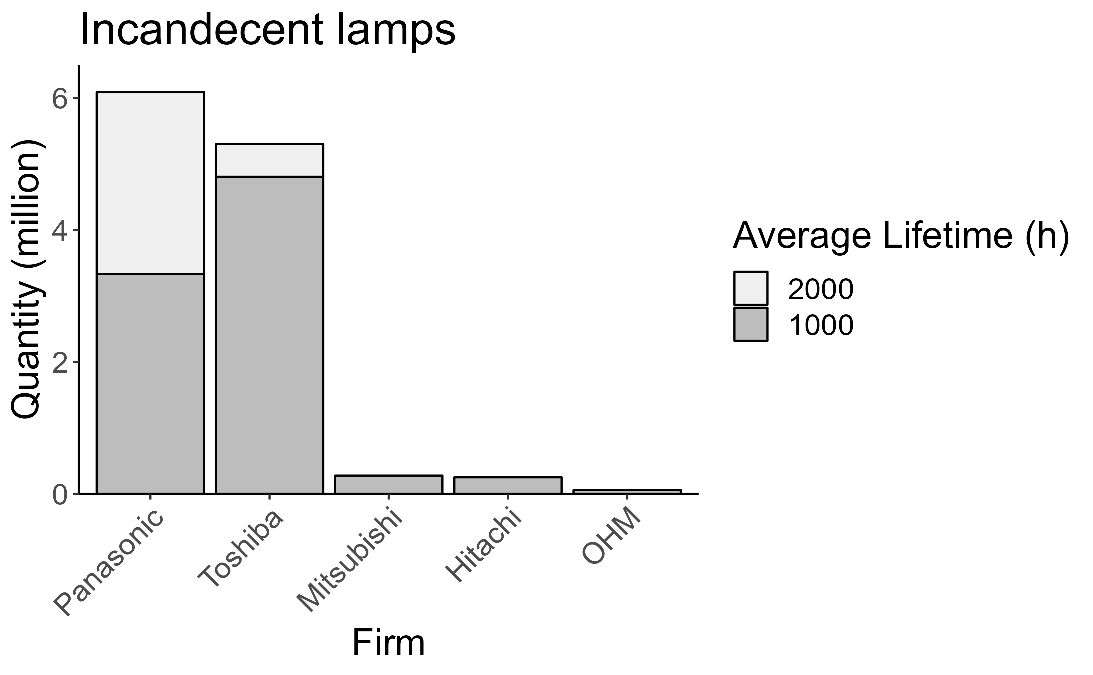}\includegraphics[scale=0.5]{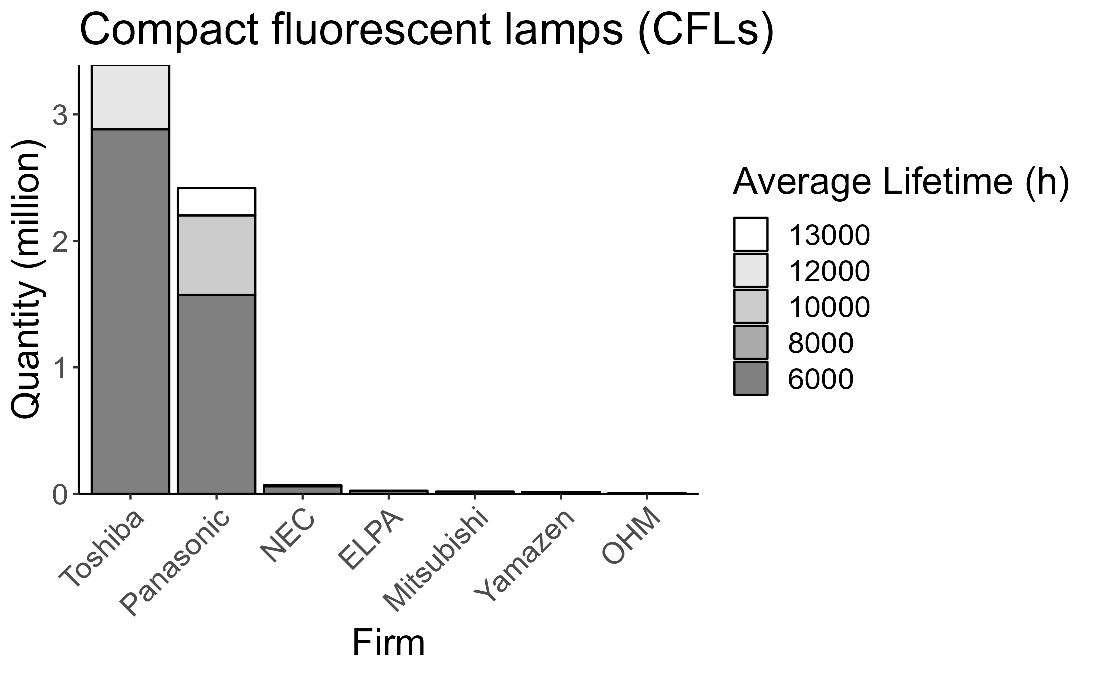}
\par\end{centering}
\caption{Quantities of light bulb products sold in the market\label{fig:quantity_sold_market}}
{\footnotesize{}Note. The figures show the cumulative quantities of light bulb products sold in the market in the sample period.}{\footnotesize\par}
\end{figure}

Figure \ref{fig:quantity_sold_market} shows the quantities of lamps sold by each manufacturer. As shown in the figure, the market was highly concentrated. There are two dominant firms, Toshiba and Panasonic. Though several other firms also produced light bulb products, their sales account for less than 10\% of the total sales in the market.

As shown in the figure, there are several durability levels for light bulb products. For incandescent lamps, average lifetimes of some products are 2000 hours, but that of others are 1000 hours. For CFLs, there are five levels of average lifetimes: 6000, 8000, 10000, 12000, and 13000 hours. The information on the average lifetimes are shown in the product packages, and consumers can easily see the durability levels and make purchase decisions.

\begin{figure}[H]
\begin{centering}
\includegraphics[scale=0.5]{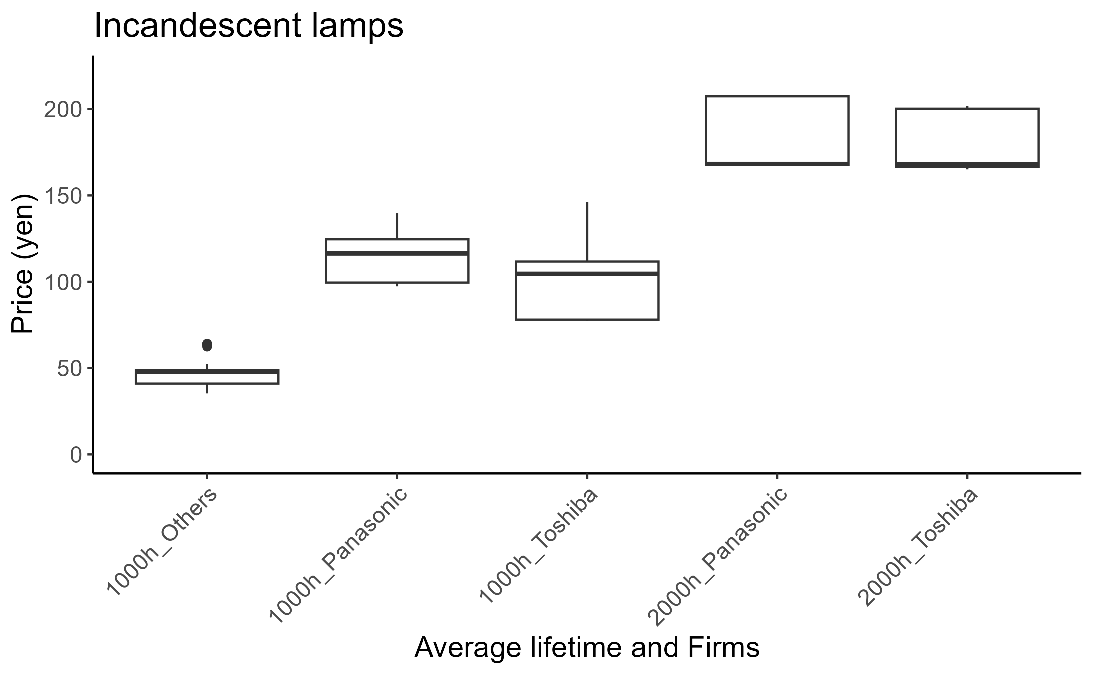}\includegraphics[scale=0.5]{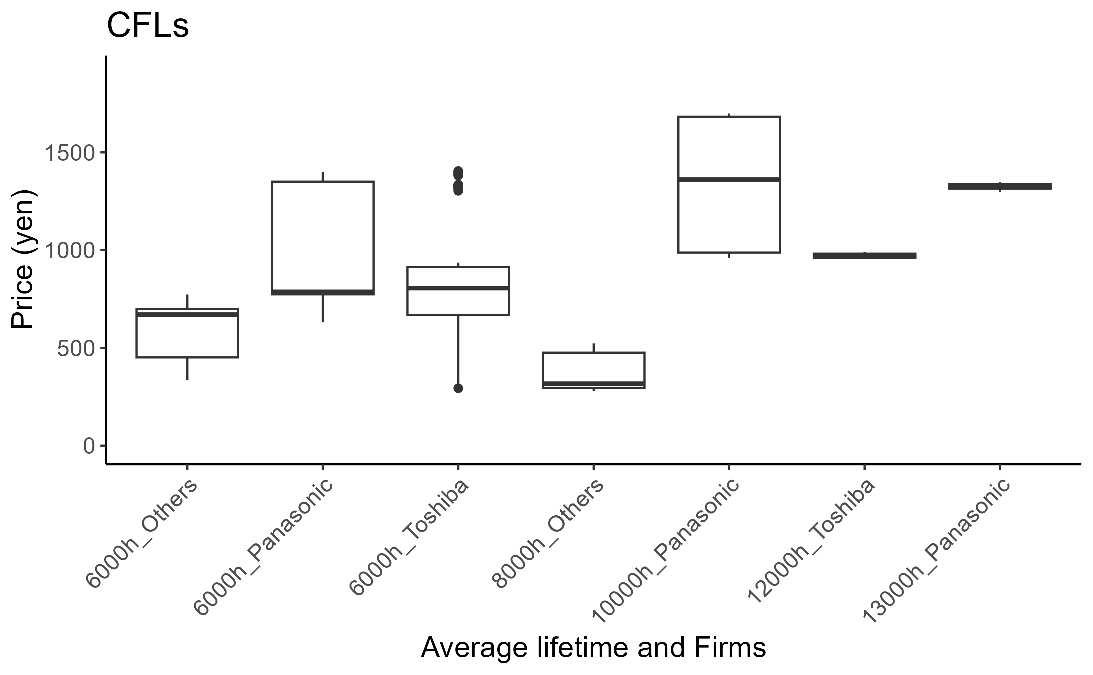}\caption{Distribution of product prices\label{fig:Prices}}
\par\end{centering}
{\footnotesize{}Notes. The figures show the distribution of the prices of light bulb products sold in the market in the sample period. As of 2009, 1 USD was roughly equivalent to 100 JPY.}{\footnotesize\par}
\end{figure}

Figure \ref{fig:Prices} shows the box plots of product prices sold by each firm. As the left of the figure shows, two dominant firms Panasonic and Toshiba set higher prices for incandescent lamps than firms with small market shares, consistent with the intuition that firms with market power set higher prices than small firms. 

Also, we can observe that the prices of 2000h lamps are roughly 50 yen higher than 1000h lamps for Panasonic and Toshiba' s products. Note that the difference of product prices between 2000h lamps and 1000h lamps comes not only from production cost differences, but also firms' dynamic incentives. Firms have dynamic incentives to set higher prices for higher durability products, for fear of less frequent replacement and losing future profits. Hence, the source of the difference in price levels is not necessarily clear without the estimation of production costs. Regarding the prices of CFLs shown in the right panel, we can also roughly observe similar tendencies. 

Note that the coexistence of different durability products has not been unique in the Japanese market in the 21st century. As shown in \citet{swan1982less},\footnote{See Tables 3 and 4 in \citet{swan1982less}.} General Electric (GE), one of the largest global firms in the industry, sold both high and low durability incandescent lamps in the U.S. and U.K. markets in 1979. Explicitly accounting for substitutions between different durability products is important for deeper understanding of firms' incentives in the real market. 

Besides, Figure \ref{fig:Prices} also shows that light bulb products are differentiated: products are different in some characteristics, such as average lifetime, colors, and electricity usage, and even the prices of the same durability products are heterogeneous across products. Accordingly, in the next section, I develop a dynamic model of differentiated products, rather than homogeneous products. 

\subsection{Phoebus Cartel\label{subsec:Phoebus-Cartel}}

In the 1920s-30s, large light bulb producers around the world had formed a cartel known as Phoebus cartel. It was known as the first global cartel,\footnote{See \citet{krajewski2014great}.} and the participating companies included General Electric (GE), Phillips, and Tokyo Electric, which was the predecessor of Toshiba. The unique feature of the cartel was the obligation on product durability. Firms participating in the cartel had to not only limit the amount of production, but also shorten the lifetime of their products to levels below 1000 hours. It was intended to increase sales by raising the frequency of replacement of products. In fact, in the letter to Phoebus in 1927, Tokyo Electric revealed that:

\medskip{}

\textsl{We have shortened the life of our lamps ... from 2500 hours to 1600 for gas-filled lamps; we could increase the sales of gas-filled lamps...}\footnote{Ex. 2131-G, letter from O. Pruessman to C. F. Johnstone, May 2, 1927, as cited in \citet{stocking1946cartels} p355. The Tokyo Electric's memorandum was enclosed in the letter.}

\medskip{}

Though the cartel was intended to last until 1955 in the cartel's 1924 agreement, it was nullified mainly because of the outbreak of World War II. In 1940s-50s, the cartel was investigated by the U.S. and U.K. competition authorities. For details of the cartel, see \citet{stocking1946cartels} and \citet{krajewski2014great}.

Based on the historical facts of the cartel, we can conclude that firms might have incentives to collude to reduce product lifetimes. We investigate the incentives in detail in Section \ref{sec:Counterfactual}.

\section{Model\label{sec:Model}}

\subsection{Consumers}

In this study, we assume each consumer considers the purchase of at most one new light bulb product only when they do not own functioning products in each period.\footnote{We assume there is no used goods market. Selling and buying products themselves incur additional transaction costs, and it would not pay to buy used light bulb products, considering their relatively cheap prices.} Though each household might purchase multiple bulbs, generally they are screwed into sockets in separate rooms, and we think of them as separate consumers. The market size corresponds to the number of sockets in houses. 

\subsubsection*{Consumers' individual state variables}

Let $x_{it}\in\chi$ be consume $i$'s state variable, which corresponds to their product holdings. Let $x_{it}=\emptyset$ be the state where consumer $i$ does not own a functioning product at the beginning of time $t$, and let $x_{it}=(j,\tau)\neq\emptyset$ be the state where consumer $i$ owns product $j$ that is purchased $\tau$ periods before and still functioning. Here, $\chi$ denotes the set of individual states, and $\chi=\{\emptyset\}\cup\left(\bigcup_{j\in\mathcal{J}}\bigcup_{\tau\geq1}(j,\tau)\right)$ holds, where $\mathcal{J}$ denotes the set of products. Besides, let $Pr_{it}(x_{it})$ be the probability that consumer $i$ is at state $x_{it}$ at time $t$.

\subsubsection*{Consideration set}

In each period, consumers make decisions. Let $a_{it}$ be the decision of consumer $i$ at time $t$, and let $\mathcal{A}_{t}(x_{it})$ be the set of alternatives that consumer $i$ can take when the consumer is at state $x_{it}$ at time $t$. This study assumes consumers use their previous products until their failure. Under the assumption, consumers consider the purchase of new products from the set of products available at time $t$ and the outside option $\mathcal{J}_{t}\cup\{0\}$ only when they do not own any functioning products.\footnote{The specification comprehends the models of perfectly durable goods without replacement demand, where consumers leave the market after making a purchase(\citet{nair2007intertemporal}, \citet{ishihara2019dynamic}). Besides, a similar specification was employed in \citet{armitage2022technology}, empirically studying the demand-side of the light bulb market to analyze the optimal timing of introducing environmental policies encouraging the purchase of energy-efficient products. Note that large difference of the current demand model with her specification is the explicit specifications of consumers' forward-looking behavior, considering the durability of products.\\
Note that we do not consider the case where consumers stockpile products in their houses. The assumption would be reasonable, because the current study uses monthly data.} Formally, the consideration set of consumers $\mathcal{A}_{t}(x_{it})$ is:

\begin{eqnarray*}
\mathcal{A}_{t}(x_{it}) & = & \begin{cases}
\mathcal{J}_{t}\cup\{0\} & \text{if}\ x_{it}=\emptyset\\
\{0\} & \text{if}\ x_{it}\in\chi-\{\emptyset\}
\end{cases}.
\end{eqnarray*}

\subsubsection*{State transition}

Transition probability of $x_{it+1}$ given $x_{it}$ and $a_{it}$ is specified as follows:

\begin{equation}
Pr(x_{it+1}|x_{it},a_{it})=\begin{cases}
\phi(i,\mu_{j},\tau=1) & \text{if}\ x_{it}=\emptyset,\ a_{it}=j,\ x_{it+1}=(j,\tau=1)\\
1-\phi(i,\mu_{j},\tau=1) & \text{if}\ x_{it}=\emptyset,\ a_{it}=j,\ x_{it+1}=\emptyset\\
1 & \text{if}\ x_{it}=\emptyset,\ a_{it}=0,\ x_{it+1}=\emptyset\\
\phi(i,\mu_{j},\tau+1|\tau) & \text{if}\ x_{it}=(j,\tau),\ a_{it}=0,\ x_{it+1}=(j,\tau+1)\\
1-\phi(i,\mu_{j},\tau+1|\tau) & \text{if}\ x_{it}=(j,\tau),\ a_{it}=0,\ x_{it+1}=\emptyset
\end{cases},\label{eq:state_transition}
\end{equation}
where $\phi(i,\mu_{j},\tau)$ denotes the probability that product $j$ with durability level $\mu_{j}$ used by consumer $i$ does not fail for $\tau$ periods. $\phi(i,\mu_{j},\tau+s|\tau)$ $(s\geq1)$ denotes the probability that product $j$ with durability level $\mu_{j}$ does not fail at age $\tau+s$ conditional on surviving at age $\tau$. More specifically, $\mu_{j}$ denotes product $j$'s average lifetime. For instance, $\mu_{j}=1000$ implies product $j$ lasts 1000 hours on average.

Besides, let $L_{ij}$ be the time until the failure of product $j$ used by consumer $i$. It is a stochastic variable characterized by survival function $\phi(i,\mu_{j},\tau)$, and $Pr(L_{ij}<\tau)=\phi(i,\mu_{j},\tau)$ holds.

\subsubsection*{Consumers' utility function}

Consumer $i$'s per-period utility is:

\begin{eqnarray*}
U_{ijt}(x_{it},\Omega_{t}^{C},\epsilon_{it})\equiv u_{ijt}(x_{it},\Omega_{t}^{C})+\epsilon_{ijt} & = & \begin{cases}
-\alpha_{i}p_{jt}+\widetilde{\delta_{jt}}+\psi_{j}+\epsilon_{ijt} & \text{if}\ x_{it}=\emptyset\ \text{and}\ j\neq0\\
\epsilon_{i0t} & \text{if\ }x_{it}=\emptyset\ \text{and}\ j=0\\
\psi_{k}+\epsilon_{i0t} & \text{if}\ x_{it}=(k,\tau)\ \text{and\ }j=0
\end{cases}.
\end{eqnarray*}

Here, $p_{jt}$ denotes the price of product $j$ sold at time $t$. $\widetilde{\delta_{jt}}$ denotes the utility consumers obtain from purchasing product $j$ at time $t$.\footnote{We later define mean utility $\delta_{jt}$, which is the sum of the utility consumers obtain from purchasing and usage of the product.} For instance, it is costly for consumers to go to stores, bring purchased products, and insert them into the sockets in their houses. Consumers might value high-durability products for environmental reasons, and such motives are reflected in this term. In contrast, the value of $\psi_{j}$ represents the utility from the usage of product $j$. Products vary in their characteristics, such as colors and electricity usage of lamps, and $\psi_{j}$ differs across products. $\epsilon_{ijt}$ denotes mean-zero idiosyncratic utility shock consumer $i$ obtains when choosing alternative $j$ at time $t$. $\Omega_{t}^{C}$ denotes market information consumers obtain at time $t$, including all the products' prices $p$, characteristics, and durability levels. Let $\Omega_{t}^{C}\equiv\left(\widetilde{\Omega_{t}^{C}},p_{t}\right)$, where $\widetilde{\Omega_{t}^{C}}$ denotes market information other than product prices consumers have.

Consumer $i$'s expected discounted utility of choosing alternative $j$ at time $t$ given the states $(x_{it},\Omega_{t}^{C})$ and idiosyncratic utility shock $\epsilon_{it}$ is:

\begin{eqnarray*}
v_{ijt}(x_{it},\Omega_{t}^{C},\epsilon_{it}) & = & U_{ijt}(x_{it},\Omega_{t}^{C},\epsilon_{it})+\beta_{C}E_{t}\left[V_{it+1}^{C}(x_{it+1},\Omega_{t+1}^{C})|x_{it},\Omega_{t}^{C},a_{it}=j\right],
\end{eqnarray*}
where $\beta_{C}$ is the discount factor of consumers, and $E_{t}\left[\cdot\right]$ denotes the expectation of future values at time $t$ conditional on the available information at time $t$. For later convenience, let $v_{ijt}(x_{it},\Omega_{t}^{C},\epsilon_{it})=\widetilde{v_{ijt}}(x_{it},\Omega_{t}^{C})+\epsilon_{ijt}\ (j\in\mathcal{A}_{t}(x_{it}))$. $V_{it}^{C}(x_{it},\Omega_{t}^{C})$ denotes the (integrated) value function of consumer $i$ at state $(x_{it},\Omega_{t}^{C})$ at time $t$, and it satisfies the following Bellman equation:

\begin{eqnarray}
V_{it}^{C}(x_{it},\Omega_{t}^{C}) & = & E_{\epsilon}\left[\max_{j\in\mathcal{A}_{t}(x_{it})}\left(u_{ijt}(x_{it},\Omega_{t}^{C})+\epsilon_{ijt}+\beta_{C}E_{t}\left[V_{it+1}^{C}(x_{it+1},\Omega_{t+1}^{C})|x_{it},\Omega_{t}^{C},a_{it}=j\right]\right)\right],\label{eq: V_C_def}
\end{eqnarray}
where $E_{\epsilon}$ denotes the expectation operator with respect to the idiosyncratic utility shocks $\epsilon$.

Then, no-inventory consumer $i$'s expected discounted utility function when buying product $j$ at time $t$ can be reexpressed as:\footnote{Since 
\begin{eqnarray*}
V_{it}^{C}(x_{it}\neq\emptyset,\Omega_{t}^{C}) & = & u_{ijt}(x_{it}\neq\emptyset,\Omega_{t}^{C})+E_{\epsilon}\left[\epsilon_{i0t}\right]+\beta_{C}E_{t}\left[V_{it+1}^{C}(x_{it+1},\Omega_{t+1}^{C})|x_{it}\neq\emptyset,\Omega_{t}^{C},a_{it}=0\right]\\
 & = & u_{ijt}(x_{it}\neq\emptyset,\Omega_{t}^{C})+\beta_{C}E_{t}\left[V_{it+1}^{C}(x_{it+1},\Omega_{t+1}^{C})|x_{it}\neq\emptyset,\Omega_{t}^{C},a_{it}=0\right].
\end{eqnarray*}

holds, we can derive:

\begin{eqnarray*}
v_{ijt}(x_{it}=\emptyset,\Omega_{t}^{C},\epsilon_{it}) & = & -\alpha_{i}p_{jt}+\widetilde{\delta_{jt}}+\sum_{\tau=0}^{\infty}\beta_{C}^{\tau}\phi(i,\mu_{j},\tau)\psi_{j}+\sum_{\tau=0}^{\infty}\beta_{C}^{\tau}f(i,\mu_{j},\tau)V_{it+L_{ij}}^{C}(x_{it+L_{ij}}=\emptyset,\Omega_{t+L_{ij}}^{C})+\epsilon_{ijt},
\end{eqnarray*}

where $f(i,\mu_{j},\tau)$ denotes the probability of failure after $\tau$period.

In equation (\ref{eq:v_ijt}), we represent $\sum_{\tau=0}^{\infty}\beta_{C}^{\tau}\phi(i,\mu_{j},\tau)\psi_{j}$ as $E_{t}\left[\sum_{\tau=0}^{L_{ij}}\beta_{C}^{\tau}\psi_{j}\right]$, and $\sum_{\tau=0}^{\infty}\beta_{C}^{\tau}f(i,\mu_{j},\tau)V_{it+L_{ij}}^{C}(x_{it+L_{ij}}=\emptyset,\Omega_{t+L_{ij}}^{C})$ as $E_{t}\left[\beta_{C}^{L_{ij}}V_{it+L_{ij}}^{C}(x_{it+L_{ij}}=\emptyset,\Omega_{t+L_{ij}}^{C})\right]$.}

\begin{eqnarray}
v_{ijt}(x_{it}=\emptyset,\Omega_{t}^{C},\epsilon_{it}) & = & -\alpha_{i}p_{jt}+\underbrace{\widetilde{\delta_{jt}}}_{\text{Utility from purchase}}+\underbrace{E_{t}\left[\sum_{\tau=0}^{L_{ij}}\beta_{C}^{\tau}\psi_{j}\right]}_{\text{Utility from usage}}+\label{eq:v_ijt}\\
 &  & \underbrace{E_{t}\left[\beta_{C}^{L_{ij}}V_{it+L_{ij}}^{C}(x_{it+L_{ij}}=\emptyset,\Omega_{t+L_{ij}}^{C})\right]}_{\text{Utility from future replacement}}+\epsilon_{ijt}.\nonumber 
\end{eqnarray}
Here, $L_{ij}$ denotes the time until the failure of product $j$. $L_{ij}$ is a stochastic variable, and consumers form expectations on the realization of $L_{ij}$ based on the information of the average lifetime of products ($\mu_{j}$) shown in product labels. $E_{t}$ denotes an expectation operator not only on $\Omega^{C}$ but also the time until product failure $L_{ij}$.

No-inventory consumer $i$'s expected discounted utility function when not buying any product at time $t$ is:

\begin{eqnarray*}
v_{i0t}(x_{it} & = & \emptyset,\Omega_{t}^{C},\epsilon_{it})=\beta_{C}E_{t}\left[V_{it+1}^{C}(x_{it+1},\Omega_{t+1}^{C})|x_{it}=\emptyset,\Omega_{t}^{C},a_{it}=0\right]+\epsilon_{i0t}.
\end{eqnarray*}

The probability that consumer $i$ chooses product $j$ at time $t$ conditional on considering the purchase is

\begin{eqnarray}
s_{ijt}^{(ccp)}(x_{it}=\emptyset,\Omega_{t}^{C}) & = & Pr\left(v_{ijt}(x_{it}=\emptyset,\Omega_{t}^{C},\epsilon_{it})>v_{ikt}(x_{it}=\emptyset,\Omega_{t}^{C},\epsilon_{it})\ \forall k\in\mathcal{J}_{t}\cup\{0\}-\{j\}\right).\label{eq:choice_prob_ccp}
\end{eqnarray}

The choice probability of the outside option conditional on considering purchase of a new product is

\begin{eqnarray}
s_{i0t}^{(ccp)}(x_{it}=\emptyset,\Omega_{t}^{C}) & = & Pr\left(v_{i0t}(x_{it}=\emptyset,\Omega_{t}^{C},\epsilon_{it})>v_{ikt}(x_{it}=\emptyset,\Omega_{t}^{C},\epsilon_{it})\ \forall k\in\mathcal{J}_{t}\right).\label{eq:outside_choice_prob_ccp}
\end{eqnarray}

\subsubsection*{Demand dynamics}

Let the probability that product $j$ purchased by consumer $i$ at time $t$ fails at time $t+\tau$ be $f(i,\mu_{j},\tau)$. By definition, $f(i,\mu_{j},\tau)+\phi(i,\mu_{j},\tau)=1$ holds. We further assume that market size $M$, namely, the number of potential consumers, is constant over time. Then, the probability that consumer $i$ considers the purchase of a product at time $t$ is:

\begin{eqnarray}
Pr0_{it} & = & \sum_{\tau=1}^{\infty}\sum_{j\in\mathcal{J}_{t-\tau}}s_{ijt-\tau}f(i,\mu_{j},\tau)+\widetilde{s_{i0t-1}},\label{eq:purchase_prob}
\end{eqnarray}
where $s_{ijt-\tau}$ denotes the fraction of type $i$ consumers purchasing product $j$ at time $t-\tau$, and $\widetilde{s_{i0t-1}}$ denotes the fraction of type $i$ consumers not owning any working product but not purchasing any product at time $t-1$.

Then, the probability that consumer $i$ purchases product $j$ at time $t$ is:

\begin{eqnarray}
s_{ijt}(\Omega_{t}^{C}) & = & Pr0_{it}\cdot s_{ijt}^{(ccp)}(x_{it}=\emptyset,\Omega_{t}^{C}).\label{eq:choice_prob_times}
\end{eqnarray}
It implies $s_{ijt}$ is the product of the probability of purchase and the choice probability conditional on purchasing. 

The probability that consumer $i$ does not purchase anything given that they do not own anything at time $t$ is:

\begin{eqnarray}
\widetilde{s_{i0t}}(\Omega_{t}^{C}) & = & Pr0_{it}\cdot s_{i0t}^{(ccp)}(x_{it}=\emptyset,\Omega_{t}^{C}).\label{eq:choice_outside_prob_times}
\end{eqnarray}

Then, the aggregate market share of product $j$ in market $t$ is:

\begin{eqnarray}
s_{jt}(\Omega_{t}^{C}) & = & \int s_{ijt}(\Omega_{t}^{C})dP(i).\label{eq: agg_market_share}
\end{eqnarray}
where $dP(i)$ represents the density of consumer $i$.

\subsection{Firms}

In each period, firms make pricing decisions, considering the future sequence of their profits. We assume the set of products and product characteristics, including product durability, are determined before the realization of demand and marginal cost shocks. 

\subsubsection*{Aggregate states}

Firms make pricing decisions based on the aggregate state variables $\Omega_{t}^{F}\equiv\left(B_{t},\widetilde{\Omega_{t}^{F}}\right)$ such that $\widetilde{\Omega_{t}^{C}}\subset\widetilde{\Omega_{t}^{F}}$. $B_{t}\equiv\left(Pr_{it}(x_{it})\right)_{x_{it}\in\chi,i\in\mathcal{I}}$ denotes the aggregate states that change depending on the demand for products. They include the fraction of consumers owning durable products that are still functioning ($Pr0_{it}\equiv Pr_{it}(x_{it}=\emptyset)$). $\mathcal{I}$ denotes the set of consumers. Note that $Pr_{it}(x_{it})$ satisfies the following transition by equation (\ref{eq:state_transition}):\footnote{We omit $\Omega_{t}^{C}$ for simplifying the exposition.}

\begin{eqnarray}
Pr_{it+1}(x_{it+1}) & = & \begin{cases}
Pr_{it}(x_{it}=\emptyset)\cdot s_{i0t}^{(ccp)}(x_{it}=\emptyset)+ & \text{if}\ x_{it+1}=\emptyset\\
\ \ \ \ \sum_{\tau\in\mathbb{N},j\in\mathcal{J}_{t-\tau}}Pr_{it}\left(x_{it}=(j,\tau)\right)\cdot\left(1-\phi(i,\mu_{j},\tau+1|\tau)\right)\\
Pr_{it}\left(x_{it}=(j,\tau-1)\right)\cdot\phi(i,\mu_{j},\tau|\tau-1) & \text{if}\ x_{it+1}=(j,\tau\geq2)\\
Pr_{it}(x_{it}=\emptyset)\cdot s_{ijt}^{(ccp)}(x_{it}=\emptyset)\cdot\phi(i,\mu_{j},\tau=1) & \text{if}\ x_{it+1}=(j,\tau=1)
\end{cases}.\label{eq:Pr_state_transition}
\end{eqnarray}

Let $B_{t+1}=B_{t+1}(s_{t},B_{t})$ be the corresponding deterministic law of motion, where $s_{t}\equiv\left(\left\{ s_{ijt}\right\} _{i,j\in\mathcal{J}_{t}}\right)$ is the vector of each type of consumers' demands at time $t$.

In contrast, $\widetilde{\Omega_{t}^{F}}$ denotes the aggregate states other than $B_{t}$ firms use. They include the set of introduced products, product marginal costs, characteristics, but do not include prices.

\subsubsection*{Firms' price setting problems}

Regarding pricing decisions, we assume firms follow Markov perfect equilibrium, and set product prices conditional on the aggregate state variables $\Omega_{t}^{F}\equiv\left(B_{t},\widetilde{\Omega_{t}^{F}}\right)$.\footnote{The assumption of Markov perfect equilibrium in the durable goods model has been used in \citet{nair2007intertemporal}, \citet{goettler2011does}, and \citet{chen2013secondary}. Market-level data shows that light bulb product prices are not constant over time, and we assume that firms did not fully commit to the future sequence of product prices in the real market. Instead, we simulate and evaluate the outcomes where firms can fully commit to set constant future product prices in Section \ref{subsec:Alternative_spec}.} 

Firm $f$'s dynamic price-setting problem is as follows:

\begin{eqnarray*}
\max_{\{p_{kt+\tau}\}_{k\in\mathcal{J}_{ft},\tau\geq0}} & E_{t}\left[\sum_{k\in\mathcal{J}_{ft}}\sum_{\tau=0}^{\infty}\beta_{F}^{\tau}\pi_{ft+\tau}(B_{t+\tau},\widetilde{\Omega_{t+\tau}^{F}},p_{t+\tau})\right],
\end{eqnarray*}
where $\pi_{ft}(B_{t},\widetilde{\Omega_{t}^{F}},p_{t})\equiv\sum_{k\in\mathcal{J}_{ft}}Ms_{kt}(B_{t},\widetilde{\Omega_{t}^{F}},p_{t})(p_{kt}-mc_{kt})$ denotes the per-period profit of firm $f$ at time $t$, and it depends on the vector of product prices $p_{t}\equiv\{p_{jt}\}_{j\in\mathcal{J}_{t}}$. $\beta_{F}$ denotes firms' discount factor, and $\mathcal{J}_{ft}$ denotes the set of products produced by firm $f$ at time $t$.

Firm $f$'s value function $V_{ft}^{F}(B_{t},\widetilde{\Omega_{t}^{F}})$ satisfies the following Bellman equation:

\begin{eqnarray}
V_{ft}^{F}(B_{t},\widetilde{\Omega_{t}^{F}}) & = & \pi_{ft}(B_{t},\widetilde{\Omega_{t}^{F}},p_{t}^{*})+\beta_{F}E_{\widetilde{\Omega}}\left[V_{ft+1}^{F}(B_{t+1}(B_{t},p_{t}^{*}(B_{t})),\widetilde{\Omega_{t+1}^{F}})|\widetilde{\Omega_{t}^{F}}\right],\label{eq:V_F}
\end{eqnarray}
where $E_{\widetilde{\Omega}}$ denotes the expectation operator concerning $\widetilde{\Omega^{F}}$, and $p_{t}^{*}(B_{t})$ denotes the equilibrium prices at time $t$. 

First order condition of firm $f$ with respect to product $j$'s price at time $t$ is:

\begin{eqnarray}
0 & = & \frac{\partial\pi_{ft}}{\partial p_{jt}}(B_{t},\widetilde{\Omega_{t}^{F}},p_{t})+\beta_{F}\frac{\partial B_{t+1}}{\partial p_{jt}}\frac{\partial E_{\widetilde{\Omega^{F}}}\left[V_{ft+1}^{F}(B_{t+1},\widetilde{\Omega_{t+1}^{F}})|\widetilde{\Omega_{t}^{F}}\right]}{\partial B_{t+1}}\label{eq:FOC_p}\\
 & = & (p_{jt}(B_{t},\widetilde{\Omega_{t}^{F}})-mc_{jt})\frac{\partial s_{jt}}{\partial p_{jt}}(p_{t},B_{t},\widetilde{\Omega_{t}^{F}})+\sum_{k\in\mathcal{J}_{ft}-\{j\}}(p_{kt}(B_{t},\widetilde{\Omega_{t}^{F}})-mc_{kt})\frac{\partial s_{kt}}{\partial p_{jt}}(p_{t},B_{t},\widetilde{\Omega_{t}^{F}})+Ms_{jt}(p_{t},B_{t},\widetilde{\Omega_{t}^{F}})+\nonumber \\
 &  & \beta_{F}\frac{\partial B_{t+1}}{\partial p_{jt}}\frac{\partial E_{\widetilde{\Omega^{F}}}\left[V_{ft+1}^{F}(B_{t+1},\widetilde{\Omega_{t+1}^{F}})|\widetilde{\Omega_{t}^{F}}\right]}{\partial B_{t+1}}.\nonumber 
\end{eqnarray}

\subsection{Stationary state}

In this study, we define the stationary state of the equilibrium by the following definition:\footnote{The idea of stationary state or steady state has been used to make the structural durable goods models tractable in previous studies, including \citet{chen2013secondary}, \citet{gavazza2014quantitative}, and \citet{gillingham2022equilibrium}.}
\begin{defn}
The market is in the stationary state, if $\Omega_{t}^{F}\equiv\left(B_{t},\widetilde{\Omega_{t}^{F}}\right)\equiv\left(\left(Pr_{it}(x_{it})\right)_{x_{it}\in\chi,i\in\mathcal{I}},\widetilde{\Omega_{t}^{F}}\right)$ , $\widetilde{\Omega_{t}^{C}}$, and equilibrium product prices $p_{t}(B_{t},\widetilde{\Omega_{t}^{F}})$ are time-invariant. 

\medskip{}

In the stationary state, equilibrium product prices $p_{jt}(B_{t},\widetilde{\Omega_{t}^{F}})$, consumers' value functions $V_{it}^{C}(x_{it},\Omega_{t}^{C})$, firms' value functions $V_{ft}^{F}(B_{t},\widetilde{\Omega_{t}^{F}})$, and firms' market shares $s_{jt}(B_{t},\Omega_{t}^{C})$ and CCPs $s_{ijt}^{(ccp)}(x_{it}=\emptyset,\Omega_{t}^{C})$ are time-invariant. Though we allow a nonstationary market environment, the idea of a stationary state is used to specify the initial state, as discussed in Section \ref{subsec:Specifications}. 
\end{defn}

\section{Estimation\label{sec:Estimation}}

\subsection{Specifications\label{subsec:Specifications}}

\subsubsection{Distributional assumption on $\epsilon_{ijt}$ and nest}

We assume that the error term $\epsilon_{ijt}$ follows GEV-type distribution to make the model tractable. In the case of light bulb products, it is plausible to assume that light bulbs in the same category (incandescent, CFL) are more highly substituted than those in other categories. The substitution pattern would not be represented just by introducing random coefficients. To accommodate the issues, I introduce nests in addition to random coefficients in the utility function as in \citet{grigolon2014nested}. Namely, we assume the following distributional assumption on $\epsilon_{ijt}$: 

\begin{eqnarray*}
\epsilon_{ijt} & = & \bar{\xi}_{igt}+(1-\rho_{g})\widetilde{\epsilon_{ijt}}\ \ \ (j\in\mathcal{J}_{gt}),
\end{eqnarray*}
where $\widetilde{\epsilon_{ijt}}$ is distributed i.i.d. mean zero type-I extreme value, and $\bar{\xi}_{igt}$ is such that $\epsilon_{ijt}$ is distributed extreme value. $\mathcal{J}_{gt}$ denotes the set of products in nest $g$ at time $t$. $\rho_{g}$ is the nest parameter, specific to nest $g$. When the values of $\{\rho_{g}\}_{g}$ are equal to zero, the problem reduces to a standard random coefficient logit model. As $\rho_{g}$ goes to 1, products in nest $g$ are perceived to be perfect substitutes.

Then, value function $V_{it}^{C}(x_{it}=\emptyset)$ is in the following form\footnote{To simplify the exposition, we omit $\Omega_{t}^{C}$.}:

\begin{eqnarray}
V_{it}^{C}(x_{it}=\emptyset) & = & \ln\left(\sum_{g\in\mathcal{G}\cup\{0\}}\exp\left(IV_{igt}^{C}\right)\right),\label{eq:V_C}
\end{eqnarray}
where $IV_{i0t}^{C}$ and $IV_{igt}^{C}\ (g\in\mathcal{G})$ denote the inclusive values in each nest:

\begin{eqnarray*}
IV_{i0t}^{C} & = & \widetilde{v_{i0t}}(x_{it}=\emptyset),\\
IV_{igt}^{C} & = & (1-\rho_{g})\ln\left(\sum_{j\in\mathcal{J}_{gt}}\exp\left(\frac{\widetilde{v_{ijt}}(x_{it}=\emptyset)}{1-\rho_{g}}\right)\right)..
\end{eqnarray*}
Here, $\mathcal{J}_{gt}$ denotes the set of products in nest $g$ at time $t$. $\mathcal{G}$ denotes the set of nests other than the outside option (incandescent lamps, CFLs).

Besides, conditional choice probabilities which appear in (\ref{eq:choice_prob_ccp}) and (\ref{eq:outside_choice_prob_ccp}) can be rewritten as:

\begin{eqnarray}
s_{ijt}^{(ccp)}(x_{it}=\emptyset) & = & \frac{\exp\left(\frac{\widetilde{v_{ijt}}(x_{it}=\emptyset)}{1-\rho_{g}}\right)}{\exp\left(\frac{IV_{igt}^{C}}{1-\rho_{g}}\right)}\frac{\exp(IV_{igt}^{C})}{\exp(V_{it}^{C}(x_{it}=\emptyset))},\ \ \ (j\in\mathcal{J}_{g})\label{eq:prob_choice_cd_nest}\\
s_{i0t}^{(ccp)}(x_{it}=\emptyset) & = & \frac{\exp(IV_{i0t}^{C})}{\exp(V_{it}^{C}(x_{it}=\emptyset))}.\label{eq:prob_outside_cd_nest}
\end{eqnarray}

\subsubsection{Failure rate}

As discussed in Section \ref{sec:Model}, $\mu_{j}$ denotes the durability level of product $j$. More specifically, we define $\mu_{j}$ to be the rated average lifetime of product $j$. Rated average lifetime is the average hours the product works, and the value indicates its durability. For instance, 1000h bulbs' rated average lifetime is 1000 hours.

Next, we assume the time until the failure $L_{ij}$ of product $j$ used by consumer $i$, whose unit is month because we use monthly data, follows discretized Weibull distribution with scale parameter $\eta(i,\mu_{j})$ and shape parameter $\lambda$.\footnote{Weibull distribution is widely used to model the reliability of products in engineering literature (\citet{bedford2001probabilistic}). In addition, the specification is broadly used in waste management and ecology literature (\citet{oguchi2010lifespan}). Furthermore, Weibull distribution accommodates some special distribution of decay patterns used in the theoretical literature on durable goods, such as exponential decay function ($\lambda=1;$ constant depreciation rate) and one-hoss shay decay function ($\lambda=\infty$; fails after $\mu$ hours without exception). } Formally, the distribution function of lifetime $L_{ij}$ is: 

\begin{eqnarray*}
F(i,\mu_{j},\tau) & \equiv Pr(L_{ij}\geq\tau)= & 1-\exp\left(-\left(\frac{\tau}{\eta(i,\mu_{j})}\right)^{\lambda}\right).\ \ \ (\tau=0,1,2,\cdots)
\end{eqnarray*}

The probability of failure after $\tau$ period is:

\begin{eqnarray*}
f(i,\mu_{j},\tau)\equiv\Pr(L_{ij}=\tau) & = & \begin{cases}
F(i,\mu_{j},\tau)-F(i,\mu_{j},\tau-1) & (\tau=1,2,\cdots)\\
0 & (\tau=0)
\end{cases}.
\end{eqnarray*}

The probability of not failing for $\tau$ period (survival rate) is:

\begin{eqnarray*}
\phi(i,\mu_{j},\tau)\equiv Pr(L_{ij}<\tau) & = & \exp\left(-\left(\frac{\tau}{\eta(i,\mu_{j})}\right)^{\lambda}\right)\ \ \ (\tau=0,1,2,\cdots).
\end{eqnarray*}

Note that $\mu_{j}$ and $\eta(i,\mu_{j})$ satisfy the following equality regarding the average lifetime: \footnote{Since the units of variables are confusing, I clarify them here. The unit of $I_{i}$ is {[}h/month{]}. The unit of $\mu_{j}$ is {[}h{]}, and that of $\tau$ is {[}month{]}. $f(i,\mu_{j},\tau)$ is a probability, and there is no unit. Then, it would be clear that the units of both sides are {[}h{]}.}

\begin{eqnarray*}
I_{i}\cdot\sum_{\tau=1}^{\infty}\tau f(i,\mu_{j},\tau) & = & \mu_{j}.
\end{eqnarray*}
Here, $I_{i}$ denotes the consumer $i$'s expected usage time per month. We assume $I_{i}$ is common for all consumer types, and set the value to 76.94 hours (2.54 hours per day), which is the average usage time based on \citet{elec_intensity2013}.\footnote{Since many of E26 sockets in Japan are located in places less frequently used, such as lavatory, corridor, average usage time is relatively low. For details, see the data in \citet{elec_intensity2013}.} The equation implies that given the values of average lifetime $\mu_{j}$ shape parameter $\lambda$, and usage time $I_{i}$, we can solve for the values of $\eta(i,\mu_{j})$ and recover the probability of failure $f(i,\mu_{j},\tau)$. In this study, the values of $\eta(i,\mu_{j})$ are numerically solved by Newton's method given the values of $I_{i}$ and $\mu_{j}$ before the demand and marginal cost estimations. We assume that consumers and firms understand the stochastic failure process, and form future expectations.

In principle, we might be able to estimate the parameter by matching the moment condition on the total number of sells of products in each period, yet it would not be accurate because we use aggregate level data. Hence, I calibrate the value to 2.4, which is the average parameter value of electronics and recommended in \citet{Oguchi2006}. Note that the estimated parameters and simulation results did not largely change even when changing the value of $\lambda$.

\subsubsection{Initial state}

As in the empirical literature on dynamic demand (\citet{hendel2006measuring,gowrisankaran2012dynamics,schiraldi2011automobile}), we need to specify the initial state, since we do not directly observe the distribution of consumer inventories at the initial period. This study assumes that the periods before the sample periods (first half of 2009) are in the stationary state, which corresponds to the average of the sample periods.\footnote{Before 2009, both incandescent lamps and CFLs coexisted, and the changes in market share of these products were not so large. Current Production Statistics Survey, published by Ministry of Economy, Trade and Industry (METI), show that mean yearly sell of incandescent lamps from 2005 to 2009 was 123.4 million with standard deviation 15.4 million, and that of CFLs was 30.3 million with standard deviation 7.2 million.}

Let $\widetilde{\delta_{j}^{(stationary)}}$ be the mean product utility of product $j$ consumers obtain from purchasing, and let $V_{i}^{C(stationary)}$ be consumer $i$'s value function, and let $mc_{j}^{(stationary)}$ be the marginal cost of product $j$ in the stationary state. We assume stationary market share of product $j$ is $S_{j}^{(stationary)}=\frac{1}{T}\sum_{t=1}^{T}S_{jt}^{(data)}$, baseline stationary price of product $j$ is $p_{j}^{(stationary)}=\frac{1}{T}\sum_{t=1}^{T}p_{jt}^{(data)}$, and assume $\widetilde{\delta_{j}^{(stationary)}},V_{i}^{C(stationary)},mc_{j}^{(stationary)}$ are consistent with stationary market shares $S_{j}^{(stationary)}$ and stationary product prices $p_{j}^{(stationary)}$.

\subsubsection{Expectations}

Basically, we assume consumers and firms have perfect foresight on the transitions of aggregate state variables:\footnote{The use of perfect foresight or rational expectations on the transitions of market-level state variables can be found in \citet{berry2000estimation}, \citet{goettler2011does}, \citet{conlon2012dynamic}, \citet{chen2013secondary}, \citet{hendel2013intertemporal}, and \citet{kalouptsidi2020linear}. The alternative specification is the use of inclusive value sufficiency, as in \citet{hendel2006measuring} and \citet{gowrisankaran2012dynamics}. Nevertheless, as discussed in Section \ref{subsec:Alternative_spec}, the use of inclusive value sufficiency is not necessarily consistent with the supply-side models and is not easy to deal with.}
\begin{itemize}
\item In the sample period $t\leq T$, consumers and firms have perfect foresight on the transitions of $B_{t}$ $\widetilde{\Omega_{t}^{F}}$, and $\Omega_{t}^{C}$:
\begin{itemize}
\item Consumers: $E_{x,\Omega^{C}}\left[V_{it+1}^{C}\left(x_{it+1},\Omega_{t+1}^{C}\right)|x_{it},\Omega_{t}^{C},a_{it}\right]=E_{x}\left[V_{it+1}^{C}\left(x_{t+1},\Omega_{t+1}^{C(realized)}\right)|x_{it},a_{it}\right]\ \forall a_{it}\in\mathcal{A}_{t}(x_{it}),\forall x_{it}\in\chi$
\item Firms: $E_{B,\Omega^{F}}\left[V_{ft+1}^{F}(B_{t+1},\widetilde{\Omega_{t+1}^{F}})|B_{t},\widetilde{\Omega_{t}^{F}}\right]=V_{ft+1}^{F}\left(B_{t+1}^{(realized)},\widetilde{\Omega_{t+1}^{F(realized)}}\right)$
\end{itemize}
\item After the last period of the data $T$ ($t>T$), $\widetilde{\Omega_{t}^{F}}=\widetilde{\Omega^{F}}^{(stationary)},\widetilde{\Omega_{t}^{C}}=\widetilde{\Omega^{C}}^{(stationary)}$, and firms set product prices fixed at the stationary level $p_{j}(B^{(stationary)})$.\footnote{Though the assumption that firms set fixed product prices might seem to be restrictive, product prices did not largely change over time. The assumption is used to separate demand and marginal cost estimations. Note that we obtain analogous simulation results even when allowing firms to adjusting prices at time $t>T$. Besides, in the dynamic investment competition model under static demand, \citet{igami2017estimating} and others employ similar specifications, where the market environment does not change after the terminal period.}
\end{itemize}
Here, $\Omega_{t+1}^{C(realized)}$ and $\widetilde{\Omega_{t+1}^{F(realized)}}$ represent the realized values of $\Omega_{t+1}^{C}$ and $\widetilde{\Omega_{t+1}^{F}}$, which are observed in the data. 

Under the specification, we do not have to solve and evaluate firms' pricing decisions at the point $\Omega_{t}^{C}\neq\Omega_{t}^{C(data)}$ and $\Omega_{t}^{F}\neq\Omega_{t}^{F(data)}$, which contributes to reducing the number of states.

Though we impose the assumption of perfect foresight to simplify the explanation, without changing the estimation procedure, we can relax the assumption to rational expectation, where consumers' expectations regarding the transitions of $\widetilde{\Omega_{t}^{C}}$ are on average correct. For details, see Appendix \ref{subsec:Assumptions_expec}.

\subsubsection{Discount factor}

It is known that discount factor is hard to identify without further exclusion restrictions (\citet{magnac2002identifying}). In this study, monthly discount factor of consumers is set to 0.99 following the literature (e.g., \citet{gowrisankaran2012dynamics}). We also assume firms and consumers share the same discount factor.\footnote{Previous studies since \citet{hausman1979individual} analyzing consumer preference for energy efficiency of durables estimated consumers' discount factor, assuming consumers are fully attentive to the information on energy efficiency. These studies used the variations in future energy costs as one source of exclusion restriction and estimated the discount factor. Nevertheless, in the light bulb market, the variations in electricity usages are small after controlling for other product characteristics. Also, it is not clear whether consumers are fully attentive to the future electricity cost. Hence, I exogenously set the value of the discount factor. Note that the responsiveness of consumers to electricity cost is reflected in the term $\psi_{j}$ in a reduced-form way, and we allow for the case where consumers are not fully attentive to electricity costs. Regarding durability, we assume consumers are fully attentive to future replacement opportunities. This assumption would be reasonable in the light bulb market, since the average lifetimes are clearly shown in product packages, and it is relatively easy to consider the expected time until the next replacement opportunity. }

\subsection{Demand Estimation\label{subsec:Demand-Estimation}}

\subsubsection{Estimation and Identification}

The estimation method is a dynamic version of \citet{berry1995automobile}'s BLP method. First, let $\delta_{jt}\equiv\widetilde{\delta_{jt}}+E_{t}\left[\sum_{\tau=0}^{L_{ij}}\beta_{C}^{\tau}\psi_{j}\right]$ be ``mean utility'' of product $j$ at time $t$. Note that $\widetilde{v_{ijt}}(x_{it}=\emptyset,\Omega_{t}^{C})=-\alpha_{i}p_{jt}+\delta_{jt}+E_{t}\left[\beta_{C}^{L_{ij}}V_{it+L_{ij}}^{C}(x_{it+L_{ij}}=\emptyset,\Omega_{t+L_{ij}}^{C})\right]$ holds. We assume equations (\ref{eq:v_ijt}), (\ref{eq:purchase_prob})-(\ref{eq: agg_market_share}), (\ref{eq:V_C})-(\ref{eq:prob_outside_cd_nest}) hold at $\Omega_{t}^{C(data)}\equiv\left(\widetilde{\Omega_{t}^{C(data)}},p_{t}^{(data)}\right)$, with the assumption of initial state, parametric assumption of failure rate, and expectation formation. We additionally assume that the values of $\delta_{jt}$ satisfy $S_{jt}^{(data)}=s_{jt}(\delta)$, as in static BLP model. The condition implies market shares predicted by the structural model are equal to the market shares observed in the data. 

To estimate the demand parameters, we assume $\widetilde{\delta_{jt}}$ and $\psi_{j}$ are in the form of $\widetilde{\delta_{jt}}=\widetilde{X_{jt}^{D}}\theta_{linear}^{D}+\xi_{jt}$ and $\psi_{j}=\overline{X_{jt}^{D}}\overline{\theta_{linear}^{D}}$, where $\widetilde{X_{jt}^{D}}$ and $\overline{X_{jt}^{D}}$ denote observed product characteristics, and $\xi_{jt}$ denotes unobserved product characteristics or demand shock. Then, mean utility $\delta_{jt}$ is in the following form: $\delta_{jt}=\widetilde{X_{jt}^{D}}\theta_{linear}^{D}+E_{L}\left[\sum_{\tau=0}^{L_{ij}}\beta_{C}^{L_{ij}}\right]\overline{X_{j}^{D}}\overline{\theta_{linear}^{D}}+\xi_{jt}$.\footnote{It depends on the assumption that the distribution of $L_{ij}$ does not depend on consumer specific factors. If we allow for such elements, we should explicitly distinguish the utility from purchasing and utility from usage, though the identification is unclear.} Hence, $\delta_{jt}$ can be reformulated as $\delta_{jt}=X_{jt}^{D}\theta_{linear}^{D}+\xi_{jt}$, where $X_{jt}^{D}$ includes the interaction terms of durability levels and observed product characteristics other than average lifetimes. As $X_{jt}^{D}$, we use color-watt equivalence dummies and watt equivalence-2000h dummies for incandescent lamps. Let $\theta_{2000h,40W},\theta_{2000h,60W},\theta_{2000h,100W}$ be the coefficients of corresponding watt-equivalence 2000h dummies. They represent the discounted sum of additional utility consumers obtain until the products fail when purchasing 2000h products. For CFLs, we use color-watt equivalence-lifetime dummies, shape-watt equivalence lifetime dummies, and electricity usage - firm -lifetime dummies.\footnote{Luminosity levels of light bulb products are shown by wattage-equivalence, such as 40W-equivalent, 60W-equivalent, and 100W-equivalent. For incandescent lamps, there are two colors (silica and clear). For CFLs, there are three colors (incandescent lamp color, daytime color and neutral white color), and two shapes (type A and T).} We also include time dummies in $X_{jt}^{D}$. 

Regarding persistent consumer heterogeneity, we assume $\alpha_{i}$ follow log-normal distribution $\alpha_{i}\sim LN(\log(\overline{\alpha}),\sigma_{\alpha}^{2})$.\footnote{Log-normal distribution, not normal distribution, is introduced to guarantee that marginal utility of money $\alpha_{i}$ is positive for all consumers.} $\overline{\alpha}$ represents the median of the distribution.

Since we consider the model where firms set product prices considering the market environment including demand shocks $\xi$, $\xi_{jt}$ and price $p_{jt}$ might be correlated. Also, we estimate additional parameters including nest parameters and random coefficients, we need instrumental variables to identify these parameters. We assume $\xi_{jt}$ is orthogonal to instrumental variables $Z_{jt}^{D}$, namely, $E\left[\xi_{jt}Z_{jt}^{D}\right]=0$, and estimate structural parameters $\theta_{linear}^{D}$ and $\theta_{nonlinear}^{D}\equiv\left(\overline{\alpha},\sigma_{\alpha},\left\{ \rho_{g}\right\} _{g\in\mathcal{G}}\right)$ by Generalized Method of Moments (GMM). $Z_{jt}^{D}$ includes exogenous product characteristics $X_{jt}^{D}$ and additional variables.\footnote{We assume that observed product characteristics $X_{jt}^{D}$ are not correlated with $\xi_{jt}$, under the assumption that product introduction choices are made before the realization of demand shocks, as in the literature of endogenous quality choice (\citet{Wollmann2018}, \citet{fan2020competition}).} As additional instruments, we use dummy variables on the number of competing products of own and rival firms in the same product category,\footnote{We assume products with the same watt-equivalence, color, shape, durability level, and type (Incandescent or CFL) are in the same product category.} which are one sort of BLP instruments. We also use energy usage-firm-lifetime dummies.\footnote{As discussed in \citet{swan1982less} in detail, durability level and energy efficiency of products affect the cost structure of the light bulbs. Also, the cost structure might be different across firms. Hence, we introduced the cross terms of durability level (lifetime), energy usage, category, and firm dummies.} To identify nest parameters, we use the number of products in the same nest (up to squared terms) and mean predicted product price in each nest\footnote{Since product price is an endogenous variable, I regress the price on other instrumental variables by OLS and use the predicted values as additional IVs, as in \citet{gandhi2019measuring}.} (up to cubic terms). Besides, to further identify the random coefficient on price sensitivity, I use differentiation IVs following \citet{gandhi2019measuring}. 

In the demand estimation, we can clearly identify the preference for durability. For instance, Toshiba sells two products with the same size, electricity usage, color, shape, and wattage equivalence in the market: LW100V54W55 and LW100V54WLL. The only difference between the observed characteristics of the two is the durability level. Average lifetime of the former product is 1000 hours, but the latter is 2000 hours. Hence, we can identify the preference for durability by the cross-sectional differences in the demand for products with different durability levels.\footnote{Though we directly estimate the preference for 2000h incandescent lamps relative to 1000h lamps, we do not introduce parameters for CFLs. This is because CFL products with different average lifetimes are also different in other characteristics, such as electricity usage or the time until getting bright, whose variations are not so large.

For instance, compared to the standard 6,000h lamps, Toshiba's 12000h lamps (Neo Ball Z Realistic PRIDE) has the additional feature of UV cut. Panasonic's 13,000h lamps get bright immediately after the turning on of the switch, though standard 6,000h lamps takes a little time to become bright.}

\subsubsection{Algorithm}

Algorithm \ref{alg:Algorithm-demand} shows the steps to estimate demand parameters.

\begin{algorithm}[H]
\begin{enumerate}
\item Set the values of nonlinear parameters $\theta_{nonlinear}^{D}\equiv\left(\overline{\alpha},\text{\ensuremath{\sigma_{\alpha}}},\left\{ \rho_{g}\right\} _{g\in\mathcal{G}}\right)$
\begin{enumerate}
\item Solve the fixed point problem and compute the value of the objective function given the values of $\theta_{nonlinear}^{D}$:
\begin{enumerate}
\item Set the initial values $V_{it}^{C(0)},IV_{igt}^{C(0)},Pr0_{it}^{(0)}$
\item Iterate the following until convergence:

\[
\left(V_{it}^{C(n+1)},IV_{igt}^{C(n+1)},Pr0_{it}^{(n+1)}\right)=\Phi^{D}\left(V_{it}^{C(n)},IV_{igt}^{C(n)},Pr0_{it}^{(n)};\theta_{nonlinear}^{D}\right)
\]

\end{enumerate}
\item Analytically compute the values of $\delta_{jt}$, given the converged values of $V_{it}^{C},IV_{igt}^{C},Pr0_{it}$ and $\theta_{nonlinear}^{D}$
\item Compute linear parameters $\theta_{linear}^{D}$ that minimize the following objective function given $\theta_{nonlinear}^{D}$:

\[
G(\theta_{linear}^{D},\theta_{nonlinear}^{D})WG(\theta_{linear}^{D},\theta_{nonlinear}^{D})
\]

where $G(\theta_{linear}^{D},\theta_{nonlinear}^{D})\equiv Z^{D}\xi(\theta_{linear}^{D},\theta_{nonlinear}^{D})=Z^{D}\left(\delta(\theta_{linear}^{D},\theta_{nonlinear}^{D})-X^{D}\theta_{linear}^{D}\right)$ and $W$ denotes the weight matrix. Let $\theta_{linear}^{D*}(\theta_{nonlinear}^{D})$ be the solution of the minimization problem, and let $m(\theta_{nonlinear}^{D})\equiv G^{\prime}\left(\theta_{linear}^{D*}(\theta_{nonlinear}^{D}),\theta_{nonlinear}^{D}\right)WG\left(\theta_{linear}^{D*}(\theta_{nonlinear}^{D}),\theta_{nonlinear}^{D}\right)$.
\end{enumerate}
\item Search for the value of $\theta_{nonlinear}^{D}$ minimizing the GMM objective $m(\theta_{nonlinear}^{D})$.
\end{enumerate}
{\footnotesize{}Notes. }{\footnotesize\par}

{\footnotesize{}$V_{it}^{C}$ denotes the values of $V_{it}^{C}(x_{it}=\emptyset,\widetilde{\Omega_{t}}^{(data)})$.}{\footnotesize\par}

\caption{Algorithm for estimating demand parameters\label{alg:Algorithm-demand}}
\end{algorithm}

Full steps of 1(a) and 1(b) in Algorithm \ref{alg:Algorithm-demand} are shown in Appendix \ref{subsec:Details_algorithm_demand}. 

To compute the value of the GMM objective, we need to solve for $\delta(\theta)$. Since $\delta$ also depends on consumers' value functions $V^{C}$ and consumer inventory $Pr0$ which are also unobserved in the data, we also have to jointly solve for these variables. As discussed in Appendix \ref{subsec:Details_algorithm_demand}, we can show that $\delta$ can be analytically represented as a function of $V^{C},IV^{C},Pr0$, and it is sufficient to solve for these variables. It implies we can avoid numerically solving for $\delta$ given other variables. As shown in the next section, the procedure largely reduced the number of iterations and computation time.

The essential idea of the proposed algorithm is that mean utility $\delta$ can be analytically represented as a function of $V^{C}$, absent nest structure.\footnote{Under the existence of nest structure, $\delta$ can be analytically represented as a function of nest-level inclusive values and value functions.} Though previous studies applying dynamic BLP models (e.g., \citet{schiraldi2011automobile}, \citet{gowrisankaran2012dynamics}) essentially solved for $\delta$ and $V^{C}$ separately, solving for two types of variables would be time-consuming. By representing $\delta$ as a function of $V^{C}$ analytically, we only have to solve for $V^{C}$, and we can avoid solving for two types of variables $\delta$ and $V^{C}$ separately.

In addition, the proposed algorithm introduces a new mapping accelerating the convergence, whose idea can be also applied to static BLP models. Though BLP contraction mapping in static BLP models update mean utility $\delta$ by $\delta_{jt}^{(n+1)}=\delta_{jt}^{(n)}+\log\left(S_{jt}^{(data)}\right)-\log\left(s_{jt}\left(\delta^{(n)}\right)\right)$, where $s_{jt}\left(\delta^{(n)}\right)$ denotes the market share of product $j$ at time $t$ predicted by the model, we alternatively consider the following updating equation: $\delta_{jt}^{(n+1)}=\delta_{jt}^{(n)}+\left[\log\left(S_{jt}^{(data)}\right)-\log\left(s_{jt}\left(\delta^{(n)}\right)\right)\right]-\left[\log\left(S_{0t}^{(data)}\right)-\log\left(s_{0t}\left(\delta^{(n)}\right)\right)\right]$. The difference with BLP contraction mapping is the existence of the term $\log\left(S_{0t}^{(data)}\right)-\log\left(s_{0t}\left(\delta^{(n)}\right)\right)$.\footnote{In the case nests exist in the demand model, we can solve for $\delta$ by the mapping $\delta_{jt}^{(n+1)}=\delta_{jt}^{(n)}+(1-\rho)\left[\log\left(S_{jt}^{(data)}\right)-\log\left(s_{jt}\left(\delta^{(n)}\right)\right)\right]+\rho\left[\log\left(S_{gt}^{(data)}\right)-\log\left(s_{gt}\left(\delta^{(n)}\right)\right)\right]-\left[\log\left(S_{0t}^{(data)}\right)-\log\left(s_{0t}\left(\delta^{(n)}\right)\right)\right]$, rather than $\delta_{jt}^{(n+1)}=\delta_{jt}^{(n)}+(1-\rho)\left[\log\left(S_{jt}^{(data)}\right)-\log\left(s_{jt}\left(\delta^{(n)}\right)\right)\right]$ used in the literature (e.g., \citet{iizuka2007experts}, \citet{grigolon2014nested}, \citet{conlon2020best}), as discussed in the Appendix of \textbackslash\citet{Fukasawa2024BLPalgorithm}. In the current empirical setting, the values of nest parameters are close to 1, as shown in the estimation results, and the the convergence of the previously applied mapping is slow. In contrast, the new mapping is relatively fast even when nest parameters are close to 1. } As discussed in \citet{Fukasawa2024BLPalgorithm} in detail, it has a good convergence property\footnote{Though there is no guarantee that the mapping is a contraction, it has a property similar to contraction, and I did not encountered non-convergence issues under the parameter settings of Monte Carlo simulation experimented in previous studies.}, and it immediately converges to the true $\delta$ if there is no random coefficients. We can also apply the idea to the mappings on $V^{C}$.

Note that we also introduce the spectral algorithm, which has been found to be effective at reducing the number of iterations until convergence for static BLP models (\citet{conlon2020best}) and dynamic models (\citet{aguirregabiria2021imposing}). It further reduces the computational time. 

The combinations of the proposed ideas would be applicable to other demand models based on the BLP framework. A separate and methodological paper \citet{Fukasawa2024BLPalgorithm} discusses in detail the essence and properties of the algorithm by developing a general static / dynamic BLP models, and shows results of numerical experiments. 

\subsection{Marginal cost Estimation\label{subsec:Marginal-cost-Estimation}}

\subsubsection{Estimation and Identification}

Given the demand-side parameters and the demand/supply- side model, we can recover the marginal costs based on firms' optimality conditions. Here, we assume firms did not collude on product prices in the sample periods.\footnote{Estimated marginal costs of most incandescent lamps are negative when we alternatively assume two dominant firms Panasonic and Toshiba colluded on product prices. In contrast, under the assumption that they did not collude on prices, estimated marginal costs take positive values. It implies it is reasonable to assume that firms set product prices competitively.} We assume marginal cost $mc_{jt}$ is in the following form:

\begin{eqnarray}
mc_{jt} & = & X_{jt}^{mc}\theta^{mc}+\nu{}_{jt}^{mc}\label{eq:mc_equation}
\end{eqnarray}

$X_{jt}^{mc}$ denotes the product characteristics of product $j$ at time $t$, and $\nu_{jt}^{mc}$ represents the unobserved cost shock of the product. We assume $X_{jt}^{mc}$ is orthogonal to the unobserved cost shock $\nu_{jt}^{mc}$, and estimate the parameters $\theta^{mc}$ by OLS.\footnote{We assume firms make product introduction and quality choices before the realization of cost shocks.} As $X^{mc}$, we use color-watt equivalence-lifetime dummies, shape-watt equivalence-lifetime dummies, energy usage-firm-lifetime dummies, and time-category dummies.

Marginal costs are recovered based on the firms' optimality conditions, and marginal cost parameters $\theta^{mc}$ are identified from the variations in product characteristics. The average lifetimes are heterogeneous across products, and it enables us to estimate marginal cost parameters regarding product durability. 

\subsubsection{Algorithm}

Under the assumption of consumers' perfect foresight, we only have to evaluate the values of $V_{it}^{C}(x_{it},\Omega_{t}^{C})$ at $V_{it}^{C}(x_{it},\widetilde{\Omega_{t}^{C}},p_{t}(B_{t}))$, which is a function of $x_{it}$ and $B_{t}$.

Suppose the equilibrium, characterized by $V_{t}^{C},V_{t}^{F},p_{t},B_{t+1},\frac{\partial B_{t+1}}{\partial p_{t}},B_{(stationary)}$, are the solution of the following fixed point problem:

\begin{equation}
\left(V_{t}^{C},V_{t}^{F},p_{t},B_{t+1},\frac{\partial B_{t+1}}{\partial p_{t}},B_{(stationary)}\right)=\Phi_{equil}\left(V_{t}^{C},V_{t}^{F},p_{t},B_{t+1},\frac{\partial B_{t+1}}{\partial p_{t}},B_{(stationary)};mc\right).\label{eq:equil_constraint}
\end{equation}
Here, $B_{(stationary)}$ denotes the value of $B_{t}$ at the stationary state.\footnote{Computing $B_{(stationary)}$ is needed because we assume firms set product prices at the stationary state after the last period of the data $T$.} Then, by choosing the mapping $\Phi_{equil}$ appropriately, we can solve the equilibrium by iteratively applying $\Phi_{equil}$ and updating the variables given the values of $mc$. Algorithm \ref{alg:solve_equil} in Appendix \ref{subsec:Details_algorithm_supply} shows the detailed algorithm.

Next, under the assumption that product prices are determined following Markov perfect equilibrium, marginal cost data $mc$ should satisfy the following constraint:

{\footnotesize{}
\begin{eqnarray}
 &  & mc_{t}\nonumber \\
 & = & p_{t}^{(data)}-\left(\Delta_{direct,t}^{(data)}(B_{t}^{(data)},p_{t}^{(data)})\right)^{-1}\nonumber \\
 &  & \ \ \ \left[s_{t}^{(data)}+\left(\Delta_{indirect,t}(B_{t}^{(data)},p_{t}(B_{t}^{(data)},mc_{jt}))\right)\left(p_{t}(B_{t}^{(data)},mc_{t})-mc_{t}\right)+\beta_{F}\frac{\partial V_{t+1}^{F}}{\partial p_{t}}(B_{t}^{(data)},mc_{t})\right]\nonumber \\
 & = & p_{t}^{(data)}-\underbrace{\left(\Delta_{direct,t}^{(data)}(B_{t}^{(data)},p_{t}^{(data)})\right)^{-1}s_{t}^{(data)}}_{\text{Static margin;\ \ensuremath{\iota_{t}^{(static)}\left(B_{t}^{(data)},p_{t}^{(data)},s_{t}^{(data)}\right)}}}-\nonumber \\
 &  & \underbrace{\left(\Delta_{direct,t}^{(data)}(B_{t}^{(data)},p_{t}^{(data)})\right)^{-1}\left[\left(\Delta_{indirect,t}(B_{t}^{(data)},p_{t}(B_{t}^{(data)},mc_{jt}))\right)\left(p_{t}(B_{t}^{(data)},mc_{t})-mc_{t}\right)+\beta_{F}\frac{\partial V_{t+1}^{F}}{\partial p_{t}}|_{own}(B_{t}^{(data)},mc_{t})\right]}_{\text{Dynamic margin;\ \ensuremath{\iota_{t}^{(dynamic)}\left(B_{t}^{(data)},p_{t},V_{t}^{C},V_{t}^{F},B_{t+1},\frac{\partial B_{t+1}}{\partial p_{t}},B_{(stationary)},mc_{t}\right)}}}\nonumber \\
 & \equiv & \Phi_{mc}\left(V_{t}^{C},V_{t}^{F},p_{t},B_{t+1},\frac{\partial B_{t+1}}{\partial p_{t}},B_{(stationary)};mc,B_{t}^{(data)},p_{t}^{(data)},s_{t}^{(data)}\right).\label{eq:price_margin_data}
\end{eqnarray}
}{\footnotesize\par}

Here, $B_{t}^{(data)}$ denotes $B_{t}$ at the observed data point, and is already recovered based on the demand estimates. $p_{t}^{(data)},s_{t}^{(data)}$ denote the vectors of observed product prices and observed market shares. $\frac{\partial V_{t+1}^{F}}{\partial p_{t}}|_{own}$ denotes $|\mathcal{J}_{t}|\times1$ vector whose $j$-th element is $\frac{\partial V_{ft+1}^{F}}{\partial p_{jt}}$ where $f$ is the producing firm of product $j$. $\Delta_{direct,t}$ and $\Delta_{indirect,t}$ are defined by $\Delta_{direct,t}\equiv\mathcal{H}_{t}\odot\left.\frac{\partial s_{t}}{\partial p_{t}}\right|_{direct}$ and $\Delta_{indirect,t}\equiv\mathcal{H}_{t}\odot\left.\frac{\partial s_{t}}{\partial p_{t}}\right|_{indirect}$, where $\odot$ denotes Hadamard product. $\mathcal{H}_{t}$ denotes the $|\mathcal{J}_{t}|\times|\mathcal{J}_{t}|$ dimensional ownership matrix, whose $(j_{1},j_{2})$-th element is equal to 1 if product $j_{1}$ and $j_{2}$ are produced by the same firm, and is equal to zero if not. Regarding $\frac{\partial s_{t}}{\partial p_{t}}$, which is a $|\mathcal{J}_{t}|\times|\mathcal{J}_{t}|$ dimensional matrix whose $(j_{1},j_{2})$-th element is equal to $\frac{\partial s_{jt}}{\partial p_{kt}}$, product prices affect product market shares not only through the change in current consumer utility given fixed values of continuation values, but also through the change in continuation values. $\left.\frac{\partial s_{t}}{\partial p_{t}}\right|_{direct}$ represents the former channel, and $\left.\frac{\partial s_{t}}{\partial p_{t}}\right|_{indirect}$ represents the latter channel.

Under the existence of firms' dynamic incentives under dynamic demand, product marginal costs cannot be directly recovered based on the firms' optimality conditions and estimated demand function unlike static demand models. As shown in equation (\ref{eq:price_margin_data}), product margins can be divided into two parts. Though static margins can be recovered based on estimated demand function and observed price data, it is not possible to directly recover dynamic margins, because they depend on firms' optimal pricing strategies as a function of $B_{t}$ and marginal costs, which are not directly observed in the data.

Nevertheless, we know that marginal costs satisfy equations (\ref{eq:equil_constraint}) and (\ref{eq:price_margin_data}), and we can recover marginal costs by jointly solving for $mc$ and $\left(V_{t}^{C},V_{t}^{F},p_{t},B_{t+1},\frac{\partial B_{t+1}}{\partial p_{t}},B_{(stationary)}\right)$ . We can solve for these variables by repeatedly applying mappings $\Phi_{equil}$ and $\Phi_{mc}$.

Note that equation (\ref{eq:equil_constraint}) is an infinite dimensional fixed point problem, because the solutions of the equation are the functions on the continuous domains of $B_{t}$. Hence, it is generally not possible to obtain the exact solutions of the equation, and some sorts of numerical approximations are necessary. To solve the problem, we use the collocation method (See \citet{judd1998numerical}). One problem of the use of the collocation method is that the aggregate state variables $B_{t}=\left(Pr_{it}(x_{it})\right)_{x_{it}\in\chi,i\in\mathcal{I}}$ are high-dimensional, because the dimension of $\chi$ depends on the maximum value of product age and the number of different durability levels. To deal with the issue, we use the idea of ``sufficient statistic'' applied in \citet{sweeting2013dynamic} and others. More specifically, I construct a new variable $\widetilde{B_{t}},$which summarizes the information of $B_{t}$ and alternatively use it as the aggregate state variable. We also incorporate the Smolyak method (\citet{smolyak1963quadrature}, \citet{judd2014smolyak}) to reduce the number of grid points further. We further combine the spectral algorithm as in the demand estimation. Note that combining the spectral algorithm not only accelerates the convergence process, but also stabilizes the convergence process (\citet{aguirregabiria2021imposing}). For details of the methods, see Appendix \ref{subsec:Details_algorithm_supply}.

The estimation algorithm I developed here is also applicable to general dynamic demand models. Typically, solving for an equilibrium is essential to assess counterfactual outcomes, and preparing the counterpart of the mapping $\Phi_{equil}$ is indispensable. At least in our setting, the computation time of marginal cost estimation was mostly the same as that of solving the equilibrium, both of which were less than 1 minute on a laptop computer. In that sense, applying a full-solution approach is not always unrealistic if algorithms are appropriately chosen. 

Algorithm \ref{alg:mc_estimation} shows the steps to solve for $mc$ and estimate marginal cost parameters.\footnote{To simplify the exposition, we show the steps without the introduction of spectral algorithm. Besides, as discussed in Appendix \ref{subsec:Details_algorithm_supply}, to be precise we cannot directly solve for $V_{t}^{C}$, if the values of $\psi_{j}$ is unknown. Hence, we alternatively define $\widetilde{V_{t}^{C}}$, and solve for the variable. Intuitively, $\widetilde{V_{t}}^{C}$ corresponds to consumers' value functions assuming consumers obtain the utility from future usage at the timing of their purchases.\\
Besides, it is possible to solve for the variables directly because I specify a nonstationary model where the firms' profits in the terminal periods are explicitly specified. If we do not want to specify the assumption, we should alternatively use the idea similar to inclusive value sufficiency to set up the model in a stationary framework.}
\begin{algorithm}[H]
\begin{enumerate}
\item Take grid points of aggregate state variables $B_{t}^{(grid)}$ and consumer level state variables $x_{it}^{(grid)}$. Set initial values of $\left\{ V_{it}^{C(0)}(x_{it}^{(grid)},B_{t}^{(grid)})\right\} _{i,x_{it}^{(grid)},B_{t}^{(grid)},t}$ (consumers' value function), $\left\{ V_{ft}^{F(0)}(B_{t}^{(grid)})\right\} _{f,B_{t}^{(grid)},t}$ (firm's value function), $\left\{ p_{jt}^{(0)}(B_{t}^{(grid)})\right\} _{j,B_{t}^{(grid)},t}$ (equilibrium price), $\left\{ B_{t+1}^{(0)}(B_{t}^{(grid)})\right\} _{B_{t}^{(grid)},t}$ (aggregate state variables in the next period), $\left\{ \frac{\partial B_{t+1}^{(0)}}{\partial p_{jt}}(B_{t}^{(grid)})\right\} _{B_{t}^{(grid)},t}$(derivative of the aggregate state variables in the next period with respect to the current price), and $B_{(stationary)}$ (stationary state). Set initial values of $\left\{ mc_{jt}^{(0)}\right\} _{j,t}$.
\item Iterate the following process until the convergence of $V_{it}^{C(n)}(x_{it}^{(grid)},B_{t}^{(grid)})$,$V_{ft}^{F(n)}(B_{t}^{(grid)})$, $p_{jt}^{(n)}(B_{t}^{(grid)}),$ $B_{t+1}^{(n)}(B_{t}^{(grid)})$, $\frac{\partial B_{t+1}^{(n)}}{\partial p_{jt}}(B_{t}^{(grid)})$, $B_{(stationary)}^{(n)}$ and $mc_{jt}^{(n)}$ ($n=0,1,2,\cdots$):
\begin{enumerate}
\item Update $V_{t}^{C},V_{t}^{F},p_{t},B_{t+1},\frac{\partial B_{t+1}}{\partial p_{t}},B_{(stationary)}$ by:

{\footnotesize{}
\begin{eqnarray*}
 &  & \left(V_{it}^{C(n+1)}(x_{it}^{(grid)},B_{t}^{(grid)}),V_{ft}^{F(n+1)}(B_{t}^{(grid)}),p_{jt}^{(n+1)}(B_{t}^{(grid)}),B_{t+1}^{(n+1)}(B_{t}^{(grid)}),\frac{\partial B_{t+1}^{(n+1)}}{\partial p_{jt}}(B_{t}^{(grid)}),B_{(stationary)}^{(n+1)}\right)\\
 & = & \Phi_{equil}\left(V_{it}^{C(n)}(x_{it}^{(grid)},B_{t}^{(grid)}),V_{ft}^{F(n)}(B_{t}^{(grid)}),p_{jt}^{(n)}(B_{t}^{(grid)}),B_{t+1}^{(n)}(B_{t}^{(grid)}),\frac{\partial B_{t+1}^{(n)}}{\partial p_{jt}}(B_{t}^{(grid)}),B_{(stationary)}^{(n)};mc_{t}^{(n)}\right).
\end{eqnarray*}
}{\footnotesize\par}
\item Update $mc$ by:

\[
mc_{t}^{(n+1)}=\Phi_{mc}\left(V_{t}^{C(n)},V_{t}^{F(n)},p_{t}^{(n)},B_{t+1}^{(n)},\frac{\partial B_{t+1}^{(n)}}{\partial p_{t}},B_{(stationary)}^{(n)};mc_{t}^{(n)},B_{t}^{(data)},p_{t}^{(data)},s_{t}^{(data)}\right).
\]

\end{enumerate}
\item Using the converged values of $mc$, estimate marginal cost parameters $\theta^{mc}$ by OLS based on equation (\ref{eq:mc_equation})
\end{enumerate}
\caption{{\footnotesize{}Algorithm for estimating product marginal costs\label{alg:mc_estimation}}}
\end{algorithm}

Note that the solutions of equations (\ref{eq:equil_constraint}) and (\ref{eq:price_margin_data}) might not be unique if multiple equilibria exist. It implies we cannot deny the possibility that multiple values of $mc_{jt}$ justifying equations (\ref{eq:equil_constraint}) and (\ref{eq:price_margin_data}) exist. To deal with the issue, I tried several initial values of the variables in the algorithm, while I have not encountered multiple solutions.\footnote{If multiple solutions exist, solutions with minimum value of the objective function in the estimation equation (\ref{eq:mc_equation}) should be adopted.} Note that the procedure is analogous to the estimation of dynamic discrete games applying a full-solution approach.

\section{Estimation results\label{sec:Estimation-results}}

\subsection{Demand\label{subsec:Estimation_results_demand}}

Table \ref{tab:Results-of-demand_est} shows the results of demand parameter estimates. In the table, I show the results without and with random coefficients on prices. Though $\sigma_{\alpha}$ is not necessarily significant, the median price coefficient is estimated to be positive, implying that marginal utility from money is positive. Estimated nest parameters show that they are heterogeneous across different nests: that of incandescent lamps is around 0.95, but that of CFLs is around 0.7. When assuming common values, estimated price elasticities are less than 1, which are not consistent with firms' profit maximization, and introducing such nest-level heterogeneity is also vital in our setting.\footnote{Results are available upon request.} The table also shows that the nest parameter of incandescent lamps is very close to 1. It implies that incandescent lamps are mostly perfect substitutes.

Besides, the bottom part of the table compares the number of iterations and computation time needed when applying the proposed algorithm and when applying the algorithm numerically solving for $\delta$, which has been typically used in the literature. As the results show, the proposed algorithm reduced the computation time by a factor of more than 20. 

\begin{table}[H]
\begin{centering}
\begin{tabular}{ccccc}
\hline 
 & \multicolumn{2}{c}{(1). No random coef.} & \multicolumn{2}{c}{(2). Random coef.}\tabularnewline
 & Est. & SE & Est. & SE\tabularnewline
\hline 
\hline 
$\overline{\alpha}:$ Median of price coef. (yen/1000) & 2.562 & 0.346 & 2.932 & 0.487\tabularnewline
$\sigma_{\alpha}$: Shape parameter of $\alpha_{i}$ & - & - & 0.096 & 0.269\tabularnewline
$\rho_{Inc}:$ nest parameter (incandescent) & 0.961 & 0.011 & 0.956 & 0.011\tabularnewline
$\rho_{CFL}:$ nest parameter (CFL) & 0.700 & 0.035 & 0.658 & 0.04\tabularnewline
$\theta_{2000h,40W}$ & 1.356 & 0.035 & 1.383 & 0.036\tabularnewline
$\theta_{2000h,60W}$ & 1.357 & 0.034 & 1.385 & 0.035\tabularnewline
$\theta_{2000h,100W}$ & 1.387 & 0.038 & 1.418 & 0.04\tabularnewline
\hline 
Number of iterations (Proposed) & \multicolumn{2}{c}{33} & \multicolumn{2}{c}{313}\tabularnewline
Number of iterations (BLP-based) & \multicolumn{2}{c}{5804} & \multicolumn{2}{c}{6237}\tabularnewline
\hline 
Comp. time (sec; Proposed) & \multicolumn{2}{c}{0.835} & \multicolumn{2}{c}{4.931}\tabularnewline
Comp. time (sec; BLP-based) & \multicolumn{2}{c}{74.684} & \multicolumn{2}{c}{87.608}\tabularnewline
\hline 
\end{tabular}
\par\end{centering}
\caption{Results of demand parameter estimates\label{tab:Results-of-demand_est}}

{\footnotesize{}Notes. }{\footnotesize\par}

{\footnotesize{}The number of observations is 433. The correlation coefficient between the values of $\delta$ with and without the term $\xi$ is 0.998. Standard deviation of the difference between the values of $\delta$ with and without the term $\xi$ is 0.0603, while the standard deviation of $\delta$ with $\xi$ is 2.7097.}{\footnotesize\par}

{\footnotesize{}The middle and bottoms part of the table compares the proposed algorithm and the algorithm based on BLP contraction mapping, regarding the number of iterations and computation time for solving the fixed point problem once, measured at the estimated parameter values. The experiment was run on the CPU AMD Ryzen 5 6600H 3.30 GHz and NVIDIA GeForce RTX 3050 Laptop GPU, 16.0 GB of RAM, Windows 11 64 bit and MATLAB 2022b. In both algorithms, operations of large arrays are computed on GPU to speed up the computation. Also, I incorporated the spectral algorithm in both algorithms. In BLP based algorithm, $V^{C},\delta,Pr0$ are jointly updated. See Appendix \ref{subsec:Details_algorithm_demand} for details.}{\footnotesize\par}
\end{table}

Aside from these specifications, I estimated the specification with random coefficients on the preference for two large firms Panasonic and Toshiba. Nevertheless, the estimated parameters were insignificant, and simulation results did not largely change even when applying the specification. The results are available upon request.

\medskip{}

In the case no random coefficients exist, we can show that we can consistently estimate main demand parameters, including $\alpha,\rho_{Inc},\rho_{CFL}$, by a linear GMM with time and group dummies without specifying the consumer expectation formation and the failure rate of products. Also, we do not have to solve the dynamic structural model to estimate the parameters. The results are shown in Appendix \ref{subsec:Estimationg-of-demand_linear_GMM}. The results yield mostly similar results to the ones estimated by fully solving the dynamic model, and it implies the estimated results are not so sensitive to the assumptions on the dynamic model.

\subsubsection*{Consumer preference for durability}

Based on the estimated demand parameters, we can compute consumers' willingness to pay (WTP) for high durability products. Table \ref{tab:WTP} shows the WTP for 2000h bulbs compared to 1000h bulbs. Here, WTP is computed by $WTP_{t}=\sum_{i}w_{i}\frac{\theta_{2000h}+\left(E_{L}\left[\beta_{C}^{L(\mu=2000)}\right]-E_{L}\left[\beta_{C}^{L(\mu=1000)}\right]\right)V_{it}^{C}(x_{it}=\emptyset)}{\alpha_{i}}$ for each watt equivalence\footnote{To compute $WTP$, we use the value of $V_{it}^{C}$ in the initial state. Note that we obtain similar results even when using the values of $V_{it}^{C}$ in the sample period.}. To compare the values of WTP, I also show 1000h and 2000h average product prices. The results show that consumers have a relatively large preference for 2000h bulbs, though they are slightly smaller than the price differences. 

\begin{table}[H]
\begin{centering}
\begin{tabular}{cccc}
\hline 
 & WTP & 1000h price & 2000h price\tabularnewline
\hline 
\hline 
40W & 62.1 & 77 & 167.1\tabularnewline
60W & 62.8 & 77.1 & 167.7\tabularnewline
100W & 74.2 & 114.1 & 204.3\tabularnewline
\hline 
\end{tabular}
\par\end{centering}
\caption{Consumers' willingness to pay (WTP) for durability\label{tab:WTP}}

{\footnotesize{}Notes. WTP shows the WTP for 2000h bulbs relative to 1000h bulbs. 1000h and 2000h prices are sales-weighted averages.}{\footnotesize\par}
\end{table}

\subsubsection*{Long-run price elasticities}

Based on the estimated demand function, we can also compute long-run price elasticities, which show how a permanent change in the price of a product affects the current demand for products.

\begin{table}[H]
\begin{centering}
\begin{tabular}{ccccccc}
\hline 
 &  & Min & 25th & Median & 75th & Max\tabularnewline
\hline 
\hline 
\multirow{2}{*}{Own} & Inc. & 2.67 & 4.11 & 7.38 & 9.56 & 13.51\tabularnewline
 & CFL & 2.67 & 5.22 & 6.63 & 8.15 & 13.62\tabularnewline
\hline 
\multirow{4}{*}{Cross} & Inc.$\rightarrow$ Inc. & 0.0048 & 0.0348 & 0.1315 & 0.3645 & 1.6503\tabularnewline
 & CFL $\rightarrow$ CFL & 0.001 & 0.0085 & 0.0433 & 0.1497 & 0.5703\tabularnewline
 & Inc. $\rightarrow$ CFL & 0.0002 & 0.0014 & 0.005 & 0.0176 & 0.1256\tabularnewline
 & CFL$\rightarrow$Inc. & -0.0929 & -0.0192 & -0.0035 & -0.0002 & 0.3393\tabularnewline
\hline 
\end{tabular}
\par\end{centering}
\caption{Long-run price elasticities of demand\label{tab:price-elasticities}}

{\footnotesize{}Note. The price elasticities are computed at the initial stationary state. }{\footnotesize\par}

{\footnotesize{}The values of own elasticities are computed so that they take positive values.}{\footnotesize\par}

{\footnotesize{}The row of ``Inc.$\rightarrow$CFL'' shows the cross elasticity of a CFL product with respect to a incandescent lamp product. Other rows are defined analogously.}{\footnotesize\par}
\end{table}

Table \ref{tab:price-elasticities} shows the results. They show the absolute values of own price elasticities are far larger than 1.\footnote{\citet{armitage2022technology}, estimating the demand of the light bulb market in the U.S., showed empirical results that median own price elasticities of incandescent lamps is around 0.5, which is less than 1. Nevertheless, her specification did not introduce a nested structure with respect to product categories (incandescent, CFL). In my preliminary analysis, I found the absolute values of estimated incandescent lamps' own elasticities get lower when not accounting for the nested structure with different heterogeneous nest parameters. The results are available upon request.} Note that it does not directly imply firms' low markups, because the light bulb market in the sample period was highly concentrated, and many of the products were sold by two dominant firms Panasonic and Toshiba.

The results on cross elasticities also show products in the same nests are more likely to be substituted. Besides, long-run cross elasticities are not necessarily positive for some products, as shown in ``CFL$\rightarrow$Inc.'' (cross elasticity of an incandescent lamp concerning a CFL product. This is caused by the substitution between different durability products. Consumers expecting higher CFL prices in the future period have a stronger preference for 2000h incandescent lamps compared to 1000h bulbs, and the demand for 1000h bulbs decreases because of the substitution among incandescent lamp products. 

\subsection{Marginal costs\label{subsec:Estimation_results_supply}}

Table \ref{tab:Results-of-mc_est} shows the part of estimated marginal cost parameters. In the table, additional marginal costs from producing 2000h incandescent lamps compared to 1000h lamps are shown. Besides, Table \ref{tab:Price-and-margins} shows the estimated margins. To assess the effect of firms' dynamic incentives, recovered margins based on static estimates are also shown.

\begin{table}[H]
\begin{centering}
\begin{tabular}{cccc}
\hline 
\multicolumn{2}{c}{Parameters} & Est. & S.E.\tabularnewline
\hline 
\hline 
\multirow{3}{*}{Panasonic} & 40W & 57.7 & 21.5\tabularnewline
 & 60W & 57.1 & 21.5\tabularnewline
 & 100W & 70.8 & 21.5\tabularnewline
\hline 
\multirow{3}{*}{Toshiba} & 40W & 73.9 & 21.5\tabularnewline
 & 60W & 74.5 & 21.5\tabularnewline
 & 100W & 87.7 & 21.5\tabularnewline
\hline 
\end{tabular}
\par\end{centering}
\caption{Results of marginal cost parameter estimates\label{tab:Results-of-mc_est}}

{\footnotesize{}Notes. }{\footnotesize\par}

{\footnotesize{}The number of observations is 433. The correlation coefficient between the values of $mc$ with and without the term $\nu^{mc}$ is 0.996. Standard deviation of the difference between the values of $mc$ with and without the term $\nu^{mc}$ is 0.0373, while the standard deviation of $\delta$ with $\xi$ is 0.4123. }{\footnotesize\par}
\end{table}

By comparing the margins based on the dynamic and static estimates in Table \ref{tab:Price-and-margins}, we can see that static estimates yield biased margins, especially for large firms' products. As discussed in \citet{Fukasawa2022}, firms' dynamic incentives is large when the CCP of the firm's product is large, and it is consistent with the discussion. 

Besides, interestingly, the sign of the biases in estimated margins, which are equivalent to the sign of firms' dynamic incentives to set higher prices, differ across different durability products. For CFLs and 2000h incandescent lamps, dynamic incentives are positive, but they are not necessarily positive for 1000h incandescent lamps (See the row of Toshiba). Regarding 1000h bulbs, setting higher prices might lead to more demand for higher durability products due to consumers' substitution, implying less future demand and less profit in the future. Consequently, firms have incentives to set lower prices for lower-durability products. 

\begin{table}[H]
\begin{centering}
\begin{tabular}{ccccc}
\hline 
 &  & Margins (Dynamic Est.) & Margins (Static est.) & Price\tabularnewline
\hline 
\hline 
\multirow{3}{*}{1000h Inc.} & Panasonic & 30 & 29.2 & 106.1\tabularnewline
 & Toshiba & 21.8 & 25.9 & 82.6\tabularnewline
 & Others & 14.6 & 14.7 & 44.8\tabularnewline
\hline 
\multirow{2}{*}{2000h Inc.} & Panasonic & 41.6 & 28.9 & 171.7\tabularnewline
 & Toshiba & 36.7 & 25.9 & 172.2\tabularnewline
\hline 
\multirow{3}{*}{CFL} & Panasonic & 179.1 & 162.4 & 880.5\tabularnewline
 & Toshiba & 214.5 & 189.5 & 657.4\tabularnewline
 & Others & 115.8 & 115.7 & 569\tabularnewline
\hline 
\end{tabular}\caption{Estimated margins and Prices\label{tab:Price-and-margins}}
\par\end{centering}
{\footnotesize{}Note: The values are sales-weighted average. ``Static est.'' is computed ignoring firms' dynamic incentives.}{\footnotesize\par}
\end{table}

\section{Counterfactual Simulation\label{sec:Counterfactual}}

Based on the estimated structural parameters, we can conduct counterfactual simulations. In Section \ref{subsec:Market_structure_durability}, we evaluate how the market structure affects firms' incentives on product durability. In Section \ref{subsec:Durability-and-Welfare}, we assess whether the product durability level is socially optimal. In Section \ref{subsec:Alternative_spec}, we simulate the outcomes under alternative model specifications, including firms' full commitment, consumers' adaptive expectations, and disregard of firms' dynamic incentives, and validate how the results change. 

Since we have assumed that firms' product introduction/quality decisions are made before the realization of unobserved demand shocks $\xi$ and $\nu^{(mc)}$ in the model, we evaluate the outcomes under the market environment without unobserved demand/cost shocks.\footnote{Besides, we consider the setting where mean product utility and marginal costs are constant over time. These values are taken to be equal to the averages of the values without unobserved shock terms in the sample periods. The procedure led to small differences even when allowing for a nonstationary market environment.} Note that the existence of shocks led to small differences, because the values of these shocks are so small in the light bulb market, as shown in the notes of Tables \ref{tab:Results-of-demand_est} and \ref{tab:Results-of-mc_est}. Besides, all the outcomes are evaluated at the same state $B^{(stationary,base)}$.\footnote{\citet{swan1970durability} showed that the durability level a monopolist chooses is socially optimal under some assumptions. In contrast, studies prior to his paper showed that the durability level a monopolist chooses is less than the socially optimal level under the same assumptions. He pointed out that these studies led to misleading results because they only evaluated the outcomes in the stationary state or long-run equilibrium. In our setting, outcomes should be evaluated in the same state to assess whether the counterfactual outcomes are profitable for firms or increase surpluses, rather than at different stationary states. See \citet{sieper1973monopoly} for further discussion.} 

\subsection{Market structure and Durability\label{subsec:Market_structure_durability}}

\subsubsection{The case without collusion on prices}

I first simulate the counterfactual outcomes where two dominant firms Panasonic and Toshiba stop producing and selling 2000h (higher durability) incandescent lamps, but continue producing and selling 1000h incandescent lamps. Here, we assume firms set product prices competitively, but not necessarily on the set of introduced products. Table \ref{tab:Durability-cartel-table} shows how the elimination of 2000h lamps affects economic variables, including firms' profits and product prices.\footnote{As shown in Table \ref{tab:welfare_Inc}, selling only 2000h lamps is not profitable for firms, and they are omitted from the table.}. The results where dominant firms eliminate higher durability CFLs are shown in Appendix \ref{subsec:CFLs-results-market-structure}. By comparing the upper part of columns (1) and (2) in the table, we can see that jointly eliminating 2000h lamps increases the joint profit of the two dominant firms by 1.39 billion yen. It also increases both firms' profits. 

Note that eliminating 2000h bulbs is not necessarily profitable for each firm. By comparing columns (1) and (3), we can see that Panasonic does not have an incentive to eliminate its own 2000h bulb products. Regarding Toshiba, eliminating its own 2000h bulb products rarely change its profit, as shown in columns (1) and (4). Hence, colluding to eliminate 2000h bulbs is not the Nash equilibrium. Nevertheless, by colluding to jointly eliminate 2000h lamps, they can attain higher profits compared to the baseline scenario where they sell 2000h bulbs. 

As discussed in Section \ref{subsubsec:Theoretical-analysis_durability} in detail by developing a theoretical model, firms have incentives to attract more customers who have preference for high durability products by selling them. Nevertheless, higher durability of own products implies lower demand for competitors' products and lower competitors' long-run profits. If they can collude on durability, firms have additional incentives to internalize such a business steeling effect, and less incentives to sell higher durability products. 
\begin{center}
{\scriptsize{}}
\begin{table}[H]
\begin{centering}
{\small{}}%
\begin{tabular}{ccccc}
\hline 
 & {\small{}(1)} & {\small{}(2)} & {\small{}(3)} & {\small{}(4)}\tabularnewline
{\small{}Panasonic} & {\small{}1000h \& 2000h} & {\small{}1000h only} & {\small{}1000h only} & {\small{}1000h \& 2000h}\tabularnewline
{\small{}Toshiba} & {\small{}1000h \& 2000h} & {\small{}1000h only} & {\small{}1000h \& 2000h} & {\small{}1000h only}\tabularnewline
\hline 
\hline 
{\small{}Joint profit} & {\small{}24.67} & {\small{}26.06} & {\small{}25.77} & {\small{}25.02}\tabularnewline
{\small{}Profit (Panasonic)} & {\small{}10.77} & {\small{}10.99} & {\small{}10.68} & {\small{}11.11}\tabularnewline
{\small{}Profit (Toshiba)} & {\small{}13.9} & {\small{}15.07} & {\small{}15.09} & {\small{}13.91}\tabularnewline
\hline 
{\small{}No inventory consumers (\%)} & {\small{}18.61} & {\small{}19.21} & {\small{}19.08} & {\small{}18.74}\tabularnewline
{\small{}Average price (1000h Inc.; yen)} & {\small{}94.73} & {\small{}98.18} & {\small{}98.08} & {\small{}94.59}\tabularnewline
{\small{}Average price (2000h Inc.; yen)} & {\small{}172.57} & {\small{}-} & {\small{}178.26} & {\small{}174.36}\tabularnewline
{\small{}Average price (CFL; yen)} & {\small{}796.53} & {\small{}799.18} & {\small{}798.87} & {\small{}796.9}\tabularnewline
{\small{}Disposal (million)} & {\small{}3.04} & {\small{}3.13} & {\small{}3.11} & {\small{}3.06}\tabularnewline
\hline 
{\small{}$\Delta$CS} & {\small{}-} & {\small{}-1.39} & {\small{}-1.09} & {\small{}-0.34}\tabularnewline
{\small{}$\Delta$PS (excluding fixed cost)} & {\small{}-} & {\small{}1.52} & {\small{}1.2} & {\small{}0.37}\tabularnewline
{\small{}$\Delta$TS (excluding Ext. / fixed costs)} & {\small{}-} & {\small{}0.13} & {\small{}0.11} & {\small{}0.03}\tabularnewline
{\small{}$\Delta$Ext. (electricity usage)} & {\small{}-} & {\small{}-1.33} & {\small{}-1.01} & {\small{}-0.31}\tabularnewline
{\small{}$\Delta$Ext. (waste disposal)} & {\small{}-} & {\small{}0.04} & {\small{}0.03} & {\small{}0.01}\tabularnewline
{\small{}$\Delta$TS (excluding fixed costs)} & {\small{}-} & {\small{}1.41} & {\small{}1.08} & {\small{}0.33}\tabularnewline
{\small{}Upper bound of Fixed cost savings} & {\small{}-} & {\small{}0.12} & {\small{}0.06} & {\small{}0.06}\tabularnewline
\hline 
\end{tabular}{\small\par}
\par\end{centering}
{\scriptsize{}\caption{Effect of eliminating 2000h incandescent lamps (The case without collusion on prices)\label{tab:Durability-cartel-table}}
}{\scriptsize\par}

    {\footnotesize{}Notes. }{\footnotesize\par}
        
    {\footnotesize{}The values at the top of the table show the discounted
    sum of variable profits of two dominant firms evaluated at $B^{(stationary,base)}$.}{\footnotesize\par}

    {\footnotesize{}The values at the bottom show the changes in surpluses relative to the base setting, and are also evaluated
    at $B^{(stationary,base)}$. Units of these values are billion yen.}{\footnotesize\par}
    
    {\footnotesize{}The values of economic variables in the middle part
    of the table are evaluated at the stationary state. Average prices are sales-weighted. The row of disposal represents the number of products disposed of in each period.}{\footnotesize\par}
    
    {\footnotesize{}$CS,PS,Ext,TS$ denote consumer surplus, producer surplus, externalties, and total surplus, respectively. $\Delta TS$ is defined by $\Delta TS\equiv\Delta CS+\Delta PS-\Delta Ext$.}{\footnotesize\par}

\end{table}
{\scriptsize\par}
\par\end{center}

\subsubsection*{Role of fixed costs}

In the discussion above, we have not considered the existence of fixed costs of the products. If 2000h products are eliminated, firms might be able to save the fixed costs of these products. Since no reliable data on fixed costs are available, we infer the values based on the revealed preference approach, assuming firms do not have incentives to deviate from the observed choices, As discussed in Section \ref{subsec:fixed_cost}, we can only estimate the bounds of the values, and in the tables I show the upper bounds of the fixed cost savings. 

The results show that fixed cost savings are much smaller than the change in producer surplus excluding fixed costs\footnote{In the light bulb market, light bulb firms produce and sell tens of products, some of which only differ in minor changes. It is consistent with the estimated results that fixed costs are relatively small.}, and allowing for fixed costs does not overturn the results. 

\subsubsection*{Role of price competition}

Then, why do dominant firms' profits increase if they jointly eliminate 2000h bulbs? We can think of two factors affecting firms' incentives: higher equilibrium prices and larger future replacement demand. 

Column (2) of Table \ref{tab:Durability-cartel-p-fixed-table} shows the case where firms do not change their pricing strategy from the one under the existence of 2000h bulbs even when 2000h lamps are eliminated from the market. Column (1) of the table shows the case where firms sell 2000h bulbs, and column (3) shows the case where firms eliminate 2000h bulbs and change pricing strategies based on the new set of products. The results show firms can increase joint profits, even when they cannot flexibly adjust product prices after eliminating 2000h bulbs. It accounts for roughly 53\% of the increase in the joint profits. The bottom part of Table \ref{tab:Durability-cartel-p-fixed-table} shows the fraction of no-inventory consumers and the amount of disposal evaluated at the stationary state in each setting. They imply consumers are more likely to experience product failure and consider new purchases in the absence of 2000h bulbs. Consequently, firms have incentives to eliminate 2000h bulbs mainly because of larger future replacement demand, and higher equilibrium prices further encourages firms to eliminate the products. 
\begin{center}
{\scriptsize{}}
\begin{table}[H]
\begin{centering}
\begin{tabular}{cccc}
\hline 
 & (1) & (2) & (3)\tabularnewline
Panasonic & 1000h \& 2000h & 1000h only & 1000h only\tabularnewline
Toshiba & 1000h \& 2000h & 1000h only & 1000h only\tabularnewline
Price & - & Fixed  & Not fixed\tabularnewline
\hline 
\hline 
Joint profit & \textsf{24.67} & \textsf{25.41} & 26.06\tabularnewline
Profit (Panasonic) & 10.77 & 10.49 & 10.99\tabularnewline
Profit (Toshiba) & \textsf{13.9} & \textsf{14.92} & 15.07\tabularnewline
\hline 
{\small{}No inventory consumers (\%)} & 18.61 & 19.28 & 19.21\tabularnewline
{\small{}Disposal (million)} & 3.04 & 3.15 & 3.13\tabularnewline
\hline 
\end{tabular}
\par\end{centering}
{\scriptsize{}\caption{Effect of eliminating 2000h incandescent lamps (Role of prices; The case without collusion on prices)\label{tab:Durability-cartel-p-fixed-table}}
}{\scriptsize\par}

{\footnotesize{}Notes.}{\footnotesize\par}

{\footnotesize{}The values at the top of the table show the discounted sum of profits of two dominant firms evaluated at $B^{(stationary,base)}$. Units of these values are billion yen.}{\footnotesize\par}

{\footnotesize{}The values of economic variables at the bottom part of the table are evaluated at the stationary state. The row of disposal represents the number of products disposed of in each period.}{\footnotesize\par}

{\footnotesize{}``Fixed'' in the row of prices implies that the column evaluates the results where product prices are fixed at the same levels as the ones in Column (1). ``Not fixed'' implies that the column evaluates the outcomes where product prices are adjusted so as to maximize each firm's profit based on the set of products.}{\footnotesize\par}

\end{table}
{\scriptsize\par}
\par\end{center}

\subsubsection{The case with collusion on prices}

So far, we have looked at the cases where oligopolistic firms set product prices competitively. Then, how do the results change if the dominant firms can also collude on prices so as to maximize their joint profits? Table \ref{tab:Durability-merger-table} shows the results.

By comparing columns (0) and (1) in the table, we can see that pricing cartel leads to higher product prices and larger profits of the firms. The joint profit of two dominant firms increases by roughly 16 billion yen. Nevertheless, further eliminating 2000h bulbs is not profitable as we can see by comparing columns (1) and (2): joint profit of the two firms decreases by 0.28 billion yen under the collusion on prices. Considering the 1.39 billion yen increase in joint profit in the absence of collusion on prices, it might seem to be counterintuitive. Nevertheless, it is not unusual, as discussed in Section \ref{subsubsec:Theoretical-analysis_durability} in detail by developing a theoretical model. 

Intuitively, firms have incentives to raise product durability under the condition where market-level consumer inventory positively affects ``service demand''. Here, ``service demand'' denotes the total amount of new and old products held by consumers. The condition would be satisfied in dynamic discrete choice models with random utility shocks where consumers can own at most one product like the current empirical model. This is because service demand is the sum of consumer inventory and new purchases, and each consumer's conditional probabilities of choosing products are not affected by market-level inventories given prices (See also Example \ref{ex:DDC} in Section \ref{subsubsec:Theoretical-analysis_durability}). Then, firms have incentives to set higher durability levels to get larger future service profit, which corresponds to the future profit from renting products. For details, see the discussion related to Proposition \ref{prop:collusion_price} in Section \ref{subsubsec:Theoretical-analysis_durability}.
\begin{center}
{\scriptsize{}}
\begin{table}[H]
\begin{centering}
{\small{}}%
\begin{tabular}{cccccc}
\hline 
 & {\small{}(0)} & {\small{}(1)} & {\small{}(2)} & {\small{}(3)} & {\small{}(4)}\tabularnewline
{\small{}Price cartel} & {\small{}No} & {\small{}Yes} & {\small{}Yes} & {\small{}Yes} & {\small{}Yes}\tabularnewline
{\small{}Panasonic} & {\small{}1000h \& 2000h} & {\small{}1000h \& 2000h} & {\small{}1000h only} & {\small{}1000h only} & {\small{}1000h \& 2000h}\tabularnewline
{\small{}Toshiba} & {\small{}1000h \& 2000h} & {\small{}1000h \& 2000h} & {\small{}1000h only} & {\small{}1000h \& 2000h} & {\small{}1000h only}\tabularnewline
\hline 
\hline 
{\small{}Joint profit} & {\small{}24.67} & {\small{}40.42} & {\small{}40.14} & {\small{}40.2} & {\small{}40.38}\tabularnewline
{\small{}Profit (Panasonic)} & {\small{}10.77} & {\small{}17.47} & {\small{}16.6} & {\small{}16.39} & {\small{}17.67}\tabularnewline
{\small{}Profit (Toshiba)} & {\small{}13.9} & {\small{}22.96} & {\small{}23.55} & {\small{}23.8} & {\small{}22.71}\tabularnewline
\hline 
{\small{}No inventory consumers (\%)} & {\small{}18.61} & {\small{}20.67} & {\small{}21.04} & {\small{}20.97} & {\small{}20.73}\tabularnewline
{\small{}Average price (1000h Inc.; yen)} & {\small{}94.73} & {\small{}126.74} & {\small{}125.37} & {\small{}125.62} & {\small{}126.53}\tabularnewline
{\small{}Average price (2000h Inc.; yen)} & {\small{}172.57} & {\small{}233.57} & - & {\small{}236.18} & {\small{}232.5}\tabularnewline
{\small{}Average price (CFL; yen)} & {\small{}796.53} & {\small{}989.49} & {\small{}989.33} & {\small{}989.35} & {\small{}989.47}\tabularnewline
{\small{}Disposal (million)} & {\small{}3.04} & {\small{}3.21} & {\small{}3.27} & {\small{}3.25} & {\small{}3.22}\tabularnewline
\hline 
{\small{}$\Delta$CS} & {\small{}-} & {\small{}-23.17} & {\small{}-23.3} & {\small{}-23.27} & {\small{}-23.19}\tabularnewline
{\small{}$\Delta$PS (excluding fixed cost)} & {\small{}-} & {\small{}17.79} & {\small{}17.62} & {\small{}17.65} & {\small{}17.76}\tabularnewline
{\small{}$\Delta$TS (excluding Ext. / fixed costs)} & {\small{}-} & {\small{}-5.38} & {\small{}-5.68} & {\small{}-5.62} & {\small{}-5.43}\tabularnewline
{\small{}$\Delta$Ext. (electricity usage)} & {\small{}-} & {\small{}-0.77} & {\small{}-1.33} & {\small{}-1.21} & {\small{}-0.86}\tabularnewline
{\small{}$\Delta$Ext. (waste disposal)} & {\small{}-} & {\small{}-0.02} & {\small{}0.00} & {\small{}0.00} & {\small{}-0.02}\tabularnewline
{\small{}$\Delta$TS (excluding fixed costs)} & {\small{}-} & {\small{}-4.59} & {\small{}-4.34} & {\small{}-4.41} & {\small{}-4.55}\tabularnewline
{\small{}Upper bound of Fixed cost savings} & {\small{}-} & {\small{}0} & {\small{}0.12} & {\small{}0.06} & {\small{}0.06}\tabularnewline
\hline 
\end{tabular}{\small\par}
\par\end{centering}
{\scriptsize{}\caption{Effect of eliminating 2000h incandescent lamps (The case with collusion on prices)\label{tab:Durability-merger-table}}
}    {\footnotesize{}Notes. }{\footnotesize\par}
        
    {\footnotesize{}The values at the top of the table show the discounted
    sum of variable profits of two dominant firms evaluated at $B^{(stationary,base)}$.}{\footnotesize\par}

    {\footnotesize{}The values at the bottom show the changes in surpluses relative to the base setting, and are also evaluated
    at $B^{(stationary,base)}$. Units of these values are billion yen.}{\footnotesize\par}
    
    {\footnotesize{}The values of economic variables in the middle part
    of the table are evaluated at the stationary state. Average prices are sales-weighted. The row of disposal represents the number of products disposed of in each period.}{\footnotesize\par}
    
    {\footnotesize{}$CS,PS,Ext,TS$ denote consumer surplus, producer surplus, externalties, and total surplus, respectively. $\Delta TS$ is defined by $\Delta TS\equiv\Delta CS+\Delta PS-\Delta Ext$.}{\footnotesize\par}

\end{table}
{\scriptsize\par}
\par\end{center}

\subsubsection{Theoretical analysis\label{subsubsec:Theoretical-analysis_durability}}

As discussed above, we obtained the following empirical results based on the simulation:
\begin{enumerate}
\item Dominant firms have incentives to collude to eliminate 2000h (high durability) incandescent lamps, though it is profitable to sell them for each firm.
\item When they can collude on prices, they don't have incentives to eliminate high durability bulbs.
\end{enumerate}
To understand the logic behind the results, we develop a simple analytical model in line with the previous theoretical studies.\footnote{The purpose of developing an analytical model is not to fully illustrate the factors affecting firms' incentives on durability in the structural model. Rather, it is intended to point out a factor not fully investigated in the previous theoretical studies.} I first describe the model settings, and then discuss the counterpart of Swan's independence result (\citet{swan1970durability}), which claims that firms choose cost-minimizing product durability levels under some assumptions, and which is regarded as the benchmark result in the literature. We then discuss more general results by relaxing some of the assumptions. Propositions \ref{prop:collusion_durability} and \ref{prop:collusion_price} discussed later are related to the two empirical results respectively. 

\subsubsection*{Model setting}

As in \citet{bulow1986economic}, we consider a two-period model where firms maximize profits over two periods. Note that we adopt a model of differentiated products, consistent with the empirical model in the current study as opposed to \citet{bulow1986economic}. Besides, we consider a setting where firms can continuously change product durability levels to make the analysis easier, unlike the empirical setting where possible durability levels are exogenously given and firms choose the set of products. Essential intuitions would not be lost with the specification. In period 1, firm $j$ produces and sells durable product $j$ which still works even in period 2 with probability $\phi_{j}$. In period 2, firm $j$ produces and sells product $j$, but it lasts only one period, consistent with the two-period model. Let $q_{jt}$ and $p_{jt}$ $(t=1,2)$ be the quantity and price of product $j$ sold at time $t$. Let $c_{jt=1}(\phi_{j})$ and $c_{jt=2}$ be the marginal cost of product $j$ at time $t=1$ and nondurable product at time $t=2$. Note that product $j$'s marginal cost, $c_{jt=1}(\phi_{j})$, depends on the durability of the product $\phi_{j}$. For convenience, let $\phi_{jt=1}=\phi_{j}$ and $\phi_{jt=2}=0$.

To make the point clear, we consider the setting where firms can commit to future product prices at period 2. We also assume firms compete in product prices, rather than quantities. At the beginning of time 1, each firm decides its own product's prices $p_{jt=1},p_{jt=2}$ and durability $\phi_{j}$. 

Besides, we define the following terms:
\begin{itemize}
\item $\Psi_{jt}$: Aggregate stock of used product $j$ at time $t$ (Consumer inventory)

In our two-period model, $\Psi_{jt=1}=0$ and $\Psi_{jt=2}=\phi_{j}q_{jt=1}$ holds.
\item $Q_{jt}=q_{jt}+\Psi_{jt}$: Total services yielded by the entire stock of product $j$ at time $t$ (Service demand)

In our two-period model, $Q_{jt=1}=q_{jt=1}$ and $Q_{jt=2}=\phi_{j}q_{jt=1}+q_{jt=2}$ holds.
\item $P_{jt}$: ``Service price'', which satisfies $p_{jt=1}=P_{jt=1}+\beta\phi_{j}P_{jt=2}$ and $p_{jt=2}=P_{jt=2}$ in our two-period model.

$P_{jt}$ corresponds to the rental price of the product if the product is rented rather than sold.
\end{itemize}
Then, the firm $j$'s profit maximization problem is:

\begin{eqnarray*}
\max_{p_{jt=1},p_{jt=2},\phi_{j}}V_{j}^{F} & \equiv & \left(p_{jt=1}-c_{jt=1}(\phi_{j})\right)q_{jt=1}+\beta(p_{jt=2}-c_{jt=2})q_{jt=2},
\end{eqnarray*}
where $\beta$ denotes the discount factor shared by firms and consumers. The problem is equivalent to:

\begin{eqnarray*}
\max_{P_{jt=1},P_{jt=2},\phi_{j}}V_{j}^{F} & = & \left(P_{jt=1}+\beta\phi_{j}P_{jt=2}-c_{jt=1}(\phi_{j})\right)Q_{jt=1}+\beta(P_{jt=2}-c_{jt=2})\left(Q_{jt=2}-\phi_{j}Q_{jt=1}\right)\\
 & = & \left(P_{jt=1}-c_{jt=1}(\phi_{j})+\beta c_{jt=2}\phi_{j}\right)Q_{jt=1}+\beta(P_{jt=2}-c_{jt=2})Q_{jt=2}.
\end{eqnarray*}

Here, we consider a model $q_{jt}$ as a function of $\left\{ \Psi_{kt}\right\} _{k\in\mathcal{J}}$, $\left\{ p_{kt}\right\} _{k\in\mathcal{J}},\left\{ p_{kt+\tau}\right\} _{k\in\mathcal{J},\tau\geq1}$, $\left\{ \phi_{kt}\right\} _{k\in\mathcal{J}}$,\footnote{In more general setting, $q_{jt}$ can be a function of $\left\{ \Psi_{ikt}\right\} _{k\in\mathcal{J}}$, where $\Psi_{ikt}$ denotes the stock of used product $j$ held by consumer $i$ at time $t$, and it can be multi-dimensional. We abstract away from these settings to simplify the analysis.} where $\mathcal{J}$ denotes the set of products. Then, $Q_{jt}$ is also a function of $\left\{ \Psi_{kt}\right\} _{k\in\mathcal{J}},\left\{ P_{kt}\right\} _{k\in\mathcal{J}},\left\{ P_{kt+\tau}\right\} _{k\in\mathcal{J},\tau\geq1},\left\{ \phi_{kt}\right\} _{k\in\mathcal{J}}$. By using service prices $P_{kt}$, $Q_{jt}$ can be represented as $Q_{jt}\left(\left\{ \Psi_{kt}\right\} _{k\in\mathcal{J}},\left\{ P_{kt}\right\} _{k\in\mathcal{J}},\left\{ P_{kt+\tau}\right\} _{k\in\mathcal{J},\tau\neq0},\left\{ \phi_{kt}\right\} _{k\in\mathcal{J}}\right)$. 
\begin{example}
(Dynamic discrete choice model)\label{ex:DDC}
\end{example}
Consider a dynamic discrete choice model with random utility shock and without persistent consumer heterogeneity. Suppose consumers do not make purchase decisions when they already own functioning products. Then, $Q_{jt=2}$ can be represented as:

\begin{eqnarray*}
Q_{jt=2} & = & \phi_{j}q_{jt=1}+q_{jt=2}\\
 & = & \phi_{j}q_{jt=1}+\left(M-\sum_{k\in\mathcal{J}}\phi_{k}q_{kt=1}\right)s_{jt=2}^{(ccp)}\\
 & = & \Psi_{jt=2}+\left(M-\sum_{k\in\mathcal{J}}\Psi_{kt=2}\right)s_{jt=2}^{(ccp)},
\end{eqnarray*}
where $M$ denotes the market size, and $s_{jt=2}^{(ccp)}$ denotes conditional choice probability that consumers choose product $j$ at time $t=2$. $s_{jt=2}^{(ccp)}$ are functions of $\left\{ P_{kt}\right\} _{k\in\mathcal{J}},\left\{ P_{kt+\tau}\right\} _{k\in\mathcal{J},\tau\neq0},\left\{ \phi_{kt}\right\} _{k\in\mathcal{J}}$, and $Q_{jt=2}$ is in the form of $Q_{jt}\left(\left\{ \Psi_{kt}\right\} _{k\in\mathcal{J}},\left\{ P_{kt}\right\} _{k\in\mathcal{J}},\left\{ P_{kt+\tau}\right\} _{k\in\mathcal{J},\tau\neq0},\left\{ \phi_{kt}\right\} _{k\in\mathcal{J}}\right)$.

Under the setting, $\frac{\partial Q_{jt=2}}{\partial\Psi_{jt=2}}=1-s_{jt=2}^{(ccp)}>0$ holds. Intuitively, each consumer not owning products only takes account of product prices and durability, not market-level consumer inventory, when choosing a product. Though larger consumer inventory $\Psi_{kt=2}$ decreases time $t=2$ demand $\left(M-\sum_{k\in\mathcal{J}}\Psi_{kt=2}\right)s_{jt=2}^{(ccp)}$, the size of the decrease is smaller than the increase in consumer inventory $\Psi_{kt=2}$, and $Q_{jt=2}=\Psi_{jt=2}+\left(M-\sum_{k\in\mathcal{J}}\Psi_{kt=2}\right)s_{jt=2}^{(ccp)}$ increases as $\Psi_{kt=2}$ increases.
\begin{example}
(Model of investment with adjustment cost)\label{ex:adjustment_cost}

Consider a simple two-period model of investment with adjustment costs, intensively studied in the macroeconomics literature (see \citet{romer2012advanced}). Here, An agent making investment decisions are treated as the consumers in the durable goods (capital goods) market. The agent maximizes its long-run profit $\Pi(K_{t=1})-P_{t=1}I_{t=1}-C(I_{t=1})+\beta\Pi(K_{t=2})$ with respect to $I_{t=1}$, such that $K_{t=2}=(1-\phi)K_{t=1}+I_{t=1}$, where $C(I_{t})$ denotes the convex adjustment cost such that $C^{\prime\prime}(I)>0$. $K_{t}$ and $I_{t}$ denote the capital stock and the amount of investment at time $t$, and $\phi$ denotes the depreciation rate of the capital. $\Pi(K_{t})$ denotes the profit given capital stocks $K_{t}$ at time $t$, such that $\Pi^{\prime}(K_{t})>0$ and $\Pi^{\prime\prime}(K_{t})<0$. Then, its first order condition is equivalent to $P_{t=1}+C^{\prime}(K_{t=2}-\phi K_{t=1})=\beta\Pi^{\prime}(K_{t=2})$, which implies $K_{t=2}$ depends on $K_{t=1}$. Note that variables $\phi K_{t=1}$, $K_{t}$, $\phi$ correspond to $\Psi_{jt=2}$, $Q_{jt}$, and $\phi_{j}$ in our model. Let $\Psi_{t=2}=\phi K_{t=1}$. 

Under the setting, $\frac{\partial K_{t=2}}{\partial\Psi_{t=2}}>0$ holds\footnote{By differentiating $P_{t=1}+C^{\prime}(K_{t=2}-\Psi_{t=2})-\beta\Pi^{\prime}(K_{t=2})=0$ with respect to $\Psi_{t=2}$, $\frac{\partial K_{t=2}}{\partial\Psi_{t=2}}\left(C^{\prime\prime}(K_{t=2}-\Psi_{t=2})-\beta\pi^{\prime\prime}(K_{t=2})\right)=C^{\prime\prime}(K_{t=2}-\Psi_{t=2})$ holds. Since $C^{\prime\prime}(K_{t=2}-\Psi_{t=2})>0$ and $\pi^{\prime\prime}(K_{t=2})<0$ hold, we obtain $\frac{\partial K_{t=2}}{\partial\Psi_{t=2}}>0$.}. Intuitively, under the adjustment cost in investment, optimal capital stock in the next period $K_{t=2}$ the agent chooses is smaller if the remaining capital stock $\Psi_{t=2}$ is smaller, because larger investment incurs larger adjustment cost. 
\end{example}

\paragraph*{Swan's independence result}

\citet{swan1970durability}, \citet{bulow1986economic}, and others considered a demand function where $Q_{jt}$ only depends on the current ``service price'' of the product. Formally, they implicitly imposed the following assumptions:

\begin{assumption}[Independence from consumers' inventory]

$Q_{jt}\left(\left\{ \Psi_{kt}\right\} _{k\in\mathcal{J}},\left\{ P_{kt}\right\} _{k\in\mathcal{J}},\left\{ P_{kt+\tau}\right\} _{k\in\mathcal{J},\tau\neq0},\left\{ \phi_{kt}\right\} _{k\in\mathcal{J}}\right)$ does not depend on consumers' inventory $\left\{ \Psi_{kt}\right\} _{k\in\mathcal{J}}$.

\label{as:inventory_indep}
\end{assumption}

\begin{assumption}[No preference for durability per se]
\label{as:no_pref_durability}

$Q_{jt}\left(\left\{ \Psi_{kt}\right\} _{k\in\mathcal{J}},\left\{ P_{kt}\right\} _{k\in\mathcal{J}},\left\{ P_{kt+\tau}\right\} _{k\in\mathcal{J},\tau\neq0},\left\{ \phi_{kt}\right\} _{k\in\mathcal{J}}\right)$ does not depend on product durability $\left\{ \phi_{kt}\right\} _{k\in\mathcal{J}}$.

\end{assumption}

\begin{assumption}[Independence from service prices in other periods]
\label{as:independence_future_service_price}

$Q_{jt}\left(\left\{ \Psi_{kt}\right\} _{k\in\mathcal{J}},\left\{ P_{kt}\right\} _{k\in\mathcal{J}},\left\{ P_{kt+\tau}\right\} _{k\in\mathcal{J},\tau\neq0},\left\{ \phi_{kt}\right\} _{k\in\mathcal{J}}\right)$ does not depend on service prices in other periods $\left\{ P_{kt+\tau}\right\} _{k\in\mathcal{J},\tau\geq1}$.

\end{assumption}

Under these assumptions, the profit maximization problem with respect to $\phi_{j}$ is equivalent to the following minimization problem:

\[
\min_{\phi_{j}}\left[c_{jt=1}(\phi_{j})-\beta c_{jt=2}\phi_{j}\right].
\]

It implies the firm chooses product durability so that it minimizes the marginal cost of providing services at time 1, $c_{jt=1}(\phi_{j})-\beta c_{jt=2}\phi_{j}$. This is in line with \citet{swan1970durability}, which showed that firms choose product durability so as to minimize the cost of providing unit services. It also implies that market structure does not affect firms' durability choices.

\subsubsection*{Results under more general settings}

If we do not assume neither of the two assumptions, we obtain different conclusions.\footnote{In the case consumers are present-biased as in \citet{Li2024}, consumers' preference does not solely depend on the service price, and may also depend on durability levels.} Proof of the propositions discussed below are shown in Appendix \ref{subsec:Proof-theory_durability}. We first prepare the following alternative assumptions:

\begin{assbis}{as:inventory_indep}[Dependence on consumer inventory]

$\frac{\partial Q_{jt}}{\partial\Psi_{kt}}\geq0$ holds for $Q_{jt}\left(\left\{ \Psi_{kt}\right\} _{k\in\mathcal{J}},\left\{ P_{kt}\right\} _{k\in\mathcal{J}},\left\{ P_{kt+\tau}\right\} _{k\in\mathcal{J},\tau\neq0},\left\{ \phi_{kt}\right\} _{k\in\mathcal{J}}\right)$.

\label{as:inventory_diff}
\end{assbis}

\begin{assbis}{as:no_pref_durability}[Preference for durability per se]

$\frac{\partial Q_{jt}}{\partial\phi_{jt}}\geq0,\ \frac{\partial Q_{kt}}{\partial\phi_{jt}}\leq0\ (k\neq j)$ hold for $Q_{jt}\left(\left\{ \Psi_{kt}\right\} _{k\in\mathcal{J}},\left\{ P_{kt}\right\} _{k\in\mathcal{J}},\left\{ P_{kt+\tau}\right\} _{k\in\mathcal{J},\tau\neq0},\left\{ \phi_{kt}\right\} _{k\in\mathcal{J}}\right)$.

\label{as:pref_for_durab_diff}
\end{assbis}

Assumption \ref{as:inventory_diff} implies larger consumer inventory leads to more service demand. $\frac{\partial Q_{jt}}{\partial\phi_{jt}}\geq0$ in Assumption \ref{as:pref_for_durab_diff} implies demand for product $j$ increases if its durability level is higher, and $\frac{\partial Q_{kt}}{\partial\phi_{jt}}\leq0\ (k\neq j)$ implies demand for products other than product $j$ decreases as product $j$'s durability gets higher.

Regarding the first simulation result, we obtain the following statement:
\begin{prop}
\label{prop:collusion_durability}Suppose that firms $j=1$ and $j=2$ collude only on durability to maximize their joint profit, and Assumption \ref{as:pref_for_durab_diff} holds. If product prices $\{P_{kt}\}_{k\in\mathcal{J},t=1,2}$ are fixed at no durability cartel levels, and additionally $\sum_{k\in\mathcal{J}}\left(\frac{\partial q_{kt=1}}{\partial\phi_{j=1}}\phi_{k}\frac{\partial q_{j=2t=2}}{\partial\Psi_{kt=2}}\right)\leq0$ and $\frac{\partial^{2}\left(V_{j=1t=1}+V_{j=2t=1}\right)}{\partial\phi_{j=1}^{2}}<0$ hold, then the collusion on durability leads to lower durability of product $j=1$.
\end{prop}
Regarding the technical assumption $\sum_{k\in\mathcal{J}}\left(\frac{\partial q_{kt=1}}{\partial\phi_{j=1}}\phi_{k}\frac{\partial q_{j=2t=2}}{\partial\Psi_{kt=2}}\right)\leq0$, if $\phi_{j}$ does not affect other products' demand $q_{jt=1}$, the left hand side is equal to $\frac{\partial q_{j=1t=1}}{\partial\phi_{j=1}}\phi_{j=1}\frac{\partial q_{j=2t=2}}{\partial\Psi_{j=1t=2}}$. The value typically takes a positive value because $\frac{\partial q_{j=1t=1}}{\partial\phi_{j=1}}\geq0$ and $\frac{\partial q_{j=2t=2}}{\partial\Psi_{j=1t=2}}\leq0$ typically hold. $\frac{\partial^{2}\left(V_{j=1t=1}+V_{j=2t=1}\right)}{\partial\phi_{j=1}^{2}}<0$ is related to the second order condition of the durability choice. The proposition implies firms colluding on durability have incentives to lower durability under consumer preference for durability. Intuitively, firms have incentives to internalize the effects of their durability choices on the competitor's profit, if they can collude on durability.

Intuitively, firms have incentives to set higher durability to attract more customers who have preference for durability, but they also have incentives to lower durability if they can collude. Though similar incentives would exist in the static models where firms choose product quality, the incentives get intensified because own firm products' higher durability leads to not only competitors' lower current demand but also lower future demand. 

\medskip{}

The next proposition is related to the second simulation result.
\begin{prop}
\label{prop:collusion_price}Suppose that product $j$'s service prices $P_{jt}$ are exogenously given, but firm $j$ can choose its product durability to maximize its profit. Under Assumptions \ref{as:inventory_diff}, \ref{as:no_pref_durability}, and \ref{as:independence_future_service_price}, and a technical assumption $\frac{\partial}{\partial\Psi_{jt=2}}\left[\frac{\partial Q_{jt=2}}{\partial P_{jt=2}}\right]\geq0$, firm $j$ raises its product durability if $P_{jt=2}$ is higher given $P_{jt=1}$.
\end{prop}
The technical assumption $\frac{\partial}{\partial\Psi_{jt=2}}\left[\frac{\partial Q_{jt=2}}{\partial P_{jt=2}}\right]\geq0$ implies that the absolute value of price sensitivity of service demand, $-\frac{\partial Q_{jt}}{\partial P_{jt=2}}$, decreases as the size of consumer inventory $\Psi_{jt}$ increases, which is likely to be satisfied in many demand functions.

Typically, service prices are raised under collusion on prices. In such a case, Proposition \ref{prop:collusion_price} implies it is optimal for each firm to set a higher durability level. It is consistent with the empirical results that firms have incentives to eliminate high durability 2000h bulbs when they can collude on durability but cannot collude on prices, though no incentive when they can. 

As shown below, firm $j$'s first order condition with respect to durability $\phi_{j}$ is:

\begin{eqnarray*}
0 & = & \frac{\partial V_{j}^{F}}{\partial\phi_{j}}\\
 & = & \left(-\frac{\partial c_{jt=1}(\phi_{j})}{\partial\phi_{j}}+\beta c_{jt=2}\right)Q_{jt=1}+\beta\left(P_{jt=2}-c_{jt=2}\right)\frac{\partial\Psi_{jt=2}}{\partial\phi_{j}}\frac{\partial Q_{jt=2}}{\partial\Psi_{jt=2}}\\
 & = & \left(-\frac{\partial c_{jt=1}(\phi_{j})}{\partial\phi_{j}}+\beta c_{jt=2}\right)Q_{jt=1}+\beta\left(P_{jt=2}-c_{jt=2}\right)Q_{jt=1}\frac{\partial Q_{jt=2}}{\partial\Psi_{jt=2}}.
\end{eqnarray*}

The equation implies the firm has incentives not only to minimize the cost of providing services at time $t=1$ (the first term), but also get more service demand (the second term). The second incentive gets large when the firm can set a higher service price $P_{jt=2}$. Intuitively, when consumer inventory positively affects its product's service demand, the firm has an incentive to set higher durability to increase future service demand through the increase in consumer inventory. Though it largely depends on the sign of $\frac{\partial Q_{jt=2}}{\partial\Psi_{jt=2}}$, it is positive in the models in Examples \ref{ex:DDC} and \ref{ex:adjustment_cost}, as discussed before. Note that the proposition might not hold if we relax Assumption \ref{as:no_pref_durability} (No preference for durability per se). Nevertheless, the discussion sheds light on firms' incentives to raise durability by affecting consumer inventory.\footnote{\citet{wu2014loyalty} theoretically showed that a monopolist has an incentive to overextend durability under consumers' brand loyalty. Under brand loyalty, consumer inventory positively affects service demand, and their result is consistent with ours.}. 

\subsection{Durability and Welfare\label{subsec:Durability-and-Welfare}}

Next, we evaluate whether the product choices firms make regarding product durability are socially optimal or not. In reality, two dominant firms Panasonic and Toshiba produce both high and low durability incandescent lamps / CFLs. We compare it with the counterfactual scenarios where they produce only high or only low durability products for each product category.\footnote{Since there are some incandescent lamp and CFL products that are not substitutable, we assume firms continue selling these products. These products account for less than 20\% of their sales.} Table \ref{tab:welfare_Inc} shows the results for incandescent lamps, and Table \ref{tab:welfare_CFL} shows the results for CFLs. Besides, the change in surpluses are also shown in Tables \ref{tab:Durability-cartel-table} and \ref{tab:Durability-merger-table} discussed in the last subsection.

Columns (A-1),(A-2),(A-3) in Tables \ref{tab:welfare_Inc} and \ref{tab:welfare_CFL} show how Panasonic's product introduction decisions affect economic variables and welfare. Similarly, columns (B-1),(B-2),(B-3) show the results of Toshiba. Table \ref{tab:welfare_Inc} shows that not selling 2000h incandescent lamps is socially optimal. Though the introduction of 2000h bulbs raises consumer surplus, it largely decreases the producer surplus. Consequently, total surplus absent externalities decreases when 2000h bulbs are introduced. 

As discussed in \citet{fan2020competition}, empirically analyzing firms' product introduction decisions under static demand, oligopolistic firms do not take account of business-steeling effect, and it might lead to excessive product offerings. In contrast, firms do not consider consumer surplus, which might lead to insufficient product offerings. In the case of our dynamic setting with durability, oligopolistic firms do not internalize their own product choice decisions concerning durability on other firms' current and future demand, leading to socially excessive durability. Though consumer surplus decreases, the former is larger than the latter in the case of incandescent lamps. Compared to static settings, the former might be larger, because firms' decisions affect not only competitors' current profits but also their future profits.

The bottom part of Table \ref{tab:Durability-cartel-table} also shows that the collusion on durability to eliminate 2000h bulbs absent price cartels increases producer surplus and total surplus, though it decreases consumer surplus. Whether such kind of collusion should be accepted or not would depend on whether the competitive authorities make decisions based on the consumer welfare standard or total surplus standard. 

Note that the elimination of 2000h bulbs should be more strongly accepted when we consider negative environmental externalities as shown in the tables, because the electricity usage of incandescent lamps are much larger than that of CFLs, and the elimination of 2000h bulbs decreases the demand for incandescent lamps. In the tables, we consider two types of externalities: CO2 emission from electricity usage through bulbs, and waste emission. For details of the calculation of externalities, see Appendix \ref{subsec:Details_algorithm_supply}.

Regarding CFLs, eliminating neither low durability (6000h) nor high durability (10000h/12000h) products decreases consumer surplus and producer surplus. It implies firms' current decisions, namely, the coexistence of high and low durability CFLs, are socially optimal. The conclusion is different from the one for incandescent lamps, and it implies the role of quantitative analysis is very large because the social optimality of durability is largely different across product categories. 
\begin{center}
{\scriptsize{}}
\begin{table}[H]
\begin{centering}
\scalebox{0.9}{{\small{}}%
\begin{tabular}{ccccccc}
\hline 
 & \multicolumn{3}{c}{{\small{}Panasonic}} & \multicolumn{3}{c}{{\small{}Toshiba}}\tabularnewline
\cline{2-7} \cline{3-7} \cline{4-7} \cline{5-7} \cline{6-7} \cline{7-7} 
 & {\small{}(A-1)} & {\small{}(A-2)} & {\small{}(A-3)} & {\small{}(B-1)} & {\small{}(B-2)} & {\small{}(B-3)}\tabularnewline
 & {\small{}1000h \& 2000h} & {\small{}1000h only} & {\small{}2000h only} & {\small{}1000h \& 2000h} & {\small{}1000h only} & {\small{}2000h only}\tabularnewline
\hline 
\hline 
{\small{}Joint profit} & {\small{}24.67} & {\small{}25.77} & {\small{}23.41} & {\small{}24.67} & {\small{}25.02} & {\small{}25.29}\tabularnewline
{\small{}Profit (Panasonic)} & {\small{}10.77} & {\small{}10.68} & {\small{}9.62} & {\small{}10.77} & {\small{}11.11} & {\small{}12.26}\tabularnewline
{\small{}Profit (Toshiba)} & {\small{}13.9} & {\small{}15.09} & {\small{}13.79} & {\small{}13.9} & {\small{}13.91} & {\small{}13.02}\tabularnewline
\hline 
{\small{}No inventory consumers (\%)} & {\small{}18.61} & {\small{}19.08} & {\small{}18} & {\small{}18.61} & {\small{}18.74} & {\small{}17.78}\tabularnewline
{\small{}Average price (1000h Inc.; yen)} & {\small{}94.73} & {\small{}98.08} & {\small{}89.43} & {\small{}94.73} & {\small{}94.59} & {\small{}112.39}\tabularnewline
{\small{}Average price (2000h Inc.; yen)} & {\small{}172.57} & {\small{}178.26} & {\small{}167.33} & {\small{}172.57} & {\small{}174.36} & {\small{}173.92}\tabularnewline
{\small{}Average price (CFL; yen)} & {\small{}796.53} & {\small{}798.87} & {\small{}796.41} & {\small{}796.53} & {\small{}796.9} & {\small{}798.09}\tabularnewline
{\small{}Disposal (million)} & {\small{}3.04} & {\small{}3.11} & {\small{}2.94} & {\small{}3.04} & {\small{}3.06} & {\small{}2.89}\tabularnewline
\hline 
{\small{}$\Delta$CS} & {\small{}-} & {\small{}-1.09} & {\small{}-0.41} & {\small{}-} & {\small{}-0.34} & {\small{}-2.03}\tabularnewline
{\small{}$\Delta$PS (excluding fixed cost)} & {\small{}-} & {\small{}1.2} & {\small{}-1.26} & {\small{}-} & {\small{}0.37} & {\small{}0.74}\tabularnewline
{\small{}$\Delta$TS (excluding Ext. / fixed costs)} & {\small{}-} & {\small{}0.11} & {\small{}-1.67} & {\small{}-} & {\small{}0.03} & {\small{}-1.29}\tabularnewline
{\small{}$\Delta$Ext. (electricity usage)} & {\small{}-} & {\small{}-1.01} & {\small{}0.92} & {\small{}-} & {\small{}-0.31} & {\small{}0.77}\tabularnewline
{\small{}$\Delta$Ext. (waste disposal)} & {\small{}-} & {\small{}0.03} & {\small{}-0.02} & {\small{}-} & {\small{}0.01} & {\small{}-0.02}\tabularnewline
{\small{}$\Delta$TS (excluding fixed costs)} & {\small{}-} & {\small{}1.08} & {\small{}-2.57} & {\small{}-} & {\small{}0.33} & {\small{}-2.04}\tabularnewline
{\small{}Upper bound of Fixed cost savings} & {\small{}-} & {\small{}0.06} & {\small{}0.83} & {\small{}-} & {\small{}0.06} & {\small{}0.83}\tabularnewline
\hline 
\end{tabular}}
\par\end{centering}
{\scriptsize{}\caption{Durability and Welfare (Incandescent lamps)\label{tab:welfare_Inc}}
}{\scriptsize\par}

    {\footnotesize{}Notes. }{\footnotesize\par}
        
    {\footnotesize{}The values at the top of the table show the discounted
    sum of variable profits of two dominant firms evaluated at $B^{(stationary,base)}$.}{\footnotesize\par}

    {\footnotesize{}The values at the bottom show the changes in surpluses relative to the base setting, and are also evaluated
    at $B^{(stationary,base)}$. Units of these values are billion yen.}{\footnotesize\par}
    
    {\footnotesize{}The values of economic variables in the middle part
    of the table are evaluated at the stationary state. Average prices are sales-weighted. The row of disposal represents the number of products disposed of in each period.}{\footnotesize\par}
    
    {\footnotesize{}$CS,PS,Ext,TS$ denote consumer surplus, producer surplus, externalties, and total surplus, respectively. $\Delta TS$ is defined by $\Delta TS\equiv\Delta CS+\Delta PS-\Delta Ext$.}{\footnotesize\par}

\end{table}
{\scriptsize\par}
\par\end{center}

\begin{center}
{\scriptsize{}}
\begin{table}[H]
\begin{centering}
\scalebox{0.9}{{\small{}}%
\begin{tabular}{ccccccc}
\hline 
 & \multicolumn{3}{c}{{\small{}Panasonic}} & \multicolumn{3}{c}{{\small{}Toshiba}}\tabularnewline
\cline{2-7} \cline{3-7} \cline{4-7} \cline{5-7} \cline{6-7} \cline{7-7} 
 & {\small{}(A-1)} & {\small{}(A-2)} & {\small{}(A-3)} & {\small{}(B-1)} & {\small{}(B-2)} & {\small{}(B-3)}\tabularnewline
 & {\small{}6000h \& 10000h} & {\small{}6000h only} & {\small{}10000h only} & {\small{}6000h \& 12000h} & {\small{}6000h only} & {\small{}12000h only}\tabularnewline
\hline 
\hline 
{\small{}Joint profit} & {\small{}24.67} & {\small{}24.53} & {\small{}24.58} & {\small{}24.67} & {\small{}24.19} & {\small{}23.32}\tabularnewline
{\small{}Profit (Panasonic)} & {\small{}10.77} & {\small{}9.65} & {\small{}8.53} & {\small{}10.77} & {\small{}11.24} & {\small{}12.27}\tabularnewline
{\small{}Profit (Toshiba)} & {\small{}13.9} & {\small{}14.88} & {\small{}16.05} & {\small{}13.9} & {\small{}12.95} & {\small{}11.05}\tabularnewline
\hline 
{\small{}No inventory consumers (\%)} & {\small{}18.61} & {\small{}19.33} & {\small{}18.16} & {\small{}18.61} & {\small{}19.55} & {\small{}18.25}\tabularnewline
{\small{}Average price (1000h Inc.; yen)} & {\small{}94.73} & {\small{}94.05} & {\small{}93.07} & {\small{}94.73} & {\small{}94.8} & {\small{}94.94}\tabularnewline
{\small{}Average price (2000h Inc.; yen)} & {\small{}172.57} & {\small{}171.81} & {\small{}170.84} & {\small{}172.57} & {\small{}172.71} & {\small{}172.74}\tabularnewline
{\small{}Average price (CFL; yen)} & {\small{}796.53} & {\small{}777.03} & {\small{}836.82} & {\small{}796.53} & {\small{}772.14} & {\small{}830.21}\tabularnewline
{\small{}Disposal (million)} & {\small{}3.04} & {\small{}3.15} & {\small{}2.95} & {\small{}3.04} & {\small{}3.19} & {\small{}2.96}\tabularnewline
\hline 
{\small{}$\Delta$CS} & {\small{}-} & {\small{}-1.68} & {\small{}-3.57} & {\small{}-} & {\small{}-0.96} & {\small{}-2.96}\tabularnewline
{\small{}$\Delta$PS (excluding fixed cost)} & {\small{}-} & {\small{}-0.1} & {\small{}0.00} & {\small{}-} & {\small{}-0.46} & {\small{}-1.27}\tabularnewline
{\small{}$\Delta$TS (excluding Ext. / fixed costs)} & {\small{}-} & {\small{}-1.78} & {\small{}-3.57} & {\small{}-} & {\small{}-1.42} & {\small{}-4.22}\tabularnewline
{\small{}$\Delta$Ext. (electricity usage)} & {\small{}-} & {\small{}0.15} & {\small{}0.37} & {\small{}-} & {\small{}0.01} & {\small{}0.41}\tabularnewline
{\small{}$\Delta$Ext. (waste disposal)} & {\small{}-} & {\small{}0.00} & {\small{}-0.05} & {\small{}-} & {\small{}0.01} & {\small{}-0.03}\tabularnewline
{\small{}$\Delta$TS (excluding fixed costs)} & {\small{}-} & {\small{}-1.93} & {\small{}-3.9} & {\small{}-} & {\small{}-1.44} & {\small{}-4.6}\tabularnewline
{\small{}Upper bound of Fixed cost savings} & {\small{}-} & {\small{}1.05} & {\small{}1.98} & {\small{}-} & {\small{}0.9} & {\small{}2.57}\tabularnewline
\hline 
\end{tabular}}
\par\end{centering}
{\scriptsize{}\caption{Durability and Welfare (CFLs)\label{tab:welfare_CFL}}
}{\scriptsize\par}

    {\footnotesize{}Notes. }{\footnotesize\par}
        
    {\footnotesize{}The values at the top of the table show the discounted
    sum of variable profits of two dominant firms evaluated at $B^{(stationary,base)}$.}{\footnotesize\par}

    {\footnotesize{}The values at the bottom show the changes in surpluses relative to the base setting, and are also evaluated
    at $B^{(stationary,base)}$. Units of these values are billion yen.}{\footnotesize\par}
    
    {\footnotesize{}The values of economic variables in the middle part
    of the table are evaluated at the stationary state. Average prices are sales-weighted. The row of disposal represents the number of products disposed of in each period.}{\footnotesize\par}
    
    {\footnotesize{}$CS,PS,Ext,TS$ denote consumer surplus, producer surplus, externalties, and total surplus, respectively. $\Delta TS$ is defined by $\Delta TS\equiv\Delta CS+\Delta PS-\Delta Ext$.}{\footnotesize\par}

\end{table}
{\scriptsize\par}
\par\end{center}

\subsection{Firms' commitment ability, consumer expectations and firms' dynamic incentives\label{subsec:Alternative_spec}}

In the base setting, we employed the model of Markov perfect equilibrium where firms use Markov pricing strategies. It implies firms in the model cannot commit to future product prices. Nevertheless, as is well known in the theoretical literature (\citet{coase1972durability}, \citet{bulow1986economic}), durable goods producers' commitment ability of future product prices largely affects their pricing: firms commiting to keep high future prices have incentives to lower their product prices once current profits are earned, and firms without commitment ability set lower prices. 

To assess the role of firms' commitment, I numerically solved the equilibrium where firms can fully commit to setting constant future prices, and compared the results with the case of Markov perfect equilibrium (MPE). For details of the solution method, see Appendix \ref{subsec:Details_algorithm_supply_commit}. 

\begin{figure}[H]
\begin{centering}
\includegraphics[scale=0.6]{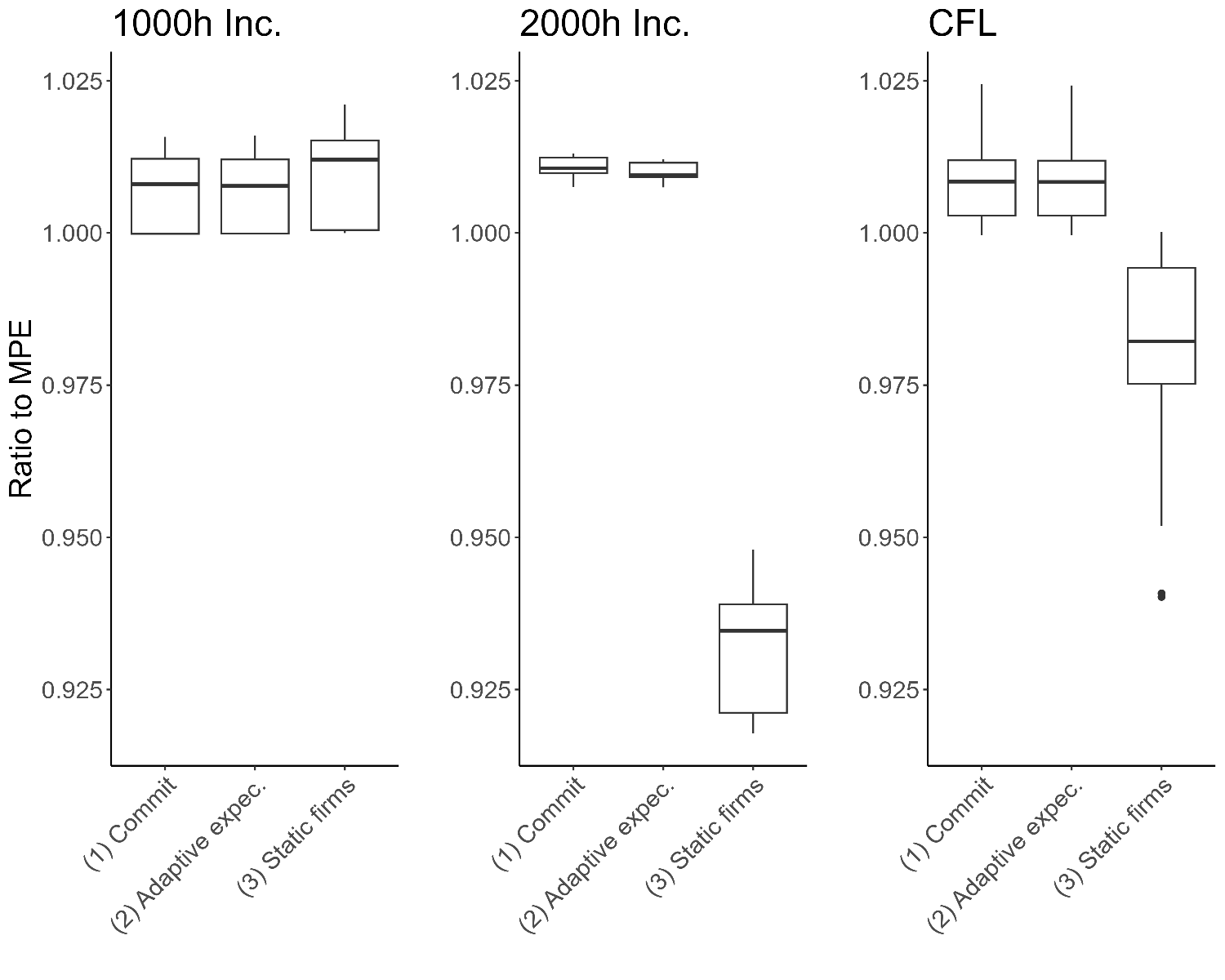}
\par\end{centering}
\caption{Distribution of equilibrium product prices relative to the base prices under MPE\label{fig:p_ratio_dist_alternative_spec}}
\end{figure}

Column (0) of Table \ref{tab:results_alternative_spec} shows the outcomes under the baseline Markov perfect equilibrium, and column (1) shows the outcomes under the setting where firms can credibly commit to setting future constant product prices. The distribution of the ratio of the prices is shown in Figure \ref{fig:p_ratio_dist_alternative_spec}. The results imply firms can set slightly higher prices, but their magnitudes are relatively small.
\begin{center}
{\scriptsize{}}
\begin{table}[H]
\begin{centering}
\begin{tabular}{ccccc}
\hline 
 & (0) MPE & (1) Full commit & (2) Adaptive expec. & (3) Static firms\tabularnewline
\hline 
\hline 
Joint profit & 24.67 & 25.66 & 25.65 & 22.14\tabularnewline
Profit (Panasonic) & 10.77 & 11.18 & 11.18 & 9.63\tabularnewline
Profit (Toshiba) & 13.9 & 14.48 & 14.47 & 12.51\tabularnewline
\hline 
No inventory consumers (\%) & 18.61 & 18.68 & 18.68 & 17.96\tabularnewline
Average price (1000h Inc.; yen) & 94.73 & 95.65 & 95.71 & 96.45\tabularnewline
Average price (2000h Inc.; yen) & 172.57 & 174.7 & 174.5 & 161.41\tabularnewline
Average price (CFL; yen) & 796.53 & 806.27 & 806.18 & 773.45\tabularnewline
Disposal (million) & 3.04 & 3.05 & 3.05 & 2.95\tabularnewline
\hline 
$\Delta$CS & - & -1.02 & -1.01 & 2.35\tabularnewline
$\Delta$PS (excluding fixed cost) & - & 1.03 & 1.02 & -2.61\tabularnewline
$\Delta$TS (excluding Ext. / fixed costs) & - & 0.01 & 0.01 & -0.27\tabularnewline
$\Delta$Ext. (electricity usage) & - & 0.04 & 0.04 & 0.71\tabularnewline
$\Delta$Ext. (waste disposal) & - & 0.00 & 0.00 & -0.02\tabularnewline
$\Delta$TS (excluding fixed costs) & - & -0.02 & -0.03 & -0.96\tabularnewline
\hline 
\end{tabular}
\par\end{centering}
{\scriptsize{}\caption{Equilibrium outcomes under alternative specifications\label{tab:results_alternative_spec}}
}{\scriptsize\par}

    {\footnotesize{}Notes. }{\footnotesize\par}
        
    {\footnotesize{}The values at the top of the table show the discounted
    sum of variable profits of two dominant firms evaluated at $B^{(stationary,base)}$.}{\footnotesize\par}

    {\footnotesize{}The values at the bottom show the changes in surpluses relative to the base setting, and are also evaluated
    at $B^{(stationary,base)}$. Units of these values are billion yen.}{\footnotesize\par}
    
    {\footnotesize{}The values of economic variables in the middle part
    of the table are evaluated at the stationary state. Average prices are sales-weighted. The row of disposal represents the number of products disposed of in each period.}{\footnotesize\par}
    
    {\footnotesize{}$CS,PS,Ext,TS$ denote consumer surplus, producer surplus, externalties, and total surplus, respectively. $\Delta TS$ is defined by $\Delta TS\equiv\Delta CS+\Delta PS-\Delta Ext$.}{\footnotesize\par}

\end{table}
{\scriptsize\par}
\par\end{center}

I also run the same simulation in Sections \ref{subsec:Market_structure_durability} and \ref{subsec:Durability-and-Welfare} under the setting where firms commit to setting constant future product prices. Nevertheless, I obtained analogous results. They are available upon request.

\medskip{}

Column (2) of Table \ref{tab:results_alternative_spec} and Figure \ref{fig:p_ratio_dist_alternative_spec} show the outcomes where firms employ Markov pricing strategy, but consumers form adaptive expectations where consumers believe future product prices are the same as the ones in the current period.\footnote{For the solution method, see Appendix \ref{subsec:Details_algorithm_supply_adaptive_expectation}.} In the baseline setting, consumers form rational expectations, and the formation of consumer expectations is the difference. Though the formation of adaptive expectations is not necessarily rational, investigating the setting sheds light on the role of consumers' expectation formations, which has not been empirically well-validated in the literature.\footnote{In the empirical literature on dynamic demand, there are two types of specifications on consumers' expectation formation. The first is the model of perfect foresight, where consumers correctly understand firms' pricing strategies and form expectations (\citet{goettler2011does}, \citet{chen2013secondary}). The second is the model where consumers expect product prices to follow the Markov process (\citet{erdem2003brand}). The model with inclusive value sufficiency (\citet{hendel2006measuring}, \citet{gowrisankaran2012dynamics}) can be categorized as the latter specification. If consumers form the expectation of the Markov transition process of prices based on the observations that prices are constant before the current period, and they do not change the belief on the process even after a price change in the current period, the latter specification is equivalent to the model with consumers' adaptive expectation. In that sense, considering the model with consumers' adaptive expectations gives suggestions on the supply side implications of the latter specification widely used in the empirical literature.. Note that \citet{fevrier2016consumers} empirically investigated whether consumers correctly expect price reductions for products with frequent promotions. Nevertheless, it is not clear how consumers form expectations for durable products. }

The results show that outcomes in columns (1) and (2) are mostly the same. It implies firms can set high prices as in the case of full commitment, if consumers form adaptive expectations. 

\medskip{}

Finally, we evaluate the role of firms' dynamic incentives. We simulate the outcomes where firms set product prices without considering the effect of current prices on consumer expectations and firms' future profits. Column (3) shows the results. As in the estimated margins shown in Table \ref{tab:Price-and-margins}, the signs of the difference in prices differ across products with different durability levels. Regarding incandescent lamps, major firms produce both 1000h and 2000h bulbs. Setting higher prices for 1000h bulbs leads to a larger demand for 2000h bulbs as substitutes for 1000h bulbs, and it leads to less future demand. Consequently, firms have incentives to set lower prices considering future profits. The results imply that generally it is hard to conclude the direction of the effect of oligopolistic firms' dynamic incentives on product prices without explicit estimation and simulation of the dynamic model. They are determined by demand structure, including the substitution patterns across products with different durability levels.

The results also show that the differences in the equilibrium outcomes due to the existence of firms' dynamic incentives are much larger than those due to the existence of firms' commitment ability on future product prices. It implies ignoring firms' dynamic incentives might lead to worse prediction on the equilibrium prices. Though we need to additionally specify the existence of firms' commitment ability when considering firms' dynamic incentives, the difference plays a smaller role, at least in the current setting. 

\section{Conclusions\label{sec:Conclusions}}

This study empirically investigates firms' incentives for not selling high durability products and overstating product durability levels to increase future replacement demand, by developing a dynamic structural equilibrium model of durable goods with forward-looking consumers and oligopolistic multi-product firms. Based on the observations of the light bulb market, it specifies a model where firms produce multiple products with different durability levels and set product prices based on dynamic incentives. It proposes novel full-solution estimation algorithms that alleviate the computational burden and data requirement for estimating demand and marginal cost parameters of dynamic demand models. Using the data of light bulb market data in Japan, structural parameters are estimated.

This study obtains the following results. First, large firms have incentives to collude to eliminate high durability incandescent lamps, though it is profitable to sell them for each firm. In contrast, when they can collude on prices, they don't have incentives to eliminate high durability bulbs. Second, eliminating high durability incandescent lamps leads to larger producer and total surplus, though it leads to lower consumer surplus. 

The model I developed here would be useful for further empirical investigations of topics related to durability, such as repairability of durables, warranties, and used goods market, which are becoming important in the society making much of the circular economy and high durability of products, from environmental perspective. Furthermore, though we have considered the setting where consumers fully understand the true durability of products, the framework developed here could be extended to analyze the role of asymmetric information between firms and consumers.

Finally, there are remarks related to the current study. While I introduced a relatively small number of discretized consumer types based on the estimated random coefficients of utility parameters, it is not unusual that models with tens or hundreds of consumer types resemble the reality well. Generally, the computational burden grows as the number of consumer types increases. Though I discuss the use of the symmetric structure for circumventing the issue in Chapter 6 of this dissertation, further investigations on the methods for analyzing dynamic demand models with many consumer types is an interesting and important topic. 

\pagebreak{}

\appendix

\section{Details of the Estimations and Counterfactuals\label{sec:Estimation_Counterfactuals_detail}}

\subsection{Algorithm of demand estimation\label{subsec:Details_algorithm_demand}}

In this subsection, I describe the algorithm to solve for mean utility $\delta$ and consumers' value functions $V^{C}$. Though it worked well, the mappings to update $V^{C}$ that appear in the algorithm might not be so intuitive due to the existence of nest structure and consumer inventory.

First, we discretize consumer types as in static BLP models, because analytical representations of the integral $s_{jt}=\int s_{ijt}dP(i)$ is not available. Here, let $w_{i}$ be type $i$ consumers' weight, namely, the fraction of type $i$ consumers among all the consumers. By appropriately choosing $w_{i}$, we can approximate $s_{jt}$ by $\sum_{i}w_{i}\cdot s_{ijt}$. In our setting, persistent consumer heterogeneity comes from the price sensitivity $\alpha_{i}\sim LN(\log(\overline{\alpha}),\sigma_{\alpha}^{2})$, and I use the Gauss-Hermite quadrature for the discretization. 

Under the discretization of consumer types, market share $S_{jt}^{(data)}$ satisfies the following: 

\begin{eqnarray*}
S_{jt}^{(data)} & = & \sum_{i}w_{i}Pr0_{it}\frac{\exp\left(\frac{\kappa_{ijt}}{1-\rho_{g}}\right)}{\exp\left(\frac{IV_{igt}^{C}}{1-\rho_{g}}\right)}\frac{\exp(IV_{igt}^{C})}{\exp(V_{it}^{C})}.
\end{eqnarray*}

Then, we obtain:

\begin{eqnarray*}
\delta_{jt} & = & (1-\rho_{g})\left[\log(S_{jt}^{(data)})-\log\left(\sum_{i}w_{i}Pr0_{it}\widehat{s_{ijt}^{(ccp)}}\right)\right],
\end{eqnarray*}
where $\widehat{s_{ijt}^{(ccp)}}=\frac{\exp\left(\frac{\kappa_{ijt}}{1-\rho_{g}}\right)}{\exp\left(\frac{IV_{igt}^{C}}{1-\rho_{g}}\right)}\frac{\exp(IV_{igt}^{C})}{\exp(V_{it}^{C})}$, which can be regarded as an up-to-scale CCP because $s_{ijt}^{(ccp)}=\widehat{s_{ijt}^{(ccp)}}\cdot\exp\left(\frac{\delta_{jt}}{1-\rho_{g}}\right)$.

Nest-level inclusive value $IV_{igt}^{C}$ satisfies the following:

\begin{eqnarray*}
IV_{igt}^{C} & = & (1-\rho_{g})\log\left(\sum_{j\in\mathcal{J}_{gt}}\exp\left(\frac{\delta_{jt}+\kappa_{ijt}}{1-\rho_{g}}\right)\right).
\end{eqnarray*}

Fraction of no-inventory consumers $Pr0_{it}$ satisfies the following: 

\begin{eqnarray*}
Pr0_{it} & = & \begin{cases}
\frac{1}{1+\sum_{\mu\in\mathcal{M}}s_{i\mu t}^{(ccp)}\cdot\sum_{\tau=0}^{\infty}\phi(i,\mu,\tau)} & \text{if\ }t:\text{in the\ stationary\ state}\\
\sum_{\tau=1}^{\infty}\sum_{\mu\in\mathcal{M}}s_{i\mu t-\tau}^{(n)}f(i,\mu,\tau)+\widetilde{s_{i0t-1}} & \text{otherwise}
\end{cases},
\end{eqnarray*}
 where $\widetilde{s_{i0t}}=Pr0_{it}\cdot\frac{\exp(\widetilde{v_{i0t}})}{\exp(V_{it}^{C})}$ and $s_{i\mu t}=Pr0_{it}\cdot\sum_{j\in\mathcal{J}_{\mu t}}s_{ijt}^{(ccp)}$. $\mathcal{J}_{\mu t}$ denotes the set of products sold at time $t$ that share the same durability level $\mu\in\mathcal{M}$, where $\mathcal{M}$ denotes the set of durability levels, and these products share the same distributions of failure. $s_{i\mu t}\equiv\sum_{j\in\mathcal{J}_{\mu t}}s_{ijt}$ represents the fraction of type $i$ consumers purchasing durability $\mu$ products at time $t$.

Besides, regarding the outside option market share $S_{0t}^{(data)}$ and nest level market share $S_{gt}^{(data)}$, the following equations hold:

\begin{eqnarray*}
S_{0t}^{(data)} & = & \sum_{i}w_{i}\widetilde{s_{i0t}}+\left(1-\sum_{i}w_{i}Pr0_{it}\right),\\
S_{gt}^{(data)} & = & \sum_{i}w_{i}Pr0_{it}\cdot\frac{\exp(IV_{igt}^{C})}{\exp(V_{it}^{C})}.
\end{eqnarray*}

Based on these equations, we can solve for the variables by Algorithm \ref{alg:Detailed-algorithm-demand} .

{\footnotesize{}}
\begin{algorithm}[H]
\begin{enumerate}
\item {\footnotesize{}Set the initial values $V_{it}^{C(0)},\log\left(IV_{igt}^{C(0)}\right),\log\left(Pr0_{it}^{(0)}\right)$}{\footnotesize\par}
\item {\footnotesize{}Iterate the following until convergence $(n=0,1,2,\cdots)$:}{\footnotesize\par}
\begin{enumerate}
\item {\footnotesize{}Compute consumer type specific utilities $\kappa_{ijt}^{(n)}\equiv E_{t}\left[-\alpha_{i}p_{jt}+\beta_{C}^{L_{ij}}V_{it+L_{ij}}^{C(n)}\right]$ and $\widetilde{v_{i0t}}^{(n)}\equiv\beta_{C}E_{t}\left[V_{it+1}^{C(n)}\right]$}{\footnotesize\par}
\item {\footnotesize{}Compute up-to-scale CCPs $\widehat{s_{ijt}^{(ccp)(n)}}=\frac{\exp\left(\frac{\kappa_{ijt}^{(n)}}{1-\rho_{g}}\right)}{\exp\left(\frac{IV_{igt}^{C(n)}}{1-\rho_{g}}\right)}\frac{\exp(IV_{igt}^{C(n)})}{\exp(V_{it}^{C(n)})}$}{\footnotesize\par}
\item {\footnotesize{}Compute $\delta_{jt}^{(n)}=(1-\rho_{g})\left[\log(S_{jt}^{(data)})-\log\left(\sum_{i}w_{i}Pr0_{it}^{(n)}\widehat{s_{ijt}^{(ccp)(n)}}\right)\right]$}{\footnotesize\par}
\item {\footnotesize{}Compute $s_{ijt}^{(ccp)(n)}=\widehat{s_{ijt}^{(ccp)(n)}}\cdot\exp\left(\frac{\delta_{jt}^{(n)}}{1-\rho_{g}}\right)$ and $s_{i0t}^{(ccp)(n)}=\frac{\exp\left(\widetilde{v_{i0t}}^{(n)}\right)}{\exp\left(V_{it}^{C(n)}\right)}$}{\footnotesize\par}
\item {\footnotesize{}Compute $s_{i\mu t}^{(n)}=Pr0_{it}^{(n)}\cdot\sum_{j\in\mathcal{J}_{\mu t}}s_{ijt}^{(ccp)(n)}$ and $\widetilde{s_{i0t}}^{(n)}=Pr0_{it}^{(n)}\cdot s_{i0t}^{(ccp)(n)}$}{\footnotesize\par}
\item {\footnotesize{}Update $IV_{igt}^{C}$ by: }{\footnotesize\par}

{\footnotesize{}
\begin{eqnarray*}
IV_{igt}^{C(n+1)} & = & (1-\rho_{g})\log\left(\sum_{j\in\mathcal{J}_{gt}}\exp\left(\frac{\delta_{jt}^{(n)}+\kappa_{ijt}^{(n)}}{1-\rho_{g}}\right)+\right.\\
 &  & \rho_{g}\left[\log(S_{gt}^{(data)})-\log\left(\sum_{i}w_{i}Pr0_{it}^{(n)}\cdot\frac{\exp(IV_{igt}^{C(n)})}{\exp(V_{it}^{C(n)})}\right)\right]-\\
 &  & \left[\log\left(S_{0t}^{(data)}\right)-\log\left(\sum_{i}w_{i}\left(1-Pr0_{it}^{(n)}+Pr0_{it}^{(n)}\cdot\frac{\exp\left(\widetilde{v_{i0t}}^{(n)}\right)}{\exp\left(V_{it}^{C(n)}\right)}\right)\right)\right]
\end{eqnarray*}
}{\footnotesize\par}
\item {\footnotesize{}Update $V_{it}^{C}$ by: $V_{it}^{C(n+1)}=\log\left(\exp\left(\widetilde{v_{i0t}}^{(n)}\right)+\sum_{g\in\mathcal{G}}\exp\left(IV_{igt}^{C(n+1)}\right)\right)$}{\footnotesize\par}
\item {\footnotesize{}Compute $\widetilde{s_{i0t}}^{*(n)}=\widetilde{s_{i0t}}^{(n)}\cdot\frac{S_{0t}^{(data)}-1+\sum_{i}w_{i}Pr0_{it}^{(n)}}{\sum_{i}w_{i}\widetilde{s_{i0t}}^{(n)}}$ }{\footnotesize\par}
\item {\footnotesize{}Update $Pr0_{it}$ by: $Pr0_{it}^{(n+1)}=\begin{cases}
\frac{1}{1+\sum_{\mu\in\mathcal{M}}s_{i\mu t}^{(ccp)(n)}\cdot\sum_{\tau=0}^{\infty}\phi(i,\mu,\tau)} & \text{if\ }t:\text{in the\ stationary\ state}\\
\sum_{\tau=1}^{\infty}\sum_{\mu\in\mathcal{M}}s_{i\mu t-\tau}^{(n)}f(i,\mu,\tau)+\widetilde{s_{i0t-1}}^{*(n)} & \text{otherwise}
\end{cases}$}{\footnotesize\par}
\end{enumerate}
\item {\footnotesize{}After the convergence, obtain $\delta$}{\footnotesize\par}
\end{enumerate}
{\footnotesize{}\caption{Detailed algorithm for solving the fixed point problem in the demand estimation\label{alg:Detailed-algorithm-demand}}
}{\footnotesize\par}

{\footnotesize{}Notes.}{\footnotesize\par}

{\footnotesize{}To simplify the exposition, we show the steps without combining the spectral algorithm.}{\footnotesize\par}

{\footnotesize{}The tolerance level of the inner loop is set to 1e-13.}{\footnotesize\par}
\end{algorithm}
{\footnotesize\par}

In the algorithm, I solve for not only $V^{C}$ but also nest level inclusive values $IV^{C}$ and the fraction of no-inventory consumers $Pr0$. Regarding nest level inclusive values $IV^{C}$, we cannot represent $\delta$ just by $V^{C}$ when the nest structure exists, and so I introduce the terms. Concerning $Pr0$, the values of $Pr0_{it}$ are not observed in the data, and we also have to solve for the variables.

Regarding Step 2(f), in principle, we can update $IV_{igt}^{C}$ just by $IV_{igt}^{C(n+1)}=(1-\rho_{g})\log\left(\sum_{j\in\mathcal{J}_{gt}}\exp\left(\frac{\delta_{jt}^{(n)}+\kappa_{ijt}^{(n)}}{1-\rho_{g}}\right)\right)$. Nevertheless, adding the terms {\footnotesize{}$\rho_{g}\left[\log(S_{gt}^{(data)})-\log\left(\sum_{i}w_{i}Pr0_{it}^{(n)}\cdot\frac{\exp(IV_{igt}^{C(n)})}{\exp\left(V_{it}^{C(n)}\right)}\right)\right]$} and {\footnotesize{}$\left[\log\left(S_{0t}^{(data)}\right)-\log\left(\sum_{i}w_{i}\left(1-Pr0_{it}^{(n)}+Pr0_{it}^{(n)}\cdot\frac{\exp\left(\widetilde{v_{i0t}}^{(n)}\right)}{\exp\left(V_{it}^{C(n)}\right)}\right)\right)\right]$} further accelerated the convergence. Please also see the discussion in \citeyear{Fukasawa2024BLPalgorithm}. Besides, Step 2(h) is not essential. Nevertheless, introducing the step further accelerated the convergence, as far as I experimented.

\subsubsection*{Spectral algorithm}

To speed up the convergence, I combine Algorithm \ref{subsec:Details_algorithm_demand} with spectral algorithm, which has been developed in the field of numerical analysis (\citet{barzilai1988two}) to solve nonlinear equations efficiently. Let $\Phi^{D}$ be the mapping represented by Algorithm \ref{alg:Detailed-algorithm-demand}. In the algorithm, variables $z\equiv(z_{(V^{C})},z_{(IV^{C})},z_{\left(\log\left(Pr0\right)\right)})\equiv\left(V^{C},IV^{C},\log\left(Pr0\right)\right)$ are updated by the following rule:

\begin{eqnarray*}
z_{(m)}^{(n+1)} & = & z_{(m)}^{(n)}-\alpha_{(m)}^{(n)}\left(z_{(m)}^{(n)}-\Phi_{(m)}^{D}(z^{(n)})\right)\ m\in\left\{ V^{C},IV^{C},\log\left(Pr0\right)\right\} 
\end{eqnarray*}

$z_{(m)}^{(n)}-\Phi_{(m)}^{D}(z^{(n)})$ represents the difference between the values of $z_{(m)}$ before and after the update. If they are sufficiently close to zero, we can assume that we obtain the solution of the fixed point problem $z=\Phi^{D}(z)$. $\alpha_{(m)}^{(n)}$ represents the step size of type $m$ variable at $n$-th iteration, and following \citet{varadhan2008simple}, we set the value by:

\begin{eqnarray*}
\alpha_{(m)}^{(n)} & = & sgn\left(s_{(m)}^{(n-1)^{\prime}}y_{(m)}^{(n-1)}\right)\frac{\left\Vert s_{(m)}^{(n-1)}\right\Vert }{\left\Vert y_{(m)}^{(n-1)}\right\Vert },
\end{eqnarray*}
where $s_{(m)}^{(n-1)}=z_{(m)}^{(n)}-z_{(m)}^{(n-1)}$ and $y_{(m)}^{(n-1)}=\Phi_{(m)}^{D}(z^{(n)})-\Phi_{(m)}^{D}(z^{(n-1)})$. I also set the value of $\alpha^{(0)}$ to 0.1.

The alternative method to solve nonlinear equations is Newton's method. If we apply Newton's method, $z$ should be updated by:

\begin{eqnarray*}
z^{(n+1)} & = & z^{(n)}-\left(\nabla_{z}\left(z^{(n)}-\Phi^{D}\left(z^{(n)}\right)\right)\right)^{-1}\left(z^{(n)}-\Phi^{D}\left(z^{(n)}\right)\right).
\end{eqnarray*}

It is known that Newton's method can solve nonlinear equations with a few iterations, if the initial values $z^{(0)}$ are appropriately chosen. Nevertheless, in our setting, computing the derivative $\nabla_{z}\left(z^{(n)}-\Phi^{D}\left(z^{(n)}\right)\right)$ is very cumbersome to code because of the complicated model structure. Furthermore, computing the inverse matrix $\left(\nabla_{z}\left(z^{(n)}-\Phi^{D}\left(z^{(n)}\right)\right)\right)^{-1}$ would be time-consuming, because of the large size of the Hessian $\nabla_{z}\left(z^{(n)}-\Phi^{D}\left(z^{(n)}\right)\right)$. Besides, Newton's method is sensitive to the initial values $z^{(0)}$.

As discussed in \citet{varadhan2008simple}, the scalar $\alpha^{(n)}$ in the spectral algorithm specified above can be regarded as an approximation of the matrix $\left(\nabla_{z}\left(z^{(n)}-\Phi^{D}\left(z^{(n)}\right)\right)\right)^{-1}$, and it works well because it inherits good properties of Newton's method.

In general, there is no guarantee that the algorithms specified above converge. Hence, \citet{varadhan2008simple} proposed using non-monotone line search to achieve global convergence. Nevertheless, in our setting, the algorithm converged without such a procedure.

\subsubsection*{Algorithm based on the traditional approach}

In Table \ref{tab:Results-of-demand_est}, I compare the performance of the proposed algorithm and the traditional algorithm based on BLP contraction mapping and value function iteration. In the latter algorithm, I jointly updated $V,\delta,Pr0$ until convergence. The algorithm is the same as the proposed algorithm, except for the following steps:
\begin{itemize}
\item 2(b,c,d):Compute $IV_{igt}^{C(n)}=\log\left(\sum_{j\in\mathcal{J}_{gt}}\exp\left(\frac{\delta_{jt}+\kappa_{ijt}^{(n)}}{1-\rho_{g}}\right)\right)$, $s_{ijt}^{(ccp)(n)}=\frac{\exp\left(\frac{\delta_{jt}+\kappa_{ijt}^{(n)}}{1-\rho_{g}}\right)}{\exp\left(\frac{IV_{igt}^{C(n)}}{1-\rho_{g}}\right)}\frac{\exp(IV_{igt}^{C(n)})}{\exp(V_{it}^{C(n)})}$, and $s_{i0t}^{(ccp)(n)}=\frac{\exp\left(\widetilde{v_{i0t}}^{(n)}\right)}{\exp\left(V_{it}^{C(n)}\right)}$
\item 2(f): Update $\delta$ by: $\delta_{jt}^{(n+1)}=\delta_{jt}^{(n)}+\log\left(S_{jt}^{(data)}\right)-\log\left(\sum_{i}w_{i}Pr0_{it}^{(n)}s_{ijt}^{(ccp)(n)}\right)$
\item 2(g): Update $V^{C}$ by: $V_{it}^{C(n+1)}=\log\left(\exp\left(\widetilde{v_{i0t}}^{(n)}\right)+\sum_{g\in\mathcal{G}}\exp\left(IV_{igt}^{C(n)}\right)\right)$
\item 2(h): Skip
\item 2(i): Use $\widetilde{s_{i0t-1}}^{*(n)}$ rather than $\widetilde{s_{i0t-1}}^{*(n)}$ 
\end{itemize}

\subsection{Solution method of the Markov perfect equilibrium\label{subsec:Details_algorithm_supply}}

To solve the equilibrium, we use the following equations derived from the model. We also use the idea of simplifying the state space to reduce the computational burden of solving the equilibrium.

\subsubsection*{Discretization of consumer types}

As in the supply-side analysis of static BLP models, we need the discretization of consumer types to solve firms' pricing problems, in the model of continuous consumer types under the existence of random coefficients. As in the demand estimation, I discretized the distribution of consumers into 4 types by using the Gauss-Hermite quadrature. 

\subsubsection*{Pricing}

By equation (\ref{eq:FOC_p}), we can derive the following equation:

\begin{equation}
p_{jt}(B_{t})=mc_{jt}-\frac{Ms_{jt}(p_{t},B_{t})+\sum_{k\in\mathcal{J}_{ft}-\{j\}}(p_{kt}(B_{t})-mc_{kt})M\frac{\partial s_{kt}}{\partial p_{jt}}(p_{t},B_{t})+\beta_{F}\frac{\partial B_{t+1}}{\partial p_{jt}}\frac{\partial V_{ft+1}^{F}(B_{t+1}(B_{t}))}{\partial B_{t+1}}}{M\frac{\partial s_{jt}}{\partial p_{jt}}(p_{t},B_{t})}.\label{eq:update_p}
\end{equation}
Note that
\begin{eqnarray}
p_{jt}(B_{t}) & = & mc_{jt}-\frac{\frac{\partial\pi_{ft}(p_{t},B_{t})}{\partial p_{jt}}+\beta_{F}\frac{\partial B_{t+1}}{\partial p_{jt}}\frac{\partial V_{ft+1}^{F}(B_{t+1}(B_{t}))}{\partial B_{t+1}}-M\frac{\partial s_{jt}}{\partial p_{jt}}(p_{t},B_{t})\cdot(p_{jt}-mc_{jt})}{M\frac{\partial s_{jt}}{\partial p_{jt}}(p_{t},B_{t})}\nonumber \\
 & = & p_{jt}(B_{t})-\frac{\frac{\partial\pi_{ft}(p_{t},B_{t})}{\partial p_{jt}}+\beta_{F}\frac{\partial B_{t+1}}{\partial p_{jt}}\frac{\partial V_{ft+1}^{F}(B_{t+1}(B_{t}))}{\partial B_{t+1}}}{M\frac{\partial s_{jt}}{\partial p_{jt}}(p_{t},B_{t})}\label{eq:update_p_equivalent}
\end{eqnarray}
holds, and updating equation (\ref{eq:update_p}) is equivalent to (\ref{eq:update_p_equivalent}). Updating equation (\ref{eq:update_p_equivalent}) implies $p_{jt}$ is updated to be larger when the derivative of the profit $\frac{\partial\pi_{ft}(p_{t},B_{t})}{\partial p_{jt}}+\beta_{F}\frac{\partial B_{t+1}}{\partial p_{jt}}\frac{\partial V_{ft+1}^{F}(B_{t+1}(B_{t}))}{\partial B_{t+1}}$ takes a positive value, since $\frac{\partial s_{jt}}{\partial p_{jt}}<0$. We use equation (\ref{eq:update_p}) to update the values of $p_{jt}(B_{t})$.\footnote{Alternatively, we can use the following equation to update the values of $p_{jt}(B_{t})$:

\begin{eqnarray*}
p_{t}(B_{t}) & = & mc_{t}+\underbrace{\left(\Delta_{direct,t}(B_{t},p_{t}(B_{t}))\right)^{-1}s_{t}(B_{t},p_{t}(B_{t}))}_{\text{Static margin}}+\\
 &  & \underbrace{\left(\Delta_{direct,t}(B_{t},p_{t}(B_{t}))\right)^{-1}\left[\left(\Delta_{indirect,t}(B_{t},p_{t}(B_{t}))\right)\left(p_{t}(B_{t})-mc_{t}\right)+\beta_{F}\frac{\partial V_{t+1}^{F}}{\partial p_{t}}(B_{t}^{(data)},p_{t}(B_{t}),mc_{t})\right].}_{\text{Dynamic margin}}
\end{eqnarray*}

If the dynamic margin term does not exist, the method corresponds to the standard approach to solve the equilibrium of static price competition (see the discussion in \citet{conlon2020best}). Nevertheless, especially in the dynamic model, the updating equation is not computationally attractive. To iteratively apply the equation above, we need to repeatedly compute the inverse matrix $\left(\Delta_{direct,t}(B_{t},p_{t})\right)^{-1}$. Since we need to compute the matrix at all the grid points $B_{t}^{(grid)}$, the computational burden is not negligible. In contrast, updating equation (\ref{eq:update_p}) does not require the computation of inverse matrices. We have not encountered convergence issues even when applying updating equation (\ref{eq:update_p}) combined with spectral algorithm.}

\subsubsection*{Consumers' value functions}

In the case where the values of $\psi_{j}$ are unknown, it is not possible to directly solve for $V_{it}^{C}$. Hence, we alternatively define a new term $\widetilde{V_{it}^{C}}$. Intuitively, $\widetilde{V_{it}}^{C}$ corresponds to consumers' value function, assuming that consumers obtain the utility from future usage at the time of purchase. Formally, the definition of $\widetilde{V_{t}^{C}}$ is as follows:\footnote{In the following, we omit $\Omega_{t}^{C}$ without loss of generality.}

\begin{eqnarray*}
\widetilde{V_{it}^{C}}(x_{it},B_{t}) & \equiv & \begin{cases}
V_{it}^{C}(x_{it},B_{t}) & \text{if}\ x_{it}=\emptyset,\\
V_{it}^{C}(x_{it},B_{t})-\left[\sum_{s=0}^{\infty}\beta_{C}^{\tau}\psi_{j}\cdot\phi\left(i,\mu_{j},\tau+s|\tau\right)\right]. & \text{if}\ x_{it}=(j,\tau).
\end{cases}
\end{eqnarray*}
Then, we obtain the following results
\begin{lem}
\label{lem:V_C_tilde}The following equations hold:
\end{lem}
(a).
\begin{eqnarray}
 &  & \widetilde{V_{it}^{C}}(x_{it},B_{t})\nonumber \\
 & = & \begin{cases}
\phi(i,\mu_{j},\tau+1|\tau)\cdot\widetilde{V_{it}^{C}}(x_{it}=(j,\tau+1),B_{t})+\left(1-\phi(i,\mu_{j},\tau+1|\tau)\right)\cdot\widetilde{V_{it}^{C}}(x_{it}=\emptyset,B_{t}) & \text{if}\ x_{it}=(j,\tau),\\
E_{\epsilon}\left[\max_{j\in\mathcal{J}_{t}\cup\{0\}}\left(-\alpha_{i}p_{jt}+\delta_{jt}+\epsilon_{ijt}+\beta_{C}E_{x}\left[\widetilde{V_{it+1}^{C}}(x_{it+1},B_{t+1}(B_{t}))|x_{it}=\emptyset,a_{it}=j\right]\right)\right] & \text{if}\ x_{it}=\emptyset.
\end{cases}\label{eq:V_C_tilde}
\end{eqnarray}

(b).

\[
\widetilde{V_{it}^{C}}(x_{it}=(j,\tau),B_{t})=\sum_{s=0}^{\infty}f(i,\mu_{j},\tau+s+1|\tau)\cdot\widetilde{V_{it+s+1}^{C}}(x_{it+s+1}=\emptyset,B_{t}).
\]

(c).

\begin{eqnarray*}
\widetilde{v_{ijt}}(x_{it}=\emptyset,\Omega_{t}^{C}) & = & -\alpha_{i}p_{jt}+\delta_{jt}+\beta_{C}E_{x}\left[\widetilde{V_{it+1}^{C}}(x_{it+1},B_{t+1}(B_{t}))|x_{it}=\emptyset,a_{it}=j\right].
\end{eqnarray*}

\begin{proof}
See Appendix \ref{subsec:Proof_Lemma_V_C_tilde}.
\end{proof}

\subsubsection*{Stationary state}

In the stationary state, the following equations hold:

\begin{eqnarray}
Pr_{i}^{(stationary)}(x_{it}) & = & \begin{cases}
\frac{s_{ij}^{(ccp)}(B^{(stationary)})\cdot\phi(i,\mu_{j},\tau)}{1+\sum_{j\in\mathcal{J}^{(stationary)}}s_{ij}^{(ccp)}(B_{(stationary)})\cdot\sum_{\tau=0}^{\infty}\phi(i,\mu_{j},\tau)} & \text{if }x_{it}=(j,\tau),\\
\frac{1}{1+\sum_{j\in\mathcal{J}^{(stationary)}}s_{ij}^{(ccp)}(B_{(stationary)})\cdot\sum_{\tau=0}^{\infty}\phi(i,\mu_{j},\tau)} & \text{if}\ x_{it}=\emptyset.
\end{cases}\label{eq:Pr_stationary}
\end{eqnarray}

\subsubsection*{Simplifying the state space of $x_{it}$ and $B_{t}$}

Under the assumption that consumers make purchase decisions only when they do not have any functioning product, the time after the purchase of a product until the next purchase decision is determined only by the durability of the product. Then, Equation (\ref{eq:Pr_state_transition}) can be rewritten as follows: 

{\footnotesize{}
\begin{eqnarray}
Pr_{it+1}(x_{it+1}) & = & \begin{cases}
Pr_{it}(x_{it}=\emptyset)\cdot s_{i0t}^{(ccp)}+\sum_{\tau\in\mathbb{N},\mu\in\mathcal{M}}Pr_{it}\left(x_{it}=(\mathcal{J}_{\mu},\tau)\right)\cdot\left(1-\phi(i,\mu,\tau+1|\tau)\right) & \text{if}\ x_{it+1}=\emptyset,\\
Pr_{it}\left(x_{it}=(\mathcal{J}_{\mu},\tau-1)\right)\cdot\phi(i,\mu,\tau|\tau-1) & \text{if}\ x_{it+1}=(\mathcal{J_{\mu}},\tau\geq2),\\
Pr_{it}(x_{it}=\emptyset)\cdot s_{i\mu t}^{(ccp)}\cdot\phi(i,\mu,\tau=1) & \text{if}\ x_{it+1}=(J_{\mu},\tau=1).
\end{cases}\label{eq:Pr_agg_state_transition}
\end{eqnarray}
}It implies firms have to care only about the distribution of consumers' inventory characterized by the durability level and product age, rather than those characterized by product name and product age. Hence, we can alternatively use $B_{t}=\left(Pr_{it}(x_{it})\right)_{x_{it}\in\{\emptyset\}\cup\bigcup_{\mu\in\mathcal{M},\tau\in\mathbb{N}}(\mathcal{J}_{\mu},\tau)}$ as the aggregate states. Let $x_{it}=(\mathcal{J}_{\mu},\tau)$ denote the state where consumer $i$ owns a product with durability level $\mu$ and age $\tau$.

In addition, Lemma \ref{lem:V_C_tilde} (b) shows that $\widetilde{V_{it}^{C}}(x_{it},B_{t})$ with different $x_{it}$ share the same value if the product age and durability level are the same. Hence, we can alternatively use $x_{it}\in\{\emptyset\}\cup\bigcup_{\mu\in\mathcal{M},\tau\in\mathbb{N}}(\mathcal{J}_{\mu},\tau)$ as the state space of $x_{it}$ used in $\widetilde{V_{it}^{C}}(x_{it},B_{t})$.

Similarly, continuation values $\beta_{C}E_{x}\left[\widetilde{V_{it+1}^{C}}(x_{it+1},B_{t+1}(B_{t}))|x_{it}=\emptyset,a_{it}=j\right]$ share the same value if the product durability level are the same. Let $h_{i\mu t}(B_{t+1})\equiv E_{x}\left[\widetilde{V_{it+1}^{C}}(x_{it+1},B_{t+1}(B_{t}))|x_{it}=\emptyset,a_{it}\in\mathcal{J}_{\mu t}\right]$ be consumer $i$'s continuation value when purchasing a product with durability level $\mu$ at time $t$.

Since the number of different durability levels (average lifetimes) is 7, which is much smaller than the number of products in each period (roughly 80), it contributes to reducing the computational burden of solving the equilibrium.

{\footnotesize{}}
\begin{algorithm}[H]
\begin{enumerate}
\item {\footnotesize{}Take grid points of aggregate state variables $B_{t}^{(grid)}$ and consumer level state variables $x_{it}^{(grid)}$. Set initial values of $\left\{ \widetilde{V_{it}^{C(0)}}(x_{it},B_{t}^{(grid)})\right\} _{i,x_{it},B_{t}}$ (consumers' value function), $\left\{ V_{ft}^{F(0)}(B_{t}^{(grid)})\right\} _{B_{t}}$ (firm's value function), $\left\{ p_{jt}^{(0)}(B_{t}^{(grid)})\right\} _{j,B_{t}}$ (equilibrium price), $\left\{ B_{t+1}^{(0)}(B_{t}^{(grid)})\right\} _{B_{t}}$ (aggregate state variables in the next period), $\left\{ \frac{\partial B_{t+1}^{(0)}}{\partial p_{jt}}(B_{t}^{(grid)})\right\} _{B_{t}}$(derivative of the aggregate state variables in the next period with respect to the current price), and $B_{(stationary)}$ (aggregate states at the stationary state) .}{\footnotesize\par}
\item {\footnotesize{}Iterate the following until the convergence of $\widetilde{V_{it}^{C(n)}}(x_{it}^{(grid)},B_{t}^{(grid)})$, $V_{ft}^{F(n)}(B_{t}^{(grid)})$, $p_{jt}^{(n)}(B_{t}^{(grid)})$, $B_{t+1}^{(n)}(B_{t}^{(grid)})$, $\frac{\partial B_{t+1}^{(n)}}{\partial p_{jt}}(B_{t}^{(grid)})$, $B_{(stationary)}$ ($n=0,1,2,\cdots$):}{\footnotesize\par}
\begin{enumerate}
\item {\footnotesize{}Given $\widetilde{V_{it}^{C(n)}}(x_{it}^{(grid)},B_{t}^{(grid)}),p_{jt}^{(n)}(B_{t}^{(grid)}),B_{t+1}^{(n)}(B_{t}^{(grid)}),\frac{\partial B_{t+1}^{(n)}}{\partial p_{jt}}(B_{t}^{(grid)}),B_{(stationary)}^{(n)}$,}{\footnotesize\par}
\begin{enumerate}
\item {\footnotesize{}Compute 
\begin{eqnarray*}
h_{i\mu t}^{(n)}(B_{t+1}^{(n)}(B_{t}^{(grid)})) & = & \phi(i,\mu,\tau=1)\cdot\widetilde{V_{it+1}^{C(n)}}\left(x_{t+1}=(\mu,\tau),B_{t+1}^{(n)}(B_{t}^{(grid)})\right)+\\
 &  & \left(1-\phi(i,\mu,\tau=1)\right)\cdot\widetilde{V_{it+1}^{C(n)}}\left(x_{it+1}=\emptyset,B_{t+1}^{(n)}(B_{t}^{(grid)})\right)
\end{eqnarray*}
}{\footnotesize\par}
\item {\footnotesize{}Compute $\widetilde{v_{ijt}}(x_{it}=\emptyset,B_{t}^{(grid)})=-\alpha_{i}p_{jt}^{(n)}(B_{t}^{(grid)})+\delta_{jt}+\beta_{C}h_{i\mu_{j}t}^{(n)}(B_{t+1}^{(n)}(B_{t}^{(grid)}))$}{\footnotesize\par}
\item {\footnotesize{}Compute $s_{ijt}^{(ccp)(n+1)}(x_{it}=\emptyset,B_{t}^{(grid)},p_{t}^{(n)}),s_{i0t}^{(ccp)(n+1)}(x_{it}=\emptyset,B_{t}^{(grid)},p_{t}^{(n)}),s_{jt}^{(n+1)}(B_{t}^{(grid)},p_{t}^{(n)})$ by (\ref{eq:choice_prob_ccp}), (\ref{eq:outside_choice_prob_ccp}), and (\ref{eq: agg_market_share})}{\footnotesize\par}
\item {\footnotesize{}Compute 
\begin{eqnarray*}
\frac{\partial h_{i\mu t}^{(n)}(B_{t+1}^{(n)}(B_{t}^{(grid)}))}{\partial B_{t+1}^{(n)}} & = & \phi(i,\mu,\tau=1)\cdot\frac{\partial\widetilde{V_{it+1}^{C}}}{\partial B_{t+1}}\left(x_{t+1}=(\mathcal{J}_{\mu t},\tau=1),B_{t+1}^{(n)}(B_{t}^{(grid)})\right)+\\
 &  & \left(1-\phi(i,\mu,\tau=1)\right)\cdot\frac{\partial\widetilde{V_{it+1}^{C}}}{\partial B_{t+1}}\left(x_{it+1}=\emptyset,B_{t+1}^{(n)}(B_{t}^{(grid)})\right)
\end{eqnarray*}
}{\footnotesize\par}
\item {\footnotesize{}Compute $\frac{\partial\widetilde{v_{ikt}}^{(n+1)}}{\partial p_{j}}=-\alpha_{i}+\beta_{C}\frac{\partial B_{t+1}^{(n)}}{\partial p_{jt}}\frac{\partial h_{ikt}^{(n)}(B_{t+1}^{(n)}(B_{t}^{(grid)}))}{\partial B_{t+1}^{(n)}}\ (k\in\mathcal{J}_{t}\cup\{0\})$}{\footnotesize\par}
\item {\footnotesize{}Compute $\frac{\partial s_{i\widetilde{\mu}t}^{(n+1)}(B_{t}^{(grid)})}{\partial p_{jt}}=\frac{\partial\widetilde{v_{ikt}}(x_{it}=\emptyset,B_{t}^{(grid)})}{\partial p_{jt}}\frac{\partial s_{i\widetilde{\mu}t}(B_{t}^{(grid)})}{\partial\widetilde{v_{ikt}}(x_{it}=\emptyset,B_{t}^{(grid)})}+\sum_{\mu\in\mathcal{M}}\frac{\partial B_{t+1}^{(n)}(B_{t}^{(grid)})}{\partial p_{jt}}\frac{\partial h_{i\mu t}(B_{t+1}^{(n)})}{\partial B_{t+1}^{(n)}}\frac{\partial s_{i\widetilde{\mu}t}(B_{t}^{(grid)})}{\partial h_{i\mu t}(B_{t+1}^{(n)})}$}{\footnotesize\par}
\item {\footnotesize{}Compute $\frac{\partial V_{ft+1}^{F}(B_{t+1}^{(n)}(B_{t}^{(grid)}))}{\partial p_{jt}}=\frac{\partial V_{ft+1}^{F}(B_{t+1}^{(n)}(B_{t}^{(grid)}))}{\partial B_{t+1}^{(n)}}\cdot\frac{\partial B_{t+1}^{(n)}}{\partial p_{jt}}(B_{t}^{(grid)})$}{\footnotesize\par}
\end{enumerate}
\item {\footnotesize{}Given $\widetilde{V_{it}^{C(n)}}(x_{it},B_{t}^{(grid)}),V_{ft}^{F(n)}(B_{t}^{(grid)}),p_{jt}^{(n)}(B_{t}^{(grid)}),B_{t+1}^{(n)}(B_{t}^{(grid)}),\frac{\partial B_{t+1}^{(n)}}{\partial p_{jt}}(B_{t}^{(grid)}),B_{(stationary)}^{(n)}$ and $s_{ijt}^{(ccp)(n+1)}(x_{it}=\emptyset,B_{t}^{(grid)})$, $s_{i0t}^{(ccp)(n+1)}(x_{it}=\emptyset,B_{t}^{(grid)}),$$s_{jt}^{(n+1)}(B_{t}^{(grid)}),\frac{\partial s_{i\mu t}^{(n+1)}(B_{t}^{(grid)})}{\partial p_{jt}}$, update variables:}{\footnotesize\par}
\begin{enumerate}
\item {\footnotesize{}Compute $p_{jt}^{(n+1)}(B_{t}^{(grid)})$ by (\ref{eq:update_p})}{\footnotesize\par}
\item {\footnotesize{}Compute $V_{ft}^{F(n+1)}(B_{t}^{(grid)})$ by (\ref{eq:V_F})}{\footnotesize\par}
\item {\footnotesize{}Compute $\widetilde{V_{it}^{C(n+1)}}(x_{it},B_{t}^{(grid)})$ by (\ref{eq:V_C_tilde})}{\footnotesize\par}
\item {\footnotesize{}Compute $B_{t+1}^{(n+1)}(B_{t}^{(grid)})$ by (\ref{eq:Pr_agg_state_transition})}{\footnotesize\par}
\item {\footnotesize{}Compute $\frac{\partial B_{t+1}^{(n+1)}}{\partial p_{jt}}(B_{t}^{(grid)})=\sum_{\mu\in\mathcal{M}}\frac{\partial s_{i\mu t}^{(n+1)}}{\partial p_{jt}}(B_{t}^{(grid)})\frac{\partial B_{t+1}}{\partial s_{i\mu t}}(B_{t}^{(grid)})$}{\footnotesize\par}
\item {\footnotesize{}Compute $B_{(stationary)}^{(n+1)}$ by (\ref{eq:Pr_stationary})}{\footnotesize\par}
\end{enumerate}
\end{enumerate}
\end{enumerate}
{\footnotesize{}\caption{{\footnotesize{}Algorithm for solving the Markov perfect equilibrium\label{alg:solve_equil}}}
}{\footnotesize\par}
\end{algorithm}
{\footnotesize\par}

As in the case of demand estimation, I also applied the spectral algorithm. Updating equations are essentially the same as the demand algorithm discussed in Section \ref{subsec:Details_algorithm_demand}.

One notable feature of the current algorithm is avoiding solving optimization problems inside loops. In the previous studies (e.g., \citet{nair2007intertemporal}), the values of optimal product prices are updated so that they maximize each firm's long-run profit, given competitors' pricing decisions and consumers' and firms' value functions. Nevertheless, solving a maximization problem of nonlinear functions typically requires nonnegligible computation time, and the computational burden gets larger when we need to repeat the process until the convergence of variables. The current algorithm avoids such an arduous procedure, by analytically representing $p_{jt}(B_{t}^{(grid)})$ in the form of equation (\ref{eq:update_p}). Even though the values of $p_{jt}(B_{t}^{(grid)})$ in the middle of the iteration might not maximize each firm's long-run profit given competitors' pricing decisions and consumers' and firms' value functions, it is guaranteed that obtained values of product prices actually satisfy the first-order conditions of profit maximization problem if the iteration converges. Note that the idea of avoiding solving maximization problems inside loops can be found in the endogenous grid method in the macroeconomics literature since \citet{carroll2006method}. A Similar idea was also applied in \citet{fukasawa2023merger} solving a dynamic oligopoly model of firms' continuous investment.

\subsubsection*{Interpolations}

To apply the collocation method, we need to interpolate the values of functions at the points other than the grid points $B_{t}^{(grid)}$. For instance, we can interpolate the values of firm $f$'s value function as a function of variables $B_{t}$ by taking appropriate functions $\Psi_{m}^{F}\ (m=1,\cdots,M_{F})$ and parameters $\theta_{ft}^{F(m)}\ (m=1,\cdots,M_{F})$. 

\[
V_{ft}^{F}(B_{t})\approx\sum_{m=1}^{M_{F}}\theta_{ft}^{F(m)}\Psi_{m}(B_{t})
\]

The values of $\theta_{ft}^{F(m)}$ can be recovered given the values of $V_{ft}^{F}(B_{t}^{(grid)})$ at $M_{F}$ grid points. However, the variables $B_{t}=\left(Pr_{it}(x_{it})\right)_{x_{it}\in\mathcal{\chi}_{t},i\in\mathcal{I}}$ are still high-dimensional, even when simplifying the state space as discussed above. The dimension depends on the maximum product age, the number of different durability levels, and the number of consumer types.

To deal with the problem, we utilize the knowledge of the economic model structure to mitigate the problem. The following variables would largely affect firms' future profits:

\[
\widetilde{B_{t}}=\left(\widetilde{B_{t}}^{(1)},\widetilde{B_{t}}^{(2)}\right)=\left(Pr0_{it},\widehat{Pr_{it}}\right),
\]
where $\widehat{Pr_{it}}\equiv\sum_{x_{it}=(\mathcal{J}_{\mu},\tau)\in\chi-\{\emptyset\}}Pr_{it}(x_{it})\cdot(1-\phi(i,\mu,\tau+1))$. The former represents the fraction of consumers who do not own any working product. It directly affects the current profit of each firm, because the current profit of firm $f$ is equal to $\sum_{k\in\mathcal{J}_{ft}}(p_{kt}-mc_{kt})s_{kt}=\sum_{k\in\mathcal{J}_{ft}}(p_{kt}-mc_{kt})\left(\sum_{i}w_{i}s_{ikt}^{(ccp)}(x_{it}=\emptyset)\cdot Pr0_{it}\right)$. The latter represents the expected fraction of consumers who own products at the beginning of time $t$ but whose products fail by the beginning of time $t+1$. It affects the profits of firms in the next period. These variables are selected because they would largely affect firms' long-run profits. 

Using the variables $\widetilde{B_{t}}$, firms' value functions are approximated by:

\begin{eqnarray*}
V_{ft}^{F}(B_{t}) & \approx & \sum_{m=1}^{\widetilde{M_{F}}}\widetilde{\theta_{ft}^{F}}^{(m)}\widetilde{\Psi_{m}}(\widetilde{B_{t}}).
\end{eqnarray*}

Typically, tensor products of Chebyshev or other polynomials are used as the basis functions $\widetilde{\Psi_{m}^{F}}(\widetilde{B_{t}})$. Nevertheless, the number of tensor products gets so large when the dimension of $\widetilde{B_{t}}$ gets large. To reduce the number of basis functions and grid points and mitigate computational burden without losing numerical accuracy, we apply the Smolyak method, where the essential basis functions and grid points are systematically selected and constructed. I set the approximation level of $\mu$ to 1 for $\widetilde{B_{t}^{(1)}}$ and $\widetilde{B_{t}^{(2)}}$.\footnote{When the approximation level $\mu$ is set to 1, $V_{ft}^{F}(B_{t})\approx\widetilde{\theta_{ft}^{F}}^{(0)}+\sum_{m=1}^{n_{B}}\widetilde{\theta_{ft}^{F}}^{(m,1)}T_{1}(B_{t}^{(m)})+\sum_{m=1}^{n_{B}}\widetilde{\theta_{ft}^{F}}^{(m,2)}T_{2}(B_{t}^{(m)})$ holds, where $n_{B}$ denotes the dimension of $B_{t}$, and $B_{t}^{(m)}$ denotes $m$-th element. $T_{1}$ and $T_{2}$ denote first and second order Chebyshev basis functions.

I also experimented the setting where $\mu=2$. but the results were mostly the same.} For details, see \citet{judd2014smolyak}.

Regarding consumers' value functions $\widetilde{V_{it}^{C}}(x_{it},B_{t})$, computing the values of them at all $x_{it}\in\chi$ is computationally demanding, because the number of elements of $\chi$ depends on the possible number of product age, which we do not restrict to be finite.\footnote{Even when setting the finite maximum product age, the value is very large, because CFL products typically survive for over 3 years (36 months).} Hence, we approximate the values of $\widetilde{V_{it}^{C}}(x_{it}=(\mathcal{J}_{\mu},\tau\geq1),B_{t})$ by:

\[
\widetilde{V_{it}^{C}}(x_{it}=(\mathcal{J}_{\mu},\tau\geq1),B_{t})\approx\sum_{m_{B}=1}^{M_{B}}\sum_{m_{x}=1}^{M_{x}}\widetilde{\theta_{i\mu t}^{C}}^{(m_{B},m_{x})}(\widetilde{B_{t}})\cdot\widetilde{\Psi_{m}^{C,B,i\mu t}}(\widetilde{B_{t}})\cdot\widetilde{\Psi_{m}^{C,x,i\mu t}}(\tau).
\]
where $\widetilde{\Psi_{m}^{C,x}}(\tau)$ denotes $m$-th Chebyshev polynomial with appropriate variable transformation.\footnote{Generally, domain of Chebyshev polynomials is $[-1,1]$. Since the age of products $\tau$ is in $[1,\infty)$, I set the maximum age to be $T_{max}\equiv2\cdot\sum_{\tau=0}^{\infty}\phi(i,\mu,\tau)$ , and apply a function $\kappa:[1,T_{max}]\rightarrow[-1,1]$ such that $\kappa(\tau)=2\cdot\frac{\tau+1}{T_{max}+1}-1$. Grid points of $\tau$ are chosen to be Chebyshev extrema in $[-1,1]$.} In contrast, the values of $\widetilde{V_{it}^{C}}(x_{it}=\emptyset,B_{t})$ are approximated by:

\begin{eqnarray*}
\widetilde{V_{it}^{C}}(x_{it}=\emptyset,B_{t}) & \approx & \sum_{m=1}^{\widetilde{M_{C}}}\widetilde{\theta_{it}^{C}}^{(m)}\widetilde{\Psi_{m}^{C,B,i0t}}(\widetilde{B_{t}}).
\end{eqnarray*}

\subsubsection*{Construction of $B_{t}^{(grid)}$}

First, we construct $\widetilde{B_{t}}^{(grid)}\equiv\left(Pr0_{it}^{(grid)},\widehat{Pr_{it}}^{(grid)}\right)$ by applying the Smolyak method. Then, 

Next, the values of $Pr_{it}^{(grid)}(x_{it})$ have to satisfy the following equations:

\begin{eqnarray*}
\sum_{\tau\in\mathbb{N}}Pr_{it}(\mathcal{J}_{\mu=1,000},\tau)+\sum_{\tau\in\mathbb{N},\mu\neq1,000}Pr_{it}(\mathcal{J}_{\mu},\tau) & = & 1-Pr0_{it}^{(grid)}\\
\sum_{\tau\in\mathbb{N}}Pr_{it}(\mathcal{J}_{\mu=1,000},\tau)\cdot f(i,\mu=1000,\tau+1|\tau)+\sum_{\tau\in\mathbb{N},\mu\neq1,000}Pr_{it}(\mathcal{J}_{\mu},\tau)\cdot f(i,\mu,\tau+1|\tau) & = & \widehat{Pr_{it}}^{(grid)}
\end{eqnarray*}

By imposing the assumption of $Pr_{it}(\mu=1000,\tau)=\text{\ensuremath{\lambda_{1}\cdot}\ensuremath{Pr_{it}^{(data,stationary)}(\mu=1000,\tau)}}$ and $Pr_{it}(\mu\neq1,000,\tau)=\text{\ensuremath{\lambda_{2}\cdot}\ensuremath{Pr_{it}^{(data,stationary)}(\mu\neq1000,\tau)}}$, we can solve for $\lambda_{1}$ and $\lambda_{2}$. $Pr_{it}^{(grid)}(x_{it})$ constructed through the process are used as the grid points.

\subsubsection*{Numerical Accuracy}

To obtain numerically accurate results, $\widetilde{B_{t+1}}$ constructed by the use of ``sufficient statistics'' approach should well approximate the true values of $\widetilde{B_{t+1}}$ computed from $\{Pr_{it+1}(x_{it+1})\}_{i\in\mathcal{I},x_{it+1}\in\chi}$. Besides, since the approximation in $V^{C}$ and $V^{F}$ are introduced, the values of observed product prices $p_{jt}^{(data)}$ and product prices based on the structural model $p_{jt}(B_{t}^{(data)})$ might not necessarily coincide. Similarly, the values of $V^{C}(x_{it}=\emptyset,B_{t}^{(data)})$ computed in the demand estimation might not coincide with the ones computed in the marginal cost estimation. 

Table \ref{tab:Distribution-of-approx_error} shows the distribution of the ratios of relative approximation errors. Here, we define the ratio of relative error for variable $z$ by: $\frac{z^{(predict)}-z^{(true)}}{z^{(true)}}$. The table shows that predicted and true values are mostly the same. It suggests the approximation rarely affected the results.

\begin{table}[H]
\begin{centering}
\begin{tabular}{cccccc}
\hline 
 & Min & 25th & Median & 75th & Max\tabularnewline
\hline 
\hline 
$\widetilde{B_{t+1}}^{(1)}$ & 0.0000 & 0.0000 & 0.0000 & 0.0000 & 0.0000\tabularnewline
$\widetilde{B_{t+1}}^{(2)}$ & -0.0006 & -0.0002 & 0.0000 & 0.0002 & 0.0006\tabularnewline
$p_{jt}$ & -0.0003 & -0.0001 & 0.0000 & 0.0000 & 0.0084\tabularnewline
$V_{it}^{C}(x_{it}=\emptyset)$ & -0.0004 & -0.0003 & -0.0001 & 0.0000 & 0.0000\tabularnewline
\hline 
\end{tabular}
\par\end{centering}
\caption{Distribution of the ratios of relative approximation errors\label{tab:Distribution-of-approx_error}}
\end{table}

\subsubsection*{Consumer surplus}

Consumer surplus $CS_{t}$ is computed by:

\begin{eqnarray*}
CS_{t} & \equiv & \int\frac{1}{\alpha_{i}}Pr_{it}(x_{it})V_{it}^{C}(x_{it})dP(i)\\
 & = & \int\frac{1}{\alpha_{i}}\left[Pr_{it}(x_{it}=\emptyset)\cdot V_{it}^{C}(x_{it}=\emptyset)+\sum_{x_{it}=(j,\tau)\in\chi-\{\emptyset\}}Pr_{it}(x_{it}=(j,\tau))\cdot V_{it}^{C}(x_{it}=(j,\tau))\right]dP(i)\\
 & = & \int\frac{1}{\alpha_{i}}\left[Pr_{it}(x_{it}=\emptyset)\cdot\widetilde{V_{it}^{C}}(x_{it}=\emptyset)+\sum_{x_{it}=(\mathcal{J}_{\mu},\tau)\in\chi-\{\emptyset\}}Pr_{it}(x_{it}=(\mathcal{J}_{\mu},\tau))\cdot\widetilde{V_{it}^{C}}(x_{it}=(\mathcal{J}_{\mu},\tau))\right]dP(i)+\\
 &  & \int\left[\frac{1}{\alpha_{i}}\sum_{x_{it}=(j,\tau)\in\chi-\{\emptyset\}}Pr_{it}(x_{it}=(j,\tau))\cdot\left[\sum_{s=0}^{\infty}\beta_{C}^{\tau}\psi_{j}\cdot\phi\left(i,\mu_{j},\tau+s|\tau\right)\right]\right]dP(i).
\end{eqnarray*}

Given the values of aggregate states $B_{t}=\left(Pr_{it}(x_{it})\right)_{x_{it}\in\chi,i\in\mathcal{I}}$, the second term does not depend on the changes in product prices or product durability after time $t$. Hence, we can treat the second term as exogenous, and it is sufficient to consider the only the first term $\int\frac{1}{\alpha_{i}}\left[Pr_{it}(x_{it}=\emptyset)\cdot\widetilde{V_{it}^{C}}(x_{it}=\emptyset)+\sum_{x_{it}=(\mathcal{J}_{\mu},\tau)}Pr_{it}(x_{it}=(\mathcal{J}_{\mu},\tau))\cdot\widetilde{V_{it}^{C}}(x_{it}=(\mathcal{J}_{\mu},\tau))\right]dP(i)$ as the consumer surplus.

\subsubsection*{Externality}

The expected discounted sum of electricity consumption through the use of bulbs purchased after time $t$ satisfies the following:

\begin{eqnarray*}
V_{t}^{elec}(B_{t}) & \equiv & \int\left\{ \sum_{\tau_{1}=0}^{\infty}\beta_{S}^{\tau_{1}}\sum_{j\in\mathcal{J}_{t+\tau_{1}}}E_{L}\left[\sum_{\tau_{2}=0}^{L_{ij}}\beta_{S}^{\tau_{2}}e_{j}I_{i}\cdot Ms_{ijt+\tau_{1}}(B_{t+\tau_{1}})\right]\right\} dP(i)\\
 & = & \int\left[\sum_{j\in\mathcal{J}_{t}}E_{L}\left[\sum_{\tau_{2}=0}^{L_{ij}}\beta_{S}^{\tau_{2}}\right]e_{j}I_{i}\cdot Ms_{ijt}(B_{t})\right]dP(i)+\beta_{S}V_{t+1}^{elec}(B_{t+1}(B_{t}))
\end{eqnarray*}
Here, $\beta_{S}$ denotes a social discount factor, and it is assumed that $\beta_{C},\beta_{F},\beta_{S}$ share the common value.

Similarly, the expected discounted sum of waste emission of product $j$ purchased after time $t$ measured by the number of units satisfies the following:

\begin{eqnarray*}
V_{jt}^{waste}(B_{t}) & \equiv & \int\sum_{\tau=0}^{\infty}\left[\beta_{S}^{\tau}E_{L}\left[\beta_{S}^{L_{ij}}Ms_{ijt+\tau_{1}}(B_{t+\tau})\right]\right]dP(i)\\
 & = & \int\left[\sum_{j\in\mathcal{J}_{t}}E_{L}\left[\beta_{S}^{L_{ij}}\right]Ms_{ijt}(B_{t})\right]dP(i)+\beta_{S}V_{jt+1}^{waste}(B_{t+1}(B_{t})).
\end{eqnarray*}
We can solve for $V_{t}^{elec}(B_{t})$ and $V_{t}^{waste}(B_{t})$ and compute the associated negative externalities after solving for other variables by Algorithm \ref{alg:solve_equil}.

Based on \citet{MLIT_CBA}, which is the guideline for cost-benefit analysis used in policymaking in Japan, we set the carbon price to be 10,600 yen/t-C, which is equivalent to 2,896 yen/t-CO2. Regarding the CO2 emission from electricity generation , CO2 emission per kWh was set to 0.412 kg-CO2/kWh, based on the data in 2009 reported by Annual Report on Energy, published by the Agency for Natural Resources and Energy. In general, CO2 is emitted in the manufacturing and disposal process. Nevertheless, regarding incandescent lamps and CFLs, they are less than 0.2\% of the CO2 emission in the usage stage, as discussed in \citet{Tabata2012Intensity}. Hence, I omitted it in the results.

Regarding the externalities from waste emissions, I compute the values based on JCLA data base\footnote{The database can be accessed in the website by Lifecycle Assessment Society of Japan (``https://lca-forum.org/database/impact/'').}, published by Research Institute of Science for Safety and Sustainability, Advanced LCA Research Group. Externalities from the disposal of incandescent lamps is set to 1.13 yen/unit, because the estimated externality of disposing of small home appliance products is 37.9 yen/kg, and the mass of each unit is around 30g. Externalities from the disposal of CFLs is set to 8.94 yen/unit. Since CFLs contain mercury, I considered the externalities from using mercury, which is estimated to be 1257.3 yen/kg. Since the mass of mercury contained in CFL products should be below 5mg by regulation, we assume each CFL product contains 5mg of mercury. I also considered the externalities from disposing of products other than from mercury, assuming the mass of CFL products is around 70g.

\subsection{Solution method of the equilibrium under firms' commitment ability\label{subsec:Details_algorithm_supply_commit}}

In this section, I describe the solution method of the equilibrium under firms' commitment ability. Here, we consider the setting where firms commit to future product prices, and the set of products and product characteristics do not change over time. Then, optimal prices should satisfy the following first-order condition:

\begin{eqnarray}
0 & = & \frac{\partial}{\partial p_{j}}\left[\sum_{\tau=0}^{\infty}\beta_{F}^{\tau}\pi(B_{t+\tau},p_{t+\tau},h_{t+\tau})\right]\nonumber \\
 & = & \sum_{\tau=0}^{\infty}\beta_{F}^{\tau}\left[\left(\frac{\partial}{\partial p_{jt+\tau}}+\frac{\partial h_{t+\tau}}{\partial p_{j}}\frac{\partial}{\partial h_{t+\tau}}\right)\pi_{ft+\tau}(B_{t+\tau};p_{t+\tau},h_{t+\tau})+\right.\nonumber \\
 &  & \ \ \ \ \left.\left(\left(\frac{\partial}{\partial p_{jt+\tau}}+\frac{\partial h_{t+\tau}}{\partial p_{j}}\frac{\partial}{\partial h_{t+\tau}}\right)s_{t+\tau}(B_{t+\tau};p_{t+\tau},h_{t+\tau})\right)\cdot\frac{\partial B_{t+\tau+1}}{\partial s_{t+\tau}}\frac{\partial\beta_{F}V_{ft+\tau+1}^{F}(B_{t+\tau+1})}{\partial B_{t+\tau+1}}\right].\label{eq:FOC_commit}
\end{eqnarray}

Here, we additionally define two new functions $V_{j}^{numer}(B_{t};p)$ and $V_{j}^{denom}(B_{t};p)$:

{\footnotesize{}
\begin{eqnarray*}
V_{j}^{numer}(B_{t};p) & \equiv & \sum_{\tau=0}^{\infty}\beta_{F}^{\tau}\left[\left(\frac{\partial}{\partial p_{jt+\tau}}+\frac{\partial h_{t+\tau}}{\partial p_{j}}\frac{\partial}{\partial h_{t+\tau}}\right)\left(\pi_{ft+\tau}(B_{t+\tau};p_{t+\tau},h_{t+\tau})-s_{jt+\tau}(B_{t+\tau};p_{t+\tau},h_{t+\tau})(p_{j}-mc_{j})\right)+\right.\\
 &  & \ \ \ \ \left.\left(\left(\frac{\partial}{\partial p_{jt+\tau}}+\frac{\partial h_{t+\tau}}{\partial p_{j}}\frac{\partial}{\partial h_{t+\tau}}\right)s_{t+\tau}(B_{t+\tau};p_{t+\tau},h_{t+\tau})\right)\cdot\frac{\partial B_{t+\tau+1}}{\partial s_{t+\tau}}\frac{\partial\beta_{F}V_{ft+\tau+1}^{F}(B_{t+\tau+1})}{\partial B_{t+\tau+1}}\right],\\
V_{j}^{denom}(B_{t};p) & \equiv & \sum_{\tau=0}^{\infty}\beta_{F}^{\tau}\left[\left(\frac{\partial}{\partial p_{jt+\tau}}+\frac{\partial h_{t+\tau}}{\partial p_{j}}\frac{\partial}{\partial h_{t+\tau}}\right)s_{jt+\tau}(B_{t+\tau};p_{t+\tau},h_{t+\tau})\right].
\end{eqnarray*}
}{\footnotesize\par}

Then, equation (\ref{eq:FOC_commit}) can be reformulated as:

\begin{eqnarray*}
0 & = & V_{j}^{numer}(B_{t};p)+(p_{j}-mc_{j})\cdot V_{j}^{denom}(B_{t};p).
\end{eqnarray*}

Hence, optimal product prices $p_{j}$ under commitment satisfies the following equation using $V_{j}^{numer}(B_{t};p)$ and $V_{j}^{denom}(B_{t};p)$, which is the counterpart of updating equation (\ref{eq:update_p}) used to solve Markov perfect equilibrium:

\begin{eqnarray}
p_{j} & = & mc_{j}-\frac{V_{j}^{numer}(B_{t};p)}{V_{j}^{denom}(B_{t};p)}.\label{eq:p_update_commit}
\end{eqnarray}

Though it is generally not possible to analytically solve for $V_{j}^{numer}(B_{t};p)$ and $V_{j}^{numer}(B_{t};p)$, they satisfy the following equations, which are the counterparts of Bellman equation:

\begin{eqnarray}
V_{j}^{numer}(B_{t};p) & = & \left(\frac{\partial}{\partial p_{jt}}+\frac{\partial h_{t}}{\partial p_{j}}\frac{\partial}{\partial h_{t}}\right)\left(\pi_{ft}(B_{t};p_{t},h_{t})-s_{jt}(B_{t};p_{t},h_{t})(p_{j}-mc_{j})\right)+\label{eq:V_numer_Bellman}\\
 &  & \left(\left(\frac{\partial}{\partial p_{jt}}+\frac{\partial h_{t}}{\partial p_{j}}\frac{\partial}{\partial h_{t}}\right)s_{t}(B_{t};p_{t},h_{t})\right)\cdot\frac{\partial B_{t+1}}{\partial s_{t}}\frac{\partial\beta_{F}V_{ft+1}^{F}(B_{t+1})}{\partial B_{t+1}}+\beta_{F}V_{j}^{numer}\left(B_{t+1}(B_{t});p\right),\nonumber \\
V_{j}^{denom}(B_{t};p) & = & \left(\frac{\partial}{\partial p_{jt}}+\frac{\partial h_{t}}{\partial p_{j}}\frac{\partial}{\partial h_{t}}\right)\left(s_{jt}(B_{t};p_{t},h_{t})\right)+\beta_{F}V_{j}^{denom}\left(B_{t+1}(B_{t});p\right).\label{eq:V_denom_Bellman}
\end{eqnarray}

Note that $\frac{\partial h_{ijt}}{\partial p_{j}}=E_{L}\left[\beta_{C}^{L_{ij}-1}\right]\cdot\frac{\partial V_{i}^{C}(x_{it}=\emptyset)}{\partial p_{j}}$ and $\frac{\partial h_{i0t}}{\partial p_{j}}=\frac{\partial V_{i}^{C}(x_{it}=\emptyset)}{\partial p_{j}}$ hold. Since the derivative of consumers' value functions $\frac{\partial V_{it}^{C}(x_{it}=\emptyset;p)}{\partial p_{j}}$ satisfies 
\begin{eqnarray*}
\frac{\partial V_{it}^{C}(x_{it}=\emptyset;p)}{\partial p_{j}} & = & \frac{\partial\widetilde{v_{ijt}}}{\partial p_{j}}\frac{\partial V_{it}^{C}}{\partial\widetilde{v_{ijt}}}+\sum_{k\in\mathcal{J}\cup\{0\}}\frac{\partial g_{ikt}}{\partial p_{j}}\frac{\partial\widetilde{v_{ijt}}}{\partial g_{ikt}}\frac{\partial V_{it}^{C}}{\partial\widetilde{v_{ijt}}}\\
 & = & -\alpha_{i}\frac{\partial V_{it}^{C}}{\partial\widetilde{v_{ijt}}}+\frac{\partial V_{it}^{C}}{\partial p_{jt}}\left(\beta_{C}\frac{\partial V_{it}^{C}}{\partial\widetilde{v_{i0t}}}+\sum_{k\in\mathcal{J}_{t}}E_{L}\left[\beta_{C}^{L_{ik}}\right]\cdot\frac{\partial V_{it}^{C}}{\partial\widetilde{v_{ijt}}}\right).,
\end{eqnarray*}

$\frac{\partial V_{it}^{C}(x_{it}=\emptyset;p)}{\partial p_{jt}}$ is in the following form:

\begin{eqnarray}
\frac{\partial V_{it}^{C}(x_{it}=\emptyset;p)}{\partial p_{jt}} & = & \frac{-\alpha_{i}\frac{\partial V_{it}^{C}}{\partial\widetilde{v_{ijt}}}}{1-\beta_{C}\frac{\partial V_{it}^{C}}{\partial\widetilde{v_{i0t}}}-\sum_{k\in\mathcal{J}_{t}}E_{L}\left[\beta_{C}^{L_{ik}}\right]\cdot\frac{\partial V_{it}^{C}}{\partial\widetilde{v_{ijt}}}}\nonumber \\
 & = & \frac{-\alpha_{i}s_{ijt}^{(ccp)}}{1-\beta_{C}s_{i0t}^{(ccp)}-\sum_{k\in\mathcal{J}_{t}}E_{L}\left[\beta_{C}^{L_{ik}}\right]\cdot s_{ijt}^{(ccp)}}.\label{eq:dVC_dp}
\end{eqnarray}

Based on these equations, full commitment equilibrium can be numerically solved by Algorithm \ref{alg:solve_equil_commit}.\footnote{\citet{chen2013secondary} numerically solved the equilibrium under the setting where the market is in the stationary state (absorbing state). The current algorithm allows for the case where the market is not in the stationary state.} The updates of variables other than $p_{jt}$ are mostly the same as the ones in Algorithm \ref{alg:solve_equil}.{\footnotesize{}}
\begin{algorithm}[H]
\begin{enumerate}
\item {\footnotesize{}Take grid points of aggregate state variables $B_{t}^{(grid)}$. Set initial values of $\left\{ \widetilde{V_{i}^{C(0)}}(x_{it}=\emptyset;p)\right\} _{i}$ (consumers' value function), $\left\{ V_{f}^{F(0)}(B_{t}^{(grid)})\right\} _{B_{t}}$ (firm's value function), $\left\{ p_{jt}^{(0)}(B_{t}^{*})\right\} _{j}$ (equilibrium price), $\left\{ B_{t+1}^{(0)}(B_{t}^{(grid)})\right\} _{B_{t}}$ (aggregate state variables in the next period), $\left\{ \frac{\partial B_{t+1}^{(0)}}{\partial p_{jt}}(B_{t}^{*})\right\} $(derivative of the aggregate state variables in the next period with respect to the current price), and $V_{j}^{numer}(B_{t}^{(grid)})$, $V_{j}^{denom}(B_{t}^{(grid)})$ .}{\footnotesize\par}
\item {\footnotesize{}Iterate the following process until the convergence of $\widetilde{V_{it}^{C(n)}}(x_{it}=\emptyset;p)$, $V_{ft}^{F(n)}(B_{t}^{(grid)})$, $p_{jt}^{(n)}(B_{t}^{*})$, $B_{t+1}^{(n)}(B_{t}^{(grid)})$, $\frac{\partial B_{lt+1}^{(n)}}{\partial p_{j}}(B_{t}^{*})$, $V_{j}^{numer(n)}(B_{t}^{(grid)})$, $V_{j}^{denom(n)}(B_{t}^{(grid)})$ ($n=0,1,2,\cdots$):}{\footnotesize\par}
\begin{enumerate}
\item {\footnotesize{}Given $\widetilde{V_{i}^{C(n)}}(x_{it}=\emptyset;p)$, $V_{f}^{F(n)}(B_{t}^{(grid)})$, $p_{j}^{(n)}(B_{t}^{*})$, $B_{lt+1}^{(n)}(B_{t}^{(grid)})$, $\frac{\partial B_{lt+1}^{(n)}}{\partial p_{jt}}(B_{t}^{*})$,}{\footnotesize\par}
\begin{enumerate}
\item {\footnotesize{}Compute $\widetilde{v_{ijt}}(x_{it}=\emptyset;p)=-\alpha_{i}p_{jt}+\delta_{jt}+E_{t}\left[\beta_{C}^{L_{ij}}\right]\cdot\widetilde{V_{l}^{C(0)}}(x_{it}=\emptyset;p)$ and $\widetilde{v_{i0t}}(x_{it}=\emptyset;p)=\beta_{C}\widetilde{V_{i}^{C(0)}}(x_{it}=\emptyset;p)$}{\footnotesize\par}
\item {\footnotesize{}Compute $s_{ijt}^{(ccp)(n+1)}(x_{it}=\emptyset,p_{t}^{(n)}),s_{i0t}^{(ccp)(n+1)}(x_{it}=\emptyset,,p_{t}^{(n)}),s_{jt}^{(n+1)}(p_{t}^{(n)})$ by (\ref{eq:choice_prob_ccp}), (\ref{eq:outside_choice_prob_ccp}), and (\ref{eq: agg_market_share})}{\footnotesize\par}
\item {\footnotesize{}Compute $\frac{\partial V_{it}^{C(n+1)}(x_{it}=\emptyset;p)}{\partial p_{j}}$ by (\ref{eq:dVC_dp})}{\footnotesize\par}
\item {\footnotesize{}Compute $\frac{\partial\widetilde{v_{ikt}}^{(n+1)}}{\partial p_{j}}=-\alpha_{i}+E_{L}\left[\beta_{C}^{L_{ij}}\right]\cdot\frac{\partial V_{i}^{C(n+1)}(x_{it}=\emptyset;p)}{\partial p_{j}}$ and $\frac{\partial\widetilde{v_{i0t}}^{(n+1)}}{\partial p_{j}}=\beta_{C}\frac{\partial V_{i}^{C}(x_{it}=\emptyset;p)}{\partial p_{j}}$}{\footnotesize\par}
\item {\footnotesize{}Compute $\frac{\partial s_{i\widetilde{\mu}t}^{(n+1)}(B_{t}^{(grid)})}{\partial p_{jt}}=\sum_{k\in\mathcal{J}_{t}\cup\{0\}}\frac{\partial\widetilde{v_{ikt}}(x_{it}=\emptyset,B_{t}^{(grid)})}{\partial p_{jt}}\frac{\partial s_{i\widetilde{\mu}t}(B_{t}^{(grid)})}{\partial\widetilde{v_{ikt}}(x_{it}=\emptyset,B_{t}^{(grid)})}+\sum_{\mu\in\mathcal{M}}\frac{\partial h_{i\mu t}(B_{t+1}^{(n)})}{\partial p_{j}}\frac{\partial s_{i\widetilde{\mu}t}(B_{t}^{(grid)})}{\partial h_{i\mu t}(B_{t+1}^{(n)})}$}{\footnotesize\par}
\item {\footnotesize{}Compute $\frac{\partial V_{ft+1}^{F}(B_{t+1}^{(n)}(B_{t}^{(grid)}))}{\partial p_{jt}}=\frac{\partial V_{ft+1}^{F}(B_{t+1}^{(n)}(B_{t}^{(grid)}))}{\partial B_{t+1}^{(n)}}\cdot\frac{\partial B_{t+1}^{(n)}}{\partial p_{jt}}(B_{t}^{(grid)})$}{\footnotesize\par}
\end{enumerate}
\item {\footnotesize{}Given $\widetilde{V_{it}^{C(n)}}(x_{it}=\emptyset,B_{t}^{(grid)}),V_{ft}^{F(n)}(B_{t}^{(grid)}),p_{jt}^{(n)}(B_{t}^{*}),B_{t+1}^{(n)}(B_{t}^{(grid)}),\frac{\partial B_{t+1}^{(n)}}{\partial p_{jt}}(B^{*}),V_{j}^{numer(n)}(B_{t}^{(grid)}),V_{j}^{denom(n)}(B_{t}^{(grid)})$ and $s_{ijt}^{(ccp)(n+1)}(x_{it}=\emptyset;p)$, $s_{i0t}^{(ccp)(n+1)}(x_{it}=\emptyset;p),$$s_{jt}^{(n+1)}(B_{t}^{^{(grid)}}),\frac{\partial s_{i\mu t}^{(n+1)}}{\partial p_{jt}}(B_{t}^{(grid)})$, update variables:}{\footnotesize\par}
\begin{enumerate}
\item {\footnotesize{}Compute $p_{jt}^{(n+1)}(B_{t}^{*})$ by (\ref{eq:p_update_commit})}{\footnotesize\par}
\item {\footnotesize{}Compute $V_{j}^{numer(n+1)}(B_{t}^{(grid)})$ by (\ref{eq:V_numer_Bellman})}{\footnotesize\par}
\item {\footnotesize{}Compute $V_{j}^{denom(n+1)}(B_{t}^{(grid)})$ by (\ref{eq:V_denom_Bellman})}{\footnotesize\par}
\item {\footnotesize{}Compute $V_{ft}^{F(n+1)}(B_{t}^{(grid)})$ by (\ref{eq:V_F})}{\footnotesize\par}
\item {\footnotesize{}Compute $\widetilde{V_{it}^{C(n+1)}}(x_{it}=\emptyset;p)$ by (\ref{eq:V_C})}{\footnotesize\par}
\item {\footnotesize{}Compute $B_{t+1}^{(n+1)}(B_{t}^{(grid)})$ by (\ref{eq:Pr_agg_state_transition})}{\footnotesize\par}
\item {\footnotesize{}Compute $\frac{\partial B_{t+1}^{(n+1)}}{\partial p_{j}}(B_{t}^{*})=\sum_{\mu\in\mathcal{M}}\frac{\partial s_{i\mu t}^{(n+1)}}{\partial p_{jt}}(B_{t}^{*})\frac{\partial B_{t+1}}{\partial s_{i\mu t}}(B_{t}^{*})$}{\footnotesize\par}
\end{enumerate}
\end{enumerate}
\end{enumerate}
{\footnotesize{}\caption{{\footnotesize{}Algorithm for solving the equilibrium (commitment case)\label{alg:solve_equil_commit}}}
}{\footnotesize\par}

{\footnotesize{}Note. $B_{t}^{*}$ denotes the value of initial $B_{t}$ at which optimal commitment prices are computed.}{\footnotesize\par}
\end{algorithm}
{\footnotesize\par}

\subsection{Solution method of the Markov perfect equilibrium under consumers' adaptive expectation\label{subsec:Details_algorithm_supply_adaptive_expectation}}

The algorithm to solve the equilibrium under consumers' adaptive expectation is close to Algorithm \ref{alg:solve_equil} under Markov perfect equilibrium with consumers' perfect foresight. The difference is that we compute $\frac{\partial h}{\partial p_{j}}$ by $\frac{\partial h_{ijt}(x_{it}=\emptyset)}{\partial p_{j}}=E_{L}\left[\beta_{C}^{L_{ij}}\right]\cdot\frac{\partial V_{i}^{C(n+1)}(x_{it}=\emptyset;p)}{\partial p_{j}}$ and $\frac{\partial h_{i0t}(x_{it}=\emptyset)}{\partial p_{j}}=\frac{\partial V_{i}^{C(n+1)}(x_{it}=\emptyset;p)}{\partial p_{j}}$, where $\frac{\partial V_{i}^{C(n+1)}(x_{it}=\emptyset;p)}{\partial p_{j}}$ is given by (\ref{eq:dVC_dp}).

\subsection{Details on the estimation of fixed costs\label{subsec:fixed_cost}}

In this section, we discuss the estimation of fixed costs using the observed data on product choices based on the moment inequality approach.

Let $V_{f}(B;\mathcal{J}_{f})$ be firm $f$'s discounted sum of profits given that it introduces the set of products $\mathcal{J}_{f}$ at aggregate states $B$. In reality, firms introduce set of products $\mathcal{J}_{f}^{(data)}$. We assume each firm does not have incentives to deviate from the current choice regarding the product introduction at the stationary aggregate state. Formally, we impose the following condition:

\begin{eqnarray*}
V_{f}^{F}\left(B^{(stationary)};\mathcal{J}_{f}^{(data)}\right) & \geq & V_{f}^{F}\left(B^{(stationary)};\mathcal{J}_{ft}^{(data)}-\{j\}\right)+F_{j}+e_{j}^{error}
\end{eqnarray*}
where $F_{j}$ denotes the fixed cost of product $j$ and $e_{j}^{error}$ represents the optimization error or approximation error of the model (cf. \citet{pakes2015moment}). The real market might not be at the stationary state, and such approximation error of the model is reflected in the term. Note that we assume $E[e_{j}^{error}]=0$ holds, which implies firms' decisions are on average optimal. We assume the products can be classified into exclusive sets of products $\mathcal{J}_{d}$, and let $F_{j}=\theta_{d}1[j\in\mathcal{J}_{d}]+e_{j}^{fc}$. $e_{j}^{fc}$ represents the expectational error or the approximation error of the model. We assume $E\left[e_{j}^{fc}\right]=0$ holds. Under these conditions, $E\left[e_{j}^{error}+e_{j}^{fc}\right]=0$ holds, and we can estimate the upper bound of the parameters $\theta_{d}$.

\begin{table}[H]
\begin{centering}
\begin{tabular}{cccc}
\hline 
 & Upper & S.D. & \# of obs.\tabularnewline
\hline 
\hline 
1000h Inc. & 0.276 & 0.17 & 6\tabularnewline
2000h. Inc. & 0.021 & 0.029 & 6\tabularnewline
6000h CFL (Panasonic) & 0.329 & 0.326 & 6\tabularnewline
6000h CFL (Toshiba) & 0.174 & 0.216 & 6\tabularnewline
10000h CFL (Panasonic) & 0.855 & 0.545 & 3\tabularnewline
10000h CFL (Toshiba) & 0.302 & 0.142 & 3\tabularnewline
\hline 
\end{tabular}
\par\end{centering}
\caption{Fixed cost estimates\label{tab:Fixed-cost-estimates}}

{\footnotesize{}Notes. }{\footnotesize\par}

{\footnotesize{}Unit of the estimates are billion yen.}{\footnotesize\par}

{\footnotesize{}The column of ``\# of obs.'' represents the number of products in each product category, which are used in the estimation.}{\footnotesize\par}
\end{table}

Table \ref{tab:Fixed-cost-estimates} shows the estimation results. Though the standard deviations are relatively large, possibly because of the small number of observations used in the estimation, the upper bound of fixed costs are estimated to be positive. 

\subsection{Assumptions on consumers' expectations\label{subsec:Assumptions_expec}}

In the estimation, we have assumed that consumers have perfect foresight to simplify the explanation. Nevertheless, without changing the estimation procedure, we can relax the assumption to rational expectation, where consumers' expectations regarding the transitions of $\widetilde{\Omega_{t}^{C}}$ are on average correct. Formally, we alternatively impose the following assumptions for $t\leq T$:

$E_{x,\Omega^{C}}\left[V_{it+1}^{C}\left(x_{t+1},\Omega_{t+1}^{C}\right)|x_{it},\Omega_{t}^{C},a_{it}=k\right]=E_{x}\left[V_{it+1}^{C}\left(x_{t+1},\Omega_{t+1}^{C(realized)}\right)|x_{it},a_{it}=k\right]+\widetilde{e_{kt}}(x_{it})\ (k\in\mathcal{A}_{t}(x_{it}))$ such that $E\left[\widetilde{e_{kt}}(x)|Z_{kt}\right]=0\ \forall x\in\mathcal{\chi},\forall k\in\mathcal{A}_{t}(x_{it})$, where $\widetilde{e_{kt}^{C}}(x)$ denotes prediction errors of consumers at state $x$ and choosing alternative $k$ at time $t$, and we assume $e_{kt}^{C}(x)$ are common for all consumers in each period. We additionally define $e_{0t}\equiv E_{t}\left[V_{it+1}^{C}\right]$ and $e_{jt}\equiv\sum_{\tau=1}^{\infty}f(i,\mu_{j},\tau)\left(E_{t}V_{it+\tau}^{C}-V_{it+\tau}^{C}\right)$, and they also satisfy $E\left[e_{0t}|Z_{kt}\right]=0$ and $E\left[e_{kt}|Z_{kt}\right]=0\ \forall k\in\mathcal{J}_{t}\ t\leq T$. Since we do not allow for prediction errors in the stationary state, $e_{kt}=e_{0t}=0$ holds for $t\geq T+1$. 

Next, in the following, without losing generality, we consider the setting where $\epsilon_{ijt}$ follows Gumbel distribution with no nest structure to simplify the explanation. Then, in the current model,

\begin{eqnarray*}
V_{it}^{C} & = & \log\left(\exp\left(\beta_{C}V_{it+1}^{C}+\beta_{C}e_{0t}\right)+\sum_{j\in\mathcal{J}_{t}}\exp\left(\delta_{jt}+\kappa_{ijt}+\sum_{\tau=1}^{\infty}f(i,\mu_{j},\tau)V_{it+\tau}^{C}+e_{jt}\right),\right)
\end{eqnarray*}
where $\kappa_{ijt}=-\alpha_{i}p_{jt}$. By defining $\widehat{V_{it}^{C}}\equiv V_{it}^{C}+\eta_{t}$ such that $\eta_{t}$ satisfies $\beta_{C}e_{0t}+\eta_{t}-\beta_{C}\eta_{t+1}=0$ and $\widehat{\delta_{jt}}\equiv\delta_{jt}-\sum_{\tau=1}^{\infty}f(i,\mu_{j},\tau)\eta_{t+\tau}+e_{jt}+\eta_{t}$, we have:

\begin{eqnarray}
\widehat{V_{it}^{C}} & = & \log\left(\exp\left(\beta_{C}\widehat{V_{it+1}^{C}}-\beta_{C}\eta_{t+1}+\eta_{t}+\beta_{C}e_{0t}\right)+\right.\nonumber \\
 &  & \left.\sum_{j\in\mathcal{J}_{t}}\exp\left(\delta_{jt}+\kappa_{ijt}+\sum_{\tau=1}^{\infty}f(i,\mu_{j},\tau)\widehat{V_{it+\tau}^{C}}-\sum_{\tau=1}^{\infty}f(i,\mu_{j},\tau)\eta_{t+\tau}+e_{jt}+\eta_{t}\right)\right)\nonumber \\
 & = & \log\left(\exp\left(\beta_{C}\widehat{V_{it+1}^{C}}\right)+\sum_{j\in\mathcal{J}_{t}}\exp\left(\widehat{\delta_{jt}}+\kappa_{ijt}+\sum_{\tau=1}^{\infty}f(i,\mu_{j},\tau)\widehat{V_{it+\tau}^{C}}\right)\right).\label{eq:V_rational_expec}
\end{eqnarray}

Since 

\begin{eqnarray}
s_{ijt}^{(ccp)} & = & \frac{\exp\left(\delta_{jt}+\kappa_{ijt}+\sum_{\tau=1}^{\infty}f(i,\mu_{j},\tau)V_{it+\tau}^{C}+e_{jt}\right)}{\exp\left(V_{it}^{C}\right)}\nonumber \\
 & = & \frac{\exp\left(\widehat{\delta_{jt}}+\kappa_{ijt}+\sum_{\tau=1}^{\infty}f(i,\mu_{j},\tau)\widehat{V_{it+\tau}^{C}}\right)}{\exp\left(\widehat{V_{it}^{C}}\right)},\label{eq:CCP_rational_expec}
\end{eqnarray}

$\widehat{\delta}$ satisfies:

\begin{eqnarray*}
S_{jt}^{(data)} & = & \int Pr0_{it}\cdot s_{ijt}^{(ccp)}dP(i)\\
 & = & \int Pr0_{it}\cdot\frac{\exp\left(\widehat{\delta_{jt}}+\kappa_{ijt}+\sum_{\tau=1}^{\infty}f(i,\mu_{j},\tau)\widehat{V_{it+\tau}^{C}}\right)}{\exp\left(\widehat{V_{it}^{C}}\right)}dP(i).
\end{eqnarray*}
These equations imply we can solve for $\delta$ and $V^{C}$ without knowing the values of $e_{jt}$ and $e_{0t}$. 

Since $\delta\equiv X^{D}\theta_{linear}^{D}+\xi$, $\widehat{\delta}=X^{D}\theta_{linear}^{D}+\widehat{\xi}$ holds if we define $\widehat{\xi_{jt}}\equiv\xi_{jt}-\sum_{\tau=1}^{\infty}f(i,\mu_{j},\tau)\eta_{t+\tau}+e_{jt}+\eta_{t}$. Under the conditions of $E\left[e_{jt},e_{0t}|Z\right]=0$, $E\left[\xi|Z^{D}\right]=0$ implies $E\left[\widehat{\xi}|Z^{D}\right]=0$, and we can alternatively impose the moment condition on $\widehat{\xi}$ to estimate parameters $\theta^{D}\equiv\left(\theta_{linear}^{D},\theta_{nonlinear}^{D}\right)$. 

In the counterfactual simulation, we evaluate the cases where the values of $\kappa_{ijt}\equiv-\alpha_{i}p_{jt}$ changes\footnote{Eliminating product $j$ is essentially equivalent to exogenously setting $p_{jt}=-\infty$.}. Even when evaluating counterfactual outcomes, we can obtain correct CCPs by (\ref{eq:V_rational_expec}) and (\ref{eq:CCP_rational_expec}) as long as the values of $\widehat{\delta}$ do not change. Regarding consumer surplus, $\widehat{V_{it}^{C}}\equiv V_{it}^{C}+\eta_{t}$ is equal to $V^{C}$ up to an additive constant, and we can correctly evaluate it as long as the values of $\eta$ do not change.

Regarding the evaluation of counterfactual outcomes, we can obtain correct values if the values of prediction errors do not change, because $\widetilde{V^{C}}$ is equal to $V$ up to an additive constant under the assumption. 

\section{Proof\label{sec:Proof}}

\subsection{Proof of the statements in Section \ref{subsubsec:Theoretical-analysis_durability}\label{subsec:Proof-theory_durability} (Theoretical analysis on firms' endogenous durability choice)}

\subsubsection*{Proof of Proposition \ref{prop:collusion_durability}}
\begin{proof}
Since $V_{j=2}^{F}=\left(p_{j=2t=1}-c_{j=2t=1}(\phi_{j=2})\right)q_{j=2t=1}+\beta(p_{j=2t=2}-c_{j=2t=2})q_{j=2t=2}$, 
\begin{eqnarray*}
\frac{\partial\left(V_{j=1}^{F}+V_{j=2}^{F}\right)}{\partial\phi_{j=1}} & = & \frac{\partial V_{j=1}^{F}}{\partial\phi_{j=1}}+\left(p_{j=2t=1}-c_{j=2t=1}(\phi_{j})\right)\cdot\frac{\partial q_{j=2t=1}}{\partial\phi_{j=1}}\\
 &  & +\beta(p_{j=2t=2}-c_{j=2t=2})\frac{\partial q_{j=2t=2}}{\partial\phi_{j=1}}.
\end{eqnarray*}

Here,

\begin{eqnarray*}
\frac{\partial q_{j=2t=2}}{\partial\phi_{j=1}} & = & \sum_{k\in\mathcal{J}}\left(\frac{\partial q_{kt=1}}{\partial\phi_{j=1}}\frac{\partial\Psi_{kt=2}}{\partial q_{kt=1}}\frac{\partial q_{j=2t=2}}{\partial\Psi_{kt=2}}\right)\\
 & = & \sum_{k\in\mathcal{J}}\left(\frac{\partial q_{kt=1}}{\partial\phi_{j=1}}\phi_{k}\frac{\partial q_{j=2t=2}}{\partial\Psi_{kt=2}}\right)\\
 & \leq & 0.
\end{eqnarray*}

By Assumption \ref{as:pref_for_durab_diff}, $\frac{\partial q_{j=2t=1}}{\partial\phi_{j=1}}\leq0$ holds. Here, let $p^{O}$ and $\phi^{O}$ be prices and durability levels in the case without collusion on durability. Then,

\begin{eqnarray*}
\left.\frac{\partial\left(V_{j=1}^{F}+V_{j=2}^{F}\right)}{\partial\phi_{j=1}}\right|_{(p^{O},\phi^{O})} & = & \left.\frac{\partial V_{j=1}^{F}}{\partial\phi_{j=1}}\right|_{(p^{O},\phi^{O})}+\left.\frac{\partial V_{j=2}^{F}}{\partial\phi_{j=1}}\right|_{(p^{O},\phi^{O})}\\
 & \leq & 0.
\end{eqnarray*}

In contrast, $\phi^{CD}(p^{O})$, durability levels under the case of durability collusion given the price levels $p^{O}$ satisfy$\left.\frac{\partial\left(V_{j=1}^{F}+V_{j=2}^{F}\right)}{\partial\phi_{j=1}}\right|_{(p^{O},\phi^{CD})}=0$. Then, under $\left.\frac{\partial^{2}\left(V_{j=1}^{F}+V_{j=2}^{F}\right)}{\partial\phi_{j=1}^{2}}\right|_{(p^{O},\phi^{CD})}<0$, $\phi^{CD}(p^{O})\leq\phi^{O}$ holds.
\end{proof}

\subsubsection*{Proof of Proposition \ref{prop:collusion_price}}
\begin{proof}
First,

\begin{eqnarray*}
V_{j}^{F} & = & \left(P_{jt=1}-c_{jt=1}(\phi_{j})+\beta c_{jt=2}\phi_{j}\right)Q_{jt=1}+\beta(P_{jt=2}-c_{jt=2})Q_{jt=2}
\end{eqnarray*}
holds. Then, Under Assumption \ref{as:no_pref_durability},

\begin{eqnarray*}
0 & = & \frac{\partial V_{j}^{F}}{\partial\phi_{j}}\\
 & = & \left(-\frac{\partial c_{jt=1}(\phi_{j})}{\partial\phi_{j}}+\beta c_{jt=2}\right)Q_{jt=1}+\beta\left(P_{jt=2}-c_{jt=2}\right)\frac{\partial\Psi_{jt=2}}{\partial\phi_{j}}\frac{\partial Q_{jt=2}}{\partial\Psi_{jt=2}}\\
 & = & \left(-\frac{\partial c_{jt=1}(\phi_{j})}{\partial\phi_{j}}+\beta c_{jt=2}\right)Q_{jt=1}+\beta\left(P_{jt=2}-c_{jt=2}\right)Q_{jt=1}\frac{\partial Q_{jt=2}}{\partial\Psi_{jt=2}}.
\end{eqnarray*}
Namely, $-\frac{\partial c_{jt=1}(\phi_{j})}{\partial\phi_{j}}+\beta c_{jt=2}+\beta\left(P_{jt=2}-c_{jt=2}\right)\frac{\partial Q_{jt=2}}{\partial\Psi_{jt=2}}=0$ holds at the profit maximizing durability level. 

Then, we have:

\begin{eqnarray*}
\frac{\partial^{2}V_{j}^{F}}{\partial\phi_{j}\partial P_{jt=2}} & = & \frac{\partial Q_{jt=1}}{\partial P_{jt=2}}\left(-\frac{\partial c_{jt=1}(\phi_{j})}{\partial\phi_{j}}+\beta c_{jt=2}+\beta\left(P_{jt=2}-c_{jt=2}\right)\frac{\partial Q_{jt=2}}{\partial\Psi_{jt=2}}\right)+\\
 &  & Q_{jt=1}\cdot\frac{\partial}{\partial P_{jt=2}}\left[\beta\left(P_{jt=2}-c_{jt=2}\right)\frac{\partial Q_{jt=2}}{\partial\Psi_{jt=2}}\right]\\
 & = & Q_{jt=1}\cdot\beta\left(\frac{\partial Q_{jt=2}}{\partial\Psi_{jt=2}}+(P_{jt=2}-c_{jt=2})\frac{\partial^{2}Q_{jt=2}}{\partial P_{jt=2}\partial\Psi_{jt=2}}\right)\\
 & > & 0\ \left(\because\frac{\partial s_{jt=2}^{(ccp)}}{\partial P_{jt=2}}<0\right).
\end{eqnarray*}

Under the assumptions of $\frac{\partial Q_{jt=2}}{\partial\Psi_{jt=2}}\geq0$ and $\frac{\partial^{2}Q_{jt=2}}{\partial P_{jt=2}\partial\Psi_{jt=2}}\geq0$, $\frac{\partial^{2}V_{j}^{F}}{\partial\phi_{j}\partial P_{jt=2}}\geq0$ holds at the profit maximizing durability level. 

By the second order condition of firm's profit maximization regarding $\phi_{j}$, $\frac{\partial^{2}V_{j}^{F}}{\partial\phi_{j}^{2}}<0$ should hold. Hence, by implicit function theorem, $\frac{\partial\phi_{j}}{\partial P_{jt=2}}=-\frac{\partial^{2}V_{j}^{F}}{\partial\phi_{j}\partial P_{jt=2}}\slash\frac{\partial^{2}V_{j}^{F}}{\partial\phi_{j}^{2}}>0$ holds. It implies firm $j$ raises its product durability when $P_{jt=2}=p_{jt=2}$, product $j$'s price at time $t=2$, is higher.
\end{proof}

\subsection{Proof of Lemma \ref{lem:V_C_tilde}\label{subsec:Proof_Lemma_V_C_tilde} in Section \ref{subsec:Details_algorithm_supply} (Equivalence of consumers' value functions)}

\paragraph*{(a).}
\begin{proof}
By the definition of the term $\widetilde{V_{it}^{C}}(x_{it},B_{t})$,

\begin{eqnarray*}
\widetilde{V_{it}^{C}}(x_{it}=(j,\tau),B_{t}) & = & V_{it}^{C}(x_{it}=(j,\tau),B_{t})-\left[\sum_{s=0}^{\infty}\beta_{C}^{\tau}\psi_{j}\cdot\phi\left(i,\mu_{j},\tau+s|\tau\right)\right]\\
 & = & \psi_{j}+\phi(i,\mu_{j},\tau+1|\tau)\cdot V_{it}^{C}(x_{it}=(j,\tau+1),B_{t})+\left(1-\phi(i,\mu_{j},\tau+1|\tau)\right)\cdot V_{it}^{C}(x_{it}=\emptyset,B_{t})\\
 &  & -\left[\sum_{s=0}^{\infty}\beta_{C}^{\tau}\psi_{j}\cdot\phi\left(i,\mu_{j},\tau+s|\tau\right)\right]\\
 & = & \psi_{j}+\phi(i,\mu_{j},\tau+1|\tau)\cdot\widetilde{V_{it}^{C}}(x_{it}=(j,\tau+1),B_{t})+\left(1-\phi(i,\mu_{j},\tau+1|\tau)\right)\cdot\widetilde{V_{it}^{C}}(x_{it}=\emptyset,B_{t})+\\
 &  & \phi(i,\mu_{j},\tau+1|\tau)\cdot\left[\sum_{s=0}^{\infty}\beta_{C}^{\tau}\psi_{j}\cdot\phi\left(i,\mu_{j},\tau+1+s|\tau+1\right)\right]-\left[\sum_{s=0}^{\infty}\beta_{C}^{\tau}\psi_{j}\cdot\phi\left(i,\mu_{j},\tau+s|\tau\right)\right].
\end{eqnarray*}

Since $\phi(i,\mu_{j},\tau+1|\tau)\cdot\phi\left(i,\mu_{j},\tau+1+s|\tau+1\right)=\phi\left(i,\mu_{j},\tau+1+s|\tau\right)$, 
\[
\psi_{j}+\phi(i,\mu_{j},\tau+1|\tau)\cdot\left[\sum_{s=0}^{\infty}\beta_{C}^{\tau}\psi_{j}\cdot\phi\left(i,\mu_{j},\tau+1+s|\tau+1\right)\right]-\left[\sum_{s=0}^{\infty}\beta_{C}^{\tau}\psi_{j}\cdot\phi\left(i,\mu_{j},\tau+s|\tau\right)\right]=0
\]
 holds. Hence, $\widetilde{V_{it}^{C}}(x_{it},B_{t})$ satisfies the following, which is the counterpart of Bellman equation:

\begin{eqnarray*}
 &  & \widetilde{V_{it}^{C}}(x_{it},B_{t})\\
 & = & \begin{cases}
\phi(i,\mu_{j},\tau+1|\tau)\cdot\widetilde{V_{it}^{C}}(x_{it}=(j,\tau+1),B_{t})+\left(1-\phi(i,\mu_{j},\tau+1|\tau)\right)\cdot\widetilde{V_{it}^{C}}(x_{it}=\emptyset,B_{t}) & \text{if}\ x_{it}=(j,\tau),\\
E_{\epsilon}\left[\max_{j\in\mathcal{J}_{t}\cup\{0\}}\left(-\alpha_{i}p_{jt}+\delta_{jt}+\epsilon_{ijt}+\beta_{C}E_{x}\left[\widetilde{V_{it+1}^{C}}(x_{it+1},B_{t+1}(B_{t}))|x_{it}=\emptyset,B_{t},a_{it}=j\right]\right)\right] & \text{if}\ x_{it}=\emptyset.
\end{cases}
\end{eqnarray*}

\end{proof}

\paragraph*{(b).}
\begin{proof}
By the results of (a),

\begin{eqnarray*}
\widetilde{V_{it}^{C}}(x_{it}=(j,\tau),B_{t}) & = & \sum_{s=0}^{\infty}\left[\prod_{q=1}^{s}\phi(i,\mu_{j},\tau+q|\tau+q-1)\cdot(1-\phi(i,\mu_{j},\tau+s+1|\tau+s))\cdot\widetilde{V_{it+s+1}^{C}}(x_{it+s+1}=\phi)\right]\\
 & = & \sum_{s=0}^{\infty}\left[\phi(i,\mu_{j},\tau+s|\tau)\cdot(1-\phi(i,\mu_{j},\tau+s+1|\tau+s))\cdot\widetilde{V_{it+s+1}^{C}}(x_{it+s+1}=\phi)\right]\\
 & = & \sum_{s=0}^{\infty}\left[\left(\phi(i,\mu_{j},\tau+s|\tau)-\phi(i,\mu_{j},\tau+s+1|\tau)\right)\cdot\widetilde{V_{it+s+1}^{C}}(x_{it+s+1}=\phi)\right]\\
 & = & \sum_{s=0}^{\infty}\left[f(i,\mu_{j},\tau+s+1|\tau)\cdot\widetilde{V_{it+s+1}^{C}}(x_{it+s+1}=\phi)\right].
\end{eqnarray*}
\end{proof}

\paragraph*{(c).}
\begin{proof}
By the definition of the term $\widetilde{V_{it}^{C}}(x_{it},B_{t})$,

\begin{eqnarray*}
\widetilde{v_{ijt}}(x_{it}=\emptyset,\Omega_{t}^{C}) & = & -\alpha_{i}p_{jt}+\widetilde{\delta_{jt}}+\psi_{j}+\beta_{C}E_{x,\Omega^{C}}\left[V_{it+1}^{C}(x_{it+1},\Omega_{t+1}^{C})|x_{it}=\emptyset,\Omega_{t}^{C},a_{it}=j\right]\\
 & = & -\alpha_{i}p_{jt}+\widetilde{\delta_{jt}}+\psi_{j}+\beta_{C}\phi(i,\mu_{j},1|0)\cdot E_{\Omega^{C}}\left[V_{it+1}^{C}\left(x_{it+1}=(j,\tau=1),\Omega_{t+1}^{C}\right)|\Omega_{t}^{C}\right]+\\
 &  & \beta_{C}\left(1-\phi(i,\mu_{j},1|0)\right)\cdot E_{\Omega^{C}}\left[V_{it+1}^{C}\left(x_{it+1}=\emptyset,\Omega_{t+1}^{C}\right)|\Omega_{t}^{C}\right]\\
 & = & -\alpha_{i}p_{jt}+\widetilde{\delta_{jt}}+\psi_{j}+\beta_{C}\phi(i,\mu_{j},1|0)\cdot E_{\Omega^{C}}\left[\widetilde{V_{it+1}^{C}}\left(x_{it+1}=(j,\tau=1),\Omega_{t+1}^{C}\right)|\Omega_{t}^{C}\right]+\\
 &  & \beta_{C}\phi(i,\mu_{j},1|0)\cdot\left[\sum_{s=0}^{\infty}\beta_{C}^{\tau}\psi_{j}\cdot\phi(i,\mu_{j},1+s|1)\right]+\\
 &  & \beta_{C}\left(1-\phi(i,\mu_{j},1|0)\right)\cdot E_{\Omega^{C}}\left[V_{it+1}^{C}\left(x_{it+1}=\emptyset,\Omega_{t+1}^{C}\right)|\Omega_{t}^{C}\right]\\
 & = & -\alpha_{i}p_{jt}+\widetilde{\delta_{jt}}+\psi_{j}+\phi(i,\mu_{j},1|0)\cdot E_{\Omega^{C}}\left[\widetilde{V_{it+1}^{C}}\left(x_{it+1}=(j,\tau=1),\Omega_{t+1}^{C}\right)|\Omega_{t}^{C}\right]+\\
 &  & \left[\sum_{s=0}^{\infty}\beta_{C}^{\tau}\psi_{j}\cdot\phi(i,\mu_{j},1+s|0)\right]+\beta_{C}\left(1-\phi(i,\mu_{j},1|0)\right)\cdot E_{\Omega^{C}}\left[V_{it+1}^{C}\left(x_{it+1}=\emptyset,\Omega_{t+1}^{C}\right)|\Omega_{t}^{C}\right]\\
 & = & -\alpha_{i}p_{jt}+\delta_{jt}+\beta_{C}E_{x,\Omega^{C}}\left[\widetilde{V_{it+1}^{C}}(x_{it+1},\Omega_{t+1}^{C})|x_{it}=\emptyset,\Omega_{t}^{C},a_{it}=j\right].
\end{eqnarray*}

\end{proof}

\section{Data details\label{sec:Data-details}}

\subsubsection*{Market size and data coverage}

We need to specify the market size of the light bulb market in Japan, as in the standard BLP method. Since light bulbs can be installed only in the place where sockets exist, the number of sockets corresponds to the market size of the light bulb market. One of the largest light bulb producers Toshiba estimated the number of E26 sockets in the household sector to be 150 million (\citet{Toshiba_proposal_2009}) . Hence, I use the value. Note that I assume the market size of light bulbs is constant in the sample period, considering the little change in population size in Japan in the sample period.

The data covers the quantity sold at electronics retail stores (coverage rate: 98\%) and home-center stores (coverage rate: 50\%). Still, some fraction of consumers might have purchased bulbs in other types of stores, such as supermarkets or online stores. In that sense, the data might not cover all the sales of light bulbs. Hence, I calculate the quantity of each bulb product sold in Japan in the following way:
\begin{enumerate}
\item Calculate the total sales of E26 bulbs in the data and those published by JLMA in 2009\footnote{The aggregate sales data published by JLMA includes bulbs in the category other than E26 socket. \citet{Toshiba_proposal_2009} estimated the share of E26 type bulbs to be 54\%. Hence, I use the ratio to calculate the total sales of E26 bulbs in Japan when using JLMA data.}
\item Calculate the coverage rate of the data in 2009
\item divide the sales quantity in the data by the coverage rate in each year
\end{enumerate}
Note that the calculated average coverage rate of the data is roughly 75\%. 

\subsubsection*{Product characteristics}

Not all the products' characteristics are recorded in the POS data. Hence, I manually collected the characteristics mainly from the websites of manufacturing firms. However, some firms do not list their products on their website. I gathered the characteristics information of these products from the websites of online retailers.

\subsubsection*{Product selection}

Some E26 light bulb products are used for special purposes, such as decorations or lighting for constructions. Since we focus on the light bulb products for daily use by households, I selected the following types of light bulb products:
\begin{itemize}
\item Incandescent: bulbs with shapes A, with colors silica and clear
\item CFL: bulbs with shapes A, and T
\end{itemize}
Other light bulbs, such as ball bulbs and chandelier light bulbs are mainly used for special purposes. Hence, I omitted the bulbs in these categories. Note that the sales of the products omitted in the process is not so large (less than 10\% in quantity).

Note that light bulbs are mainly used by households. According to \citet{light_survey_2019}, more than 70 \% of them are used by them. Since I omit the lamps which are mainly used in other settings, we can assume light bulbs in our sample would be mainly used by household. 

In addition, I omitted the products whose sales are less than 1000 in each month. The purpose of this manipulation is the removal of products with too few demand.

\subsubsection*{Multiple units}

In the light bulb market, not all the bulbs are sold separately, and some of them are bundled as one product. For instance, two bulbs are sold as one product. However, to model the demand for multiple-unit products, we need to explicitly model the consumers' stockpiling behavior or interdependent utility of bulbs. This complicates the empirical model, and I abstract away from the point in this paper. In this study, we assume 2 bulbs are sold separately when a bundled product containing 2 bulbs is sold.

\section{Additional results\label{sec:Additional-results}}

\subsection{Estimation of demand parameters by linear GMM\label{subsec:Estimationg-of-demand_linear_GMM}}

In the case no consumer heterogeneity exists, we can consistently estimate the main parameters, including price coefficient, nest parameters, and consumer preference for durability, by a linear GMM with time dummies without solving the dynamic model.\footnote{Note that we cannot correctly compute price elasticities of demand without fully developing the dynamic structural model, though we can estimate the main utility parameters.} In this case, we don't have to impose assumptions neither on the state transition process nor on the formation of consumer expectations.

In the case where no consumer heterogeneity exists,\footnote{Since only one consumer type exists, we omit the subscript $i$ in the following. We also omit $\Omega_{t}^{C}$ and $x_{it}=\emptyset$ in $V^{C}$ for simple exposition.} 

\begin{eqnarray*}
\ln\left(S_{jt}\right) & = & \ln\left(s_{jt}^{(ccp)}\right)+\ln\left(Pr0_{t}\right)\\
 & = & \frac{\widetilde{v_{jt}}}{1-\rho_{g}}-\frac{\rho_{g}}{1-\rho_{g}}IV_{gt}^{C}-V_{t}^{C}+\ln\left(Pr0_{t}\right).
\end{eqnarray*}

We also have:

\begin{eqnarray*}
\ln\left(S_{j|g,t}\right) & = & \frac{\widetilde{v_{jt}}}{1-\rho_{g}}-\frac{IV_{gt}^{C}}{1-\rho_{g}}.
\end{eqnarray*}

Using these equations,

\[
\ln\left(S_{jt}\right)-\rho_{g}\ln\left(S_{j|g,t}\right)=\widetilde{v_{jt}}-V_{t}^{C}+\ln\left(Pr0_{t}\right).
\]

Then, we can derive the following estimation equation:

\begin{eqnarray*}
\ln\left(S_{jt}\right) & = & -\alpha p_{jt}+\delta_{jt}+E_{t}\left[\beta_{C}^{L_{j}}V_{t+L_{j}}^{C}\right]-V_{t}^{C}+\ln\left(Pr0_{t}\right)+\rho_{g}\ln\left(S_{j|g,t}\right).
\end{eqnarray*}

Under the additional assumption that $E_{t}\left[\beta_{C}^{L_{j}}V_{t+L_{j}}^{C}\right]\approx\overline{V^{C}}E_{t}\left[\beta_{C}^{L_{j}}\right]$,\footnote{In principle, we can consistently estimate the parameters by treating $E_{t}\left[\beta_{C}^{L_{j}}V_{t+L_{j}}^{C}\right]$ as time durability dummies. Nevertheless, under the specification, the estimated values of nest parameters were more than 1 though insignificant, which is not consistent with the model. It might be due to the insufficient number of products sharing the same durability levels and period.}we obtain the following equation:

\begin{eqnarray*}
\ln\left(S_{jt}\right) & = & -\alpha p_{jt}+\delta_{jt}+E_{t}\left[\beta_{C}^{L_{j}}\right]\overline{V^{C}}-V_{t}^{C}+\ln\left(Pr0_{t}\right)+\rho_{g}\ln\left(S_{j|g,t}\right)\\
 & = & -\alpha p_{jt}+X_{jt}^{D}\theta_{linear}^{D}+E_{t}\left[\beta_{C}^{L_{j}}\right]\overline{V^{C}}-V_{t}^{C}+\rho_{g}\ln\left(S_{j|g,t}\right)+\xi_{jt}\\
 & = & -\alpha p_{jt}+\widehat{X_{jt}^{D}}\widehat{\theta_{linear}^{D}}+\widetilde{c_{t}}+\rho_{g}\ln\left(S_{j|g,t}\right)+\xi_{jt}.
\end{eqnarray*}

Here, $\widetilde{c_{t}}$ is defined by $\widetilde{c_{t}}\equiv-V_{t}^{C}+\ln\left(Pr0_{t}\right)$, and $\widehat{X_{jt}^{D}}\widehat{\theta_{linear}^{D}}$ satisfies $\widehat{X_{jt}^{D}}\widehat{\theta_{linear}^{D}}=\delta_{jt}+E_{t}\left[\beta_{C}^{L_{j}}V_{t+L_{j}}^{C}\right]+\overline{V^{C}}E_{t}\left[\beta_{C}^{L_{j}}\right]$ and $\widehat{X_{jt}^{D}}$ include durability dummies. We can consistently estimate main parameters $\alpha,\{\rho_{g}\}_{g\in\mathcal{G}}$ by linear GMM method.

Table \ref{tab:static_demand_est} shows the results, and they show static estimates are similar to the results fully specifying dynamic demand structures in the case without random coefficients.

\begin{table}[H]
\begin{centering}
\begin{tabular}{ccccc}
\hline 
 & \multicolumn{2}{c}{Static Est.} & \multicolumn{2}{c}{Dynamic est.}\tabularnewline
 & Est. & SE & Est. & SE\tabularnewline
\hline 
\hline 
$\alpha:$ price coef. (yen/1000) & 2.558 & 0.346 & 2.562 & 0.346\tabularnewline
$\rho_{Inc}:$ nest parameter (incandescent) & 0.961 & 0.011 & 0.961 & 0.011\tabularnewline
$\rho_{CFL}:$ nest parameter (CFL) & 0.701 & 0.035 & 0.700 & 0.035\tabularnewline
\hline 
\end{tabular}
\par\end{centering}
\caption{Parameter Estimates (The case without random coefficients)\label{tab:static_demand_est}}
\end{table}

\subsection{Results of CFLs (market structure)\label{subsec:CFLs-results-market-structure}}
\begin{center}
{\scriptsize{}}
\begin{table}[H]
\small
\begin{centering}
{\small{}}%
\begin{tabular}{ccccc}
\hline 
 & {\small{}(1)} & {\small{}(2)} & {\small{}(3)} & {\small{}(4)}\tabularnewline
{\small{}Panasonic} & {\small{}6000h \& 10000h} & {\small{}6000h only} & {\small{}6000h only} & {\small{}6000h \& 10000h}\tabularnewline
{\small{}Toshiba} & {\small{}6000h \& 12000h} & {\small{}6000h only} & {\small{}6000h \& 12000h} & {\small{}12000h only}\tabularnewline
\hline 
\hline 
{\small{}Joint profit} & {\small{}24.67} & {\small{}23.91} & {\small{}24.53} & {\small{}24.19}\tabularnewline
{\small{}Profit (Panasonic)} & {\small{}10.77} & {\small{}10.06} & {\small{}9.65} & {\small{}11.24}\tabularnewline
{\small{}Profit (Toshiba)} & {\small{}13.9} & {\small{}13.86} & {\small{}14.88} & {\small{}12.95}\tabularnewline
\hline 
{\small{}No inventory consumers (\%)} & {\small{}18.61} & {\small{}20.43} & {\small{}19.33} & {\small{}19.55}\tabularnewline
{\small{}Average price (1000h Inc.; yen)} & {\small{}94.73} & {\small{}94.15} & {\small{}94.05} & {\small{}94.8}\tabularnewline
{\small{}Average price (2000h Inc.; yen)} & {\small{}172.57} & {\small{}171.96} & {\small{}171.81} & {\small{}172.71}\tabularnewline
{\small{}Average price (CFL; yen)} & {\small{}796.53} & {\small{}747.61} & {\small{}777.03} & {\small{}772.14}\tabularnewline
{\small{}Disposal (million)} & {\small{}3.04} & {\small{}3.32} & {\small{}3.15} & {\small{}3.19}\tabularnewline
\hline 
{\small{}$\Delta$CS} & {\small{}-} & {\small{}-2.65} & {\small{}-1.68} & {\small{}-0.96}\tabularnewline
{\small{}$\Delta$PS (excluding fixed cost)} & {\small{}-} & {\small{}-0.69} & {\small{}-0.1} & {\small{}-0.46}\tabularnewline
{\small{}$\Delta$TS (excluding Ext. / fixed costs)} & {\small{}-} & {\small{}-3.34} & {\small{}-1.78} & {\small{}-1.42}\tabularnewline
{\small{}$\Delta$Externality (electricity usage)} & {\small{}-} & {\small{}0.16} & {\small{}0.15} & {\small{}0.01}\tabularnewline
{\small{}$\Delta$Externality (waste disposal)} & {\small{}-} & {\small{}0.02} & {\small{}0.00} & {\small{}0.01}\tabularnewline
{\small{}$\Delta$TS (excluding fixed costs)} & {\small{}-} & {\small{}-3.52} & {\small{}-1.93} & {\small{}-1.44}\tabularnewline
{\small{}Upper bound of Fixed cost savings} & {\small{}-} & {\small{}1.95} & {\small{}1.05} & {\small{}0.9}\tabularnewline
\hline 
\end{tabular}{\small\par}
\par\end{centering}
\begin{centering}
{\scriptsize{}\caption{Effect of eliminating high durability CFLs (The case without collusion on prices)\label{tab:Durability-cartel-table-2}}
}{\scriptsize\par}
\par\end{centering}
    {\footnotesize{}Notes. }{\footnotesize\par}
        
    {\footnotesize{}The values at the top of the table show the discounted
    sum of variable profits of two dominant firms evaluated at $B^{(stationary,base)}$.}{\footnotesize\par}

    {\footnotesize{}The values at the bottom show the changes in surpluses relative to the base setting, and are also evaluated
    at $B^{(stationary,base)}$. Units of these values are billion yen.}{\footnotesize\par}
    
    {\footnotesize{}The values of economic variables in the middle part
    of the table are evaluated at the stationary state. Average prices are sales-weighted. The row of disposal represents the number of products disposed of in each period.}{\footnotesize\par}
    
    {\footnotesize{}$CS,PS,Ext,TS$ denote consumer surplus, producer surplus, externalties, and total surplus, respectively. $\Delta TS$ is defined by $\Delta TS\equiv\Delta CS+\Delta PS-\Delta Ext$.}{\footnotesize\par}

\end{table}
{\scriptsize\par}
\par\end{center}

\begin{center}
{\scriptsize{}}
\begin{table}[H]
\begin{centering}
\begin{tabular}{ccc}
\hline 
 & (1) & (2)\tabularnewline
Panasonic & 6000h \& 10000h & 6000h only\tabularnewline
Toshiba & 6000h \& 12000h & 6000h only\tabularnewline
\hline 
\hline 
Joint profit & \textsf{24.67} & \textsf{24.07}\tabularnewline
Profit (Panasonic) & 10.77 & 10.04\tabularnewline
Profit (Toshiba) & \textsf{13.9} & \textsf{14.03}\tabularnewline
\hline 
No inventory consumers (\%) & 18.61 & 20.45\tabularnewline
Disposal (million) & 3.04 & 3.33\tabularnewline
\hline 
\end{tabular}
\par\end{centering}
{\scriptsize{}\caption{Effect of eliminating high durability CFLs (Role of prices; The case without collusion on prices)\label{tab:Durability-cartel-p-fixed-table-2}}
}{\scriptsize\par}

{\footnotesize{}Notes.}{\footnotesize\par}

{\footnotesize{}The values at the top of the table show the discounted sum of profits of two dominant firms evaluated at $B^{(stationary,base)}$. Units of these values are billion yen.}{\footnotesize\par}

{\footnotesize{}The values of economic variables at the bottom part of the table are evaluated at the stationary state. The row of disposal represents the number of products disposed of in each period.}{\footnotesize\par}

{\footnotesize{}``Fixed'' in the row of prices implies that the column evaluates the results where product prices are fixed at the same levels as the ones in Column (1). ``Not fixed'' implies that the column evaluates the outcomes where product prices are adjusted so as to maximize each firm's profit based on the set of products.}{\footnotesize\par}

\end{table}
{\scriptsize\par}
\par\end{center}

\begin{center}
{\scriptsize{}}
\begin{table}[H]
\small

{\small{}}%
\begin{tabular}{cccccc}
\hline 
 & {\small{}(0)} & {\small{}(1)} & {\small{}(2)} & {\small{}(3)} & {\small{}(4)}\tabularnewline
{\small{}Price cartel} & {\small{}No} & {\small{}Yes} & {\small{}Yes} & {\small{}Yes} & {\small{}Yes}\tabularnewline
{\small{}Panasonic} & {\small{}6000h \& 10000h} & {\small{}6000h \& 10000h} & {\small{}6000h only} & {\small{}6000h only} & {\small{}6000h \& 10000h}\tabularnewline
{\small{}Toshiba} & {\small{}6000h \& 12000h} & {\small{}6000h \& 12000h} & {\small{}6000h only} & {\small{}6000h \& 12000h} & {\small{}6000h only}\tabularnewline
\hline 
\hline 
{\small{}Joint profit} & {\small{}24.67} & {\small{}40.42} & {\small{}38.05} & {\small{}39.45} & {\small{}39.2}\tabularnewline
{\small{}Profit (Panasonic)} & {\small{}10.77} & {\small{}17.47} & {\small{}15.95} & {\small{}14.69} & {\small{}18.91}\tabularnewline
{\small{}Profit (Toshiba)} & {\small{}13.9} & {\small{}22.96} & {\small{}22.1} & {\small{}24.76} & {\small{}20.29}\tabularnewline
\hline 
{\small{}No inventory consumers (\%)} & {\small{}18.61} & {\small{}20.67} & {\small{}22.7} & {\small{}21.19} & {\small{}21.85}\tabularnewline
{\small{}Average price (1000h Inc.; yen)} & {\small{}94.73} & {\small{}126.74} & {\small{}127.02} & {\small{}126.84} & {\small{}126.85}\tabularnewline
{\small{}Average price (2000h Inc.; yen)} & {\small{}172.57} & {\small{}233.57} & {\small{}231.37} & {\small{}232.71} & {\small{}232.32}\tabularnewline
{\small{}Average price (CFL; yen)} & {\small{}796.53} & {\small{}989.49} & {\small{}907.7} & {\small{}952.41} & {\small{}956.26}\tabularnewline
{\small{}Disposal (million)} & {\small{}3.04} & {\small{}3.21} & {\small{}3.51} & {\small{}3.28} & {\small{}3.39}\tabularnewline
\hline 
{\small{}$\Delta$CS} & {\small{}-} & {\small{}-23.17} & {\small{}-24.7} & {\small{}-23.82} & {\small{}-23.93}\tabularnewline
{\small{}$\Delta$PS (excluding fixed cost)} & {\small{}-} & {\small{}17.79} & {\small{}15.55} & {\small{}16.87} & {\small{}16.64}\tabularnewline
{\small{}$\Delta$TS (excluding Ext. / fixed costs)} & {\small{}-} & {\small{}-5.38} & {\small{}-9.14} & {\small{}-6.95} & {\small{}-7.3}\tabularnewline
{\small{}$\Delta$Externality (electricity usage)} & {\small{}-} & {\small{}-0.77} & {\small{}-0.71} & {\small{}-0.73} & {\small{}-0.74}\tabularnewline
{\small{}$\Delta$Externality (waste disposal)} & {\small{}-} & {\small{}-0.02} & {\small{}0.01} & {\small{}-0.01} & {\small{}0.00}\tabularnewline
{\small{}$\Delta$TS (excluding fixed costs)} & {\small{}-} & {\small{}-4.59} & {\small{}-8.45} & {\small{}-6.2} & {\small{}-6.55}\tabularnewline
{\small{}Upper bound of Fixed cost savings} & {\small{}-} & {\small{}0} & {\small{}1.95} & {\small{}1.05} & {\small{}0.9}\tabularnewline
\hline 
\end{tabular}{\scriptsize{}\caption{Effect of eliminating high durability CFLs (The case with collusion on prices)\label{tab:Durability-merger-table-2}}
}{\scriptsize\par}

    {\footnotesize{}Notes. }{\footnotesize\par}
        
    {\footnotesize{}The values at the top of the table show the discounted
    sum of variable profits of two dominant firms evaluated at $B^{(stationary,base)}$.}{\footnotesize\par}

    {\footnotesize{}The values at the bottom show the changes in surpluses relative to the base setting, and are also evaluated
    at $B^{(stationary,base)}$. Units of these values are billion yen.}{\footnotesize\par}
    
    {\footnotesize{}The values of economic variables in the middle part
    of the table are evaluated at the stationary state. Average prices are sales-weighted. The row of disposal represents the number of products disposed of in each period.}{\footnotesize\par}
    
    {\footnotesize{}$CS,PS,Ext,TS$ denote consumer surplus, producer surplus, externalties, and total surplus, respectively. $\Delta TS$ is defined by $\Delta TS\equiv\Delta CS+\Delta PS-\Delta Ext$.}{\footnotesize\par}

\end{table}
{\scriptsize\par}
\par\end{center}

\section{Considerations on the specifications\label{sec:Considerations-specifications}}

\subsubsection*{Future expectations on the periods after the sample periods}

Though LED lamps, which are largely different from incandescent lamps and CFLs, appeared in the market in the latter half of 2009, it was observed that the diffusion of LED lamps was slow, especially before the Great East Japan Earthquake in 2011 (METI (2011)\footnote{https://www.meti.go.jp/committee/summary/0004296/pdf/001\_05\_00.pdf}). Hence, the assumption of stationarity after the terminal period (June 2009) would be justifiable.

\subsubsection*{Continuous durability choice}

Many previous theoretical studies have specified models where firms continuously adjust their product durability levels. Nevertheless, based on the light bulb market data, we cannot see clear evidence that firms continuously adjust product durability. Rather, product durability levels seem to be discrete, and the decisions of which products to introduce would be more important in the market. Though we cannot rule out the possibility that firms continuously adjust product durability, it is not possible to estimate the cost structure using the observed data, since small continuous variation in product characteristics exists. Consequently, we cannot reliably evaluate firms' incentives on continuous durability choices, and I did not evaluate them in this study. 

\subsubsection*{Specifications of the utility function: Nested structure}

In this study, I employ the demand model with nested structure, and allow for the case where the values of nest parameters $\rho_{g}$ are not the same across nests. As a preliminary analysis, I estimated the following equation for incandescent lamps and CFLs separately by IV method:

\[
\ln S_{jt}^{(data)}=-\alpha p_{jt}+\widehat{X_{jt}}\theta+c_{t}+\xi_{jt}.
\]

where $c_{t}$ denotes time fixed effect. The results show that price coefficients are largely heterogeneous across the two product groups: price coefficient for incandescent lamps was around -50, but that for CFLs was around -5. To allow for such heterogeneous price sensitivity across different nests, I introduced heterogeneous values of nest parameters $\rho_{g}$.

\bibliographystyle{apalike}
\bibliography{literature_light_bulb}

\end{document}